\documentclass[letterpaper, 11pt]{article}
\usepackage{graphicx}
\usepackage{etoolbox}
\usepackage{amsmath}
\usepackage{amsfonts}
\usepackage{amsthm}
 \usepackage{amscd}
\usepackage{mathrsfs}
\usepackage{caption}
\usepackage{subcaption}
\usepackage{float}
\usepackage{hyperref}
\usepackage[top=1in,left=1in,right=1in,bottom=1in]{geometry}
\usepackage{amsmath,amsthm,amssymb,accents}

\newtheorem{theorem}{Theorem}[section]
\newtheorem{proposition}[theorem]{Proposition}

\newtheorem{corollary}[theorem]{Corollary}
\newtheorem{definition}[theorem]{Definition}
\newtheorem{example}[theorem]{Example}

\numberwithin{equation}{section}
\hypersetup{
	colorlinks = true,
	linkcolor  = red,
	citecolor  = green,
	filecolor  = cyan,
	urlcolor   = magenta,}

\title{Factorization for the matrix-valued general Jacobi system on the full-line lattice}
\author{Tuncay Aktosun\\
Department of Mathematics,
University of Texas at Arlington\\
Arlington, TX 76019-0408, USA\\
\\
Abdon E. Choque-Rivero\\
Instituto de F\'isica y Matem\'aticas,
Universidad Michoacana de San Nicolás de Hidalgo\\
Morelia, Michoac\'an, M\'exico\\
\\
Vassilis G. Papanicolaou\\
Department of Mathematics,
National Technical University of Athens, Zografou Campus\\
157 80 Athens, Greece\\
\\
Mehmet Unlu\\
Department of Mathematics,
Recep Tayyip Erdogan University\\
53100 Rize, Turkey\\
\\
Ricardo Weder\thanks{corresponding author: weder@unam.mx}\\
Departamento de F\'\i sica Matem\'atica\\
Instituto de Investigaciones en
Matem\'aticas Aplicadas y en Sistemas\\
Universidad Nacional Aut\'onoma de M\'exico,
Apartado Postal 20-126, IIMAS-UNAM\\
 Ciudad de M\'exico, CP 01000, M\'exico}

\date{}

\begin{document}

\maketitle

\begin{abstract}

\noindent The Jacobi system with matrix-valued
coefficients and with the spectral parameter
depending on a matrix-valued weight factor is considered on the full-line lattice.
The scattering from the full-line lattice is expressed in terms of the scattering from the fragments
of the whole lattice by developing a factorization formula for the corresponding transition
matrices.
In particular, the matrix-valued transmission and reflection coefficients
for the full-line lattice are explicitly expressed in terms of the scattering coefficients
for the left and right lattice fragments.
Since the matrix-valued scattering coefficients are easier to determine for
the fragments than for the full-line lattice, the factorization formula presented provides a method to determine
the scattering coefficients for full-line lattices. The theory presented is illustrated with various explicit examples, including an example
demonstrating that the matrix-valued left transmission coefficient in general is not equal to  
the matrix-valued right transmission coefficient for a lattice.

\end{abstract}



\newpage

\section{Introduction}\label{section1}
Consider the matrix-valued general Jacobi system
\begin{equation}\label{1.1}
a(n+1)\,\psi(n+1)+b(n)\,\psi(n)+a(n)^\dagger\,\psi(n-1)=\lambda\,w(n)\,\psi(n),
\qquad n\in\mathbb Z,
\end{equation}
where the spacial coordinate $n$ takes values on the set $\mathbb Z$ of integers,
$\lambda$ is the spectral parameter, the dagger denotes the matrix adjoint (matrix transpose and complex conjugation),
and the coefficients $a(n),$ $b(n),$ and $w(n)$ are some $q\times q$ matrix-valued functions of $n$ with complex-valued entries. Here, $q$ is a fixed 
positive integer. We use solutions to \eqref{1.1} that are either $q \times q$ matrices or column vectors with $q$ components.

The scalar case occurs when $q=1.$ There are two important special versions of \eqref{1.1} in the scalar case.
The first is the Jacobi system on the full-line lattice with $w(n)\equiv 1,$ and it is given by 
\begin{equation}\label{1.2}
a(n+1)\,\psi(n+1)+b(n)\,\psi(n)+a(n)\,\psi(n-1)=\lambda\,\psi(n),
\qquad n\in\mathbb Z,
\end{equation}
where the coefficients $a(n)$ and $b(n)$ are real valued. The second special version in the scalar case is the discrete 
Schr\"odinger equation on the full-line lattice, and it is given by 
\begin{equation}\label{1.3}
-\psi(n+1)+\left[2+V(n)\right]\psi(n)- \psi(n-1)=\lambda\,\psi(n),
\qquad n\in\mathbb Z,
\end{equation}
where $V(n)$ is a real-valued potential. Due to the presence of the matrix $w(n)$ in \eqref{1.1},
we refer to \eqref{1.1} as the general Jacobi system. 

Our goal in this paper is to analyze certain aspects related to the direct scattering problem for \eqref{1.1}. That direct 
problem consists of the determination of the matrix-valued scattering coefficients and the bound-state information when the 
coefficients in \eqref{1.1} are specified in an appropriate class. In our paper we do not consider the determination of the 
bound-state information, and we only concentrate on the determination of the scattering coefficients.
In particular, we are interested in partitioning the full-line lattice $\mathbb Z$ into a finite 
number of fragments and then relating the scattering from the full-line lattice to the scattering from those individual fragments.
There are various reasons to analyze such a lattice partitioning. One reason is that it is easier to determine the scattering from the 
individual fragments than the scattering from the whole lattice. This is because it is easier to determine the relevant 
particular solutions associated with the fragments
than the full-line lattice.
Another reason is that the partitioning helps us understand and visualize how the whole scattering is obtained from the accumulation of the scattering from the 
individual fragments.

For the factorization of the scattering data for
the one-dimensional Schr\"odinger equation in the continuous case, we refer the reader
to \cite{A1992}. The factorization results in the continuous case
for  one-dimensional Schr\"odinger-type equations are available in
\cite{AKV1996}. For the full-line lattice factorization results in the scalar case
with real-valued coefficients for the general Jacobi system, we refer the reader to
\cite{AC2017}. In the continuous case, the factorization result
 for the one-dimensional continuous Schr\"odinger equation with a position-dependent mass
can be found in
\cite{SV1995}. In the continuous case for the radial Schr\"odinger equation, we refer
the reader to \cite{A2000} for
the factorization of the scattering data.
The factorization results  
for continuous quantum graphs without potentials
can be found in \cite{KS2001}.
 We refer the reader to \cite{AW2023} for the factorization results in the continuous case 
for the matrix-valued Schr\"odinger equations on the full line and on the half line and
for the resulting unitary equivalence between the full-line and the half-line cases.  
A general introduction to Jacobi systems can be found in \cite{T2000}.
For some recent results on Jacobi equations
related to the threshold behavior and Levinson's theorems
we refer the reader to
\cite{BFS2021,BFGS2022,BFNS2024}, 
and we refer the reader to
\cite{SSVW2025} for the study of  the direct scattering problem for
the Jacobi system with operator-valued coefficients.

We assume that the $q \times q$ matrix-valued coefficients appearing in \eqref{1.1} have their spacial asymptotics given by 
\begin{equation}\label{1.4}
\lim_{n\to \pm\infty}a(n)=a_\infty\, I,
\quad
\lim_{n\to \pm\infty}b(n)=b_\infty\, I,
\quad
\lim_{n\to \pm\infty}w(n)=w_\infty\, I,
\end{equation}
where $I$ denotes the $q \times q$ identity matrix and the quantities $a_\infty,$ $b_\infty,$ $w_\infty$ are some scalars.
In order to have a manageable scattering theory for \eqref{1.1}, we impose certain restrictions on the coefficient matrices $a(n),$
$b(n),$ $w(n)$ and their spacial asymptotics. For that reason, we consider \eqref{1.1} when the
the matrix-valued coefficients 
in \eqref{1.1} belong to the class $\mathcal A$ introduced in the following definition.

\begin{definition}\label{definition1.1}
The $q\times q$ matrix-valued coefficients $a(n),$ $b(n),$ $w(n)$ in \eqref{1.1} with their spacial asymptotics determined by the
scalars $a_\infty,$ $b_\infty,$ $w_\infty$ in \eqref{1.4} belong to the class $\mathcal A$ if the following restrictions are satisfied: 

\begin{enumerate}
\item[\text{\rm(a)}] The
matrices $b(n)$ and $w(n)$ are selfadjoint, i.e. we have
\begin{equation}
\label{1.5}
b(n)^\dagger=b(n),\quad w(n)^\dagger=w(n),
\qquad n\in\mathbb Z.
\end{equation}
On the other hand, the matrix $a(n)$ is not necessarily selfadjoint.

\item[\text{\rm(b)}] The matrix $a(n)$ is invertible and the matrix $w(n)$ is positive for
all $n\in\mathbb Z,$ where by a positive matrix we mean a positive definite matrix.

\item[\text{\rm(c)}] The scalar $w_\infty$ is positive, $b_\infty$ is real, and $a_\infty$ is real and nonzero. Based on the comparison of
\eqref{1.1} with \eqref{1.2} and \eqref{1.3}, we refer to \eqref{1.1} as Jacobi-like when $a_\infty>0$ and 
as Schr\"odinger-like when $a_\infty<0,$ respectively.

\item[\text{\rm(d)}] The matrix-valued coefficients $a(n),$ $b(n),$ $w(n)$ and the scalars $a_\infty,$ $b_\infty,$ $w_\infty$
satisfy the summability condition given by
\begin{equation}
\label{1.6}
\displaystyle\sum_{n=-\infty}^\infty  |n|  \left( ||\mathbf P(n)||+ ||\mathbf Q(n)|| \right)<+\infty,
\end{equation}
where we have defined
\begin{equation}
\label{1.7}
\mathbf P(n):=\displaystyle\frac{w_\infty}{a_\infty}\, w(n-1)^{-1/2} a(n)\, w(n)^{-1/2}- I,
\end{equation}
\begin{equation}
\label{1.8}
\mathbf Q(n):=\displaystyle\frac{w_\infty}{a_\infty}\, w(n)^{-1/2} b(n)\, w(n)^{-1/2}-\displaystyle\frac{b_\infty}{a_\infty}\,I,
\end{equation}
and we use the operator norm $||\cdot||$ in \eqref{1.6}  even though any other norm for a $q\times q$ matrix can be used there.
This is because all norms for a $q\times q$ matrix are equivalent.

\end{enumerate}
\end{definition}

Since $w(n)$ is assumed to be positive, the matrix $w(n)^{1/2}$ is uniquely defined as the positive matrix satisfying
\begin{equation*}
w(n)^{1/2}\,w(n)^{1/2}=w(n),\qquad n\in\mathbb Z.
\end{equation*}
Because the $q\times q$ matrix $w(n)^{1/2}$ is positive, its matrix inverse exists, and we use $w(n)^{-1/2}$ to denote that 
matrix inverse.

Our paper is organized as follows. 
In Section~\ref{section2} we transform the Jacobi system \eqref{1.1} into the simpler Jacobi system given in \eqref{2.9}.
We prove that the linear operators in those two Jacobi systems are bounded and selfadjoint when the coefficients 
in \eqref{1.1} belong to the class $\mathcal A$ or equivalently when the coefficients in \eqref{2.9}
belong to the class $\tilde{\mathcal A}$ described in Definition~\ref{definition2.1}.
We also introduce the auxiliary spectral parameter $z$ and
express the Jacobi systems \eqref{1.1} and \eqref{2.9} in terms of $z$ in the forms given 
in \eqref{2.17} and \eqref{2.15}, respectively.
In Section~\ref{section3}, the use of $z$ enables us to 
introduce the Jost solutions,
the scattering coefficients, and the scattering matrix
and to
describe their basic properties.
Furthermore, it allows us to
relate the Jost solutions to \eqref{2.15} and \eqref{2.17} to each other in a simple way as listed 
in \eqref{3.8} and \eqref{3.9}.
Consequently, as shown in \eqref{3.14} the scattering coefficients for
\eqref{2.15} and those for \eqref{2.17} become equivalent. 
In Section~\ref{section4} we introduce
the left and right transition matrices for \eqref{2.17} 
in terms of the corresponding scattering coefficients.
We also present the relevant properties
of those transition matrices.
In Section~\ref{section5} we state and prove our factorization results for the transition matrices 
when the full-line lattice is partitioned into two or more fragments.
The factorization formulas presented in Section~\ref{section5} also allow us
to express each scattering coefficient for the full-line lattice in terms of the
scattering coefficients corresponding to the fragments.
Finally, in Section~\ref{section6} we 
provide some explicit examples 
of the transition matrices corresponding to a nonhomogeneity
concentrated at a single lattice point.
We also illustrate some cases where the determinants
of the left and right transmission coefficients are equal or unequal.

The establishment of the factorization formulas presented in our paper requires the development of many preliminary results, especially due to the fact that the matrix-valued coefficient $a(n)$ for $n\in\mathbb Z$ is not assumed to be 
selfadjoint. To help our reader concentrate on the factorization result with less emphasis on the preliminary background material, we provide the following guidance. We write the matrix-valued general Jacobi system \eqref{1.1} in the form \eqref{2.17} by using the auxiliary spectral parameter $z,$ where
$z$ is related to the spectral parameter $\lambda$ as in \eqref{2.16}. This is needed because it is simpler to express the scattering coefficients in $z$ rather than in $\lambda.$ We partition the full-line lattice $\mathbb Z$ into its two fragments $\mathbb Z_1$ and $\mathbb Z_2$ as indicated in \eqref{5.1}--\eqref{5.4}.
Associated with the fragments $\mathbb Z_1$ and $\mathbb Z_2,$ we introduce the left transition matrices $\Lambda_1(z)$ and
$\Lambda_2(z)$ defined in \eqref{5.12} and \eqref{5.13}, respectively. The left transition matrix $\Lambda(z)$
associated with the whole lattice $\mathbb Z$ is already defined in \eqref{4.8}. Each of these left transition matrices is uniquely determined by the corresponding scattering coefficients.
We then present our main factorization result in \eqref{5.41} by showing that the matrix $\Lambda(z)$
is equal to the ordered matrix product $\Lambda_1(z)\,\Lambda_2(z).$
This factorization result is generalized in Corollary~\ref{corollary5.4} when the full-line lattice is partitioned into an arbitrary number of fragments. The physical significance of the factorization formula is that it shows how the scattering from the whole lattice is composed of the scattering from its fragments in an orderly way as we move from the left to the right.
Similarly, the factorization formula \eqref{5.42} involving the right transition matrices shows how the scattering from the whole lattice is composed of the scattering from its fragments in an orderly way as we move from the right to the left.
The result in \eqref{4.12} shows that a right transition matrix is the matrix inverse of the corresponding left transition matrix. Hence, the two factorization formulas in \eqref{5.41} and \eqref{5.42} are complementary.
We remark that an elementary fragment of the whole lattice $\mathbb Z$ consists of a single lattice point.
Thus, the knowledge of the transition matrix corresponding to an elementary fragment enables us to determine the transition matrix for a nonelementary fragment by using our factorization formulas. In Section~\ref{section6} we present some
explicit examples of transition matrices corresponding to some elementary fragments.
Let us mention that, as seen from \eqref{4.8}, a left transition matrix is constructed by using the corresponding left transmission and left reflection coefficients. Similarly, as seen from \eqref{4.9}, a right transition matrix is constructed by using the corresponding right transmission and right reflection coefficients. On the other hand, the determinant of either transition matrix, as indicated in \eqref{4.13} and \eqref{4.14}, is constructed by using only the ratio of the determinants of the left and right transmission coefficients. This prompts us to investigate how the determinants of the left and right transmission coefficients are related to each other.
This is investigated in Section~\ref{section4} and also illustrated in some examples in Section~\ref{section6}.

We finally remark that the Jacobi systems are extensively used in physics. Two such examples of the uses are in tight-binding models in solid-state physics describing electrons tightly bound to atoms in crystals and the Jaynes--Cummings model in quantum optics
describing a two-level atomic system interacting with a quantized mode of an optical cavity. Our method can be used to compute the scattering coefficients in such models and in other similar applications. A survey of applications of Jacobi systems in various areas of physics
can be found in \cite{CFKS1987}.

\section{Preliminaries}
\label{section2}

In this section, we transform the general Jacobi system \eqref{1.1} into the Jacobi system \eqref{2.9} in which the coefficients are simpler.
We provide the explicit transformations for the coefficients and the spectral parameters by using a tilde to identify the quantities related to the transformed Jacobi system.
In analogy with the class $\mathcal A$ described in Definition~\ref{definition1.1}, we introduce the class $\tilde{\mathcal A}$ in Definition~\ref{definition2.1}.
We show that the coefficients in \eqref{1.1} belong to the class $\mathcal A$ if and only if
the coefficients in \eqref{2.9} belong to the class $\tilde{\mathcal A}.$ For those two classes, we establish the boundedness and selfadjointness of the linear operators associated with \eqref{1.1} and \eqref{2.9}, respectively.
We then introduce the auxiliary spectral parameter $z$ and express 
\eqref{1.1} and \eqref{2.9} as the Jacobi systems \eqref{2.17} and \eqref{2.15}, respectively, in $z.$
The use of $z$ enables us later in Section~\ref{section3} to introduce the relevant scattering solutions and scattering
coefficients for those two Jacobi systems.

We can write \eqref{1.1} as the eigenvalue equation
$\mathbf H \psi=\lambda\psi,$ 
with the action of the operator 
$\mathbf H$ on
$\psi$ described by
\begin{equation*}
(\mathbf H\psi)(n)=w(n)^{-1}\left[a(n+1)\,\psi(n+1)+b(n)\,\psi(n)+a(n)^\dagger\,\psi(n-1)\right],  \qquad n\in\mathbb Z,
\end{equation*}
where the operator $\mathbf H$ acts on the weighted Hilbert space $\ell_w^2(\mathbb Z)$ consisting of
complex-valued sequences $\psi$ with values $\psi(n)$ for $n\in\mathbb Z.$ The inner product on that Hilbert space is given by
\begin{equation*}
\langle \phi,\psi\rangle:=\sum_{n=-\infty}^\infty \phi(n)^\dagger w(n)\,\psi(n),
\end{equation*}
with the implied norm given by
\begin{equation*}
||\psi||:=\sqrt{\sum_{n=-\infty}^\infty \psi(n)^\dagger\,w(n)\,\psi(n)}.
\end{equation*}
It can be verified directly that $\mathbf H$ is bounded and selfadjoint
when the matrices $b(n)$ are selfadjoint for $n\in\mathbb Z,$
the sequences $\{||a(n)||\}_{n\in\mathbb Z}$ and 
$\{||b(n)||\}_{n\in\mathbb Z}$ are bounded, and
the matrices $w(n)$ for $n\in\mathbb Z$ satisfy
$c_1 I-w(n)\ge 0$ and $w(n)-c_2 I\ge 0$
for some positive constants $c_1$ and $c_2,$
where we recall that a square matrix is nonnegative if and only if
each of its eigenvalues is positive or zero.
In particular, when the coefficients in \eqref{1.1} belong to the class $\mathcal A$
described in
Definition~\ref{definition1.1}, the operator
$\mathbf H$ is bounded and selfadjoint.

It is possible to transform \eqref{1.1} into a simpler system where the weight factor $w(n)$ no longer appears and where
$a_\infty=1$ and
$b_\infty=0.$ For this, 
we proceed as follows. We transform the $q\times q$ matrix-valued coefficients in \eqref{1.1} to the $q\times q$ matrix-valued 
coefficients $\tilde a(n)$ and $\tilde b(n),$ transform the wavefunction $\psi(n)$ to the wavefunction $\tilde\psi(n),$ and transform the 
spectral parameter $\lambda$ to the spectral parameter $\tilde\lambda$ through the assignments
\begin{equation}
\label{2.4} 
\tilde a(n)=\displaystyle\frac{w_\infty}{a_\infty}\, w(n-1)^{-1/2}\, a(n)\, w(n)^{-1/2}, \qquad n\in\mathbb Z,
\end{equation}
\begin{equation}
\label{2.5} 
\tilde b(n)=\displaystyle\frac{w_\infty}{a_\infty}\, w(n)^{-1/2}\, b(n)\, w(n)^{-1/2}-\displaystyle\frac{b_\infty}{a_\infty}\,I, \qquad n\in\mathbb Z,
\end{equation}
\begin{equation}
\label{2.6} 
\tilde\psi(n)=\displaystyle\frac{w(n)^{1/2}}{w_\infty^{1/2}}\, \psi(n), \qquad n\in\mathbb Z,
\end{equation}
\begin{equation}
\label{2.7} 
\tilde\lambda=\displaystyle\frac{w_\infty\,\lambda-b_\infty}{a_\infty}.
\end{equation}
Using \eqref{1.4} in \eqref{2.4} and \eqref{2.5}, we see that
\begin{equation}\label{2.8}
\lim_{n\to\pm\infty}\tilde a(n)=I,
\quad
\lim_{n\to\pm\infty}\tilde b(n)=0,
\end{equation}
where it is understood that the quantity $0$ denotes the $q\times q$ zero matrix.
Under the assignments described in \eqref{2.4}--\eqref{2.7}, we transform \eqref{1.1} to the $q\times q$ matrix-valued 
Jacobi system given by
\begin{equation}\label{2.9}
\tilde a(n+1)\,\tilde\psi(n+1)+\tilde b(n)\,\tilde\psi(n)+\tilde a(n)^\dagger\,\tilde\psi(n-1)=\tilde\lambda\,\tilde\psi(n),  \qquad n\in\mathbb Z.
\end{equation}
Furthermore, the restriction on the coefficients $a(n),$ $b(n),$ $w(n)$ given in \eqref{1.6} becomes equivalent to the simpler
restriction
\begin{equation}\label{2.10}
\displaystyle\sum_{n=-\infty}^\infty  |n|  \left( ||\tilde a(n)-I ||+ ||\tilde b(n)|| \right)<+\infty,
\end{equation}
where the norm denotes the standard norm. We note that
the restriction in \eqref{2.10} is similar to the restriction used in \cite{S1985,S1987}.
We consider \eqref{2.9} when the coefficients there belong to the class $\tilde{\mathcal A},$ which is described in the following definition.

\begin{definition}\label{definition2.1}
The $q\times q$ matrix-valued coefficients $\tilde a(n)$ and $\tilde b(n)$ in \eqref{2.9} with their spacial asymptotics in \eqref{2.8}
belong to the class $\tilde{\mathcal A}$ if the following restrictions are satisfied: 

\begin{enumerate}
\item[\text{\rm(a)}] The
matrix $\tilde b(n)$ is selfadjoint for $n\in\mathbb Z.$
On the other hand, the matrix $\tilde a(n)$ is not necessarily selfadjoint.

\item[\text{\rm(b)}] The matrix $\tilde a(n)$ is invertible for
$n\in\mathbb Z.$

\item[\text{\rm(c)}] The matrices $\tilde a(n)$ and
 $\tilde b(n)$ satisfy the summability condition given in \eqref{2.10}.

\end{enumerate}
\end{definition}

When the coefficients in \eqref{1.1} belong to the class $\mathcal A$
described in
Definition~\ref{definition1.1},
with the help of \eqref{2.4}--\eqref{2.8} and \eqref{2.10}, 
we can verify that the coefficients in
\eqref{2.9} belong to the class
 $\tilde{\mathcal A}.$
 Conversely, if the coefficients in \eqref{2.9} belong to the class $\tilde{\mathcal A}$ then it follows that
 the coefficients in \eqref{1.1} belong to the class $\mathcal A.$

We can write \eqref{2.9} as 
the eigenvalue equation
$\tilde{\mathbf H}\tilde\psi=\tilde\lambda\tilde\psi,$ 
with the action of the operator 
$\tilde{\mathbf H}$ on
$\tilde\psi$ described by
\begin{equation*}
(\tilde{\mathbf H}\tilde\psi)(n)=\tilde a(n+1)\,\tilde\psi(n+1)+\tilde b(n)\,\tilde\psi(n)+\tilde a(n)^\dagger\,\tilde\psi(n-1),  \qquad n\in\mathbb Z,
\end{equation*}
where the operator $\tilde{\mathbf H}$ acts on the Hilbert space $\ell^2(\mathbb Z)$ consisting of
complex-valued sequences $\tilde\psi$ with values $\tilde\psi(n)$ for $n\in\mathbb Z.$ The inner product on that Hilbert space is the usual
inner product given by
\begin{equation*}
\langle \tilde\phi,\tilde\psi\rangle:=\sum_{n=-\infty}^\infty \tilde\phi(n)^\dagger\,\tilde\psi(n),
\end{equation*}
with the implied norm defined as
\begin{equation*}
||\tilde\psi||:=\sqrt{\sum_{n=-\infty}^\infty \tilde\psi(n)^\dagger\,\tilde\psi(n)},
\end{equation*}
which is the norm used in \eqref{2.10}.
It can be verified directly that 
the operator $\tilde{\mathbf H}$ is bounded and selfadjoint 
when the matrices $\tilde b(n)$
are selfadjoint for $n\in\mathbb Z$ and
the sequences $\{||\tilde a(n)||\}_{n\in\mathbb Z}$
and $\{||\tilde b(n)||\}_{n\in\mathbb Z}$ are bounded. In particular, 
the operator $\tilde{\mathbf H}$ is bounded and selfadjoint when 
\eqref{2.10} holds and 
 the matrices $\tilde b(n)$
are selfadjoint for $n\in\mathbb Z.$

Even though \eqref{1.1} can be transformed into \eqref{2.9} with simpler asymptotics for
its coefficients, we continue considering \eqref{1.1} for its importance in applications where
its coefficients and its solutions have some appropriate physical relevance.
It is convenient to parameterize $\tilde\lambda$ appearing in \eqref{2.9}
in terms of an auxiliary parameter $z$ as
\begin{equation}\label{2.14}
\tilde\lambda=z+\displaystyle\frac{1}{z},
\qquad z\in\mathbb C\setminus\{0\},
\end{equation}
where we use $\mathbb C$ to denote the complex plane.
By replacing $\tilde\lambda$ in \eqref{2.9} by its equivalent in terms of $z$ given in \eqref{2.14}, we can write \eqref{2.9} in terms
of the auxiliary spectral parameter $z$ as
\begin{equation}\label{2.15}
\tilde a(n+1)\,\tilde\psi(n+1)+\tilde b(n)\,\tilde\psi(n)+\tilde a(n)^\dagger\,\tilde\psi(n-1)=(z+z^{-1})\,\tilde\psi(n),  \qquad n\in\mathbb Z.
\end{equation}
From \eqref{2.7} and \eqref{2.14}, it follows that the spectral parameter $\lambda$ is expressed
in terms of the auxiliary parameter $z$ as
\begin{equation}\label{2.16}
\lambda=\displaystyle\frac{a_\infty(z+z^{-1})+b_\infty}{w_\infty}.
\end{equation}
Using \eqref{2.16} in \eqref{1.1}, we also express \eqref{1.1} in terms of the auxiliary spectral parameter $z$ as
\begin{equation}\label{2.17}
a(n+1)\,\psi(n+1)+b(n)\,\psi(n)+a(n)^\dagger\,\psi(n-1)=
\displaystyle\frac{a_\infty(z+z^{-1})+b_\infty}{w_\infty}\,w(n)\,\psi(n),
\qquad n\in\mathbb Z.
\end{equation}
Even though \eqref{1.1} and \eqref{2.17} are equivalent, it is convenient to use
\eqref{2.17} to obtain solutions expressed in terms of the auxiliary spectral parameter $z$ rather than the spectral parameter $\lambda.$ 
In fact, those particular solutions
satisfying certain spacial asymptotics, which are usually known as the Jost solutions, are
conveniently expressed in terms of $z$ rather than in terms of $\lambda.$ Furthermore,
the scattering coefficients associated with \eqref{1.1} are also conveniently expressed
in terms of $z$ rather than $\lambda.$ Hence, in the analysis of \eqref{1.1} in our paper, 
it is more convenient to use \eqref{2.17} rather than \eqref{1.1}.
Furthermore, \eqref{2.15} is more closely related to \eqref{2.17} than to \eqref{1.1} because
each of \eqref{2.15} and \eqref{2.17} is expressed in terms of $z.$
We can use the known results for the simpler equation \eqref{2.15}, 
and by using the transformations listed in \eqref{2.4}--\eqref{2.6}, we can establish
the corresponding results for \eqref{2.17}.

Since \eqref{2.15} and \eqref{2.17} are both linear homogeneous systems, any constant scalar multiple of
a solution to each of those two systems of equations remains a solution.
Thus, we can multiply either side of \eqref{2.6} by a constant scalar without affecting 
the transformation from \eqref{2.15} to \eqref{2.17}.
With the special choice of the transformation given in \eqref{2.6}, we see that
$\psi(n)$ and $\tilde\psi(n)$ have the same leading spacial asymptotics as $n\to\pm\infty.$

Let us remark on the transformation $\lambda\mapsto z$ described in \eqref{2.16}.
We
use $\mathbb C^+$ for the upper-half complex
plane, $\mathbb C^-$ for the lower-half complex plane, $\overline {\mathbb C^+}$ for the closure of
$\mathbb C^+$ so that we have $\overline{\mathbb C^+}:=\mathbb C^+\cup\mathbb R,$ and we use
 $\overline {\mathbb C^-}$ for the closure of
$\mathbb C^-$ so that we have $\overline{\mathbb C^-}:=\mathbb C^-\cup\mathbb R.$
Let us also use $\mathbb T$ to denote the unit circle $|z|=1$ in the complex $z$-plane,
 $\mathbb D$ for the open unit disc $|z|<1,$ and $\overline {\mathbb D}$ for the
closed unit disc $|z|\le 1$ so that we have
$\overline{\mathbb D}:=\mathbb D\cup \mathbb T.$
We let
\begin{equation*}
\lambda_{\min}:=\displaystyle\frac{-2\,|a_\infty|+b_\infty}{w_\infty},
\quad
\lambda_{\max}:=\displaystyle\frac{2\,|a_\infty|+b_\infty}{w_\infty}.
\end{equation*}
If \eqref{1.1} is Schr\"odinger-like, i.e. when we have $a_\infty<0,$ the transformation  $\lambda\mapsto z$ in \eqref{2.16}
maps the real $\lambda$-axis to the boundary of the upper portion of $\mathbb D$
in such a way that
the interval
$\lambda\in(-\infty,\lambda_{\min})$
is mapped to the real interval
$z\in (0,1),$ the
interval $\lambda\in(\lambda_{\min},\lambda_{\max})$
is mapped to the upper portion of $\mathbb T,$
the interval $\lambda\in(\lambda_{\max},+\infty)$
is mapped to the real interval $z\in(-1,0),$
while $\lambda=\lambda_{\min}$ is mapped to $z=1$ and
$\lambda=\lambda_{\max}$ is mapped to $z=-1.$
Thus, the upper-half complex $\lambda$-plane is mapped to the upper portion of
$\mathbb D$ by preserving the orientation.
On the other hand, if \eqref{1.1} is Jacobi-like, i.e. when we have
$a_\infty>0,$ the transformation $\lambda\mapsto z$
maps the real $\lambda$-axis to the boundary of the lower portion of $\mathbb D$
in such a way that
the interval
$\lambda\in(-\infty,\lambda_{\min})$
is mapped to the real interval
$z\in (-1,0),$ the
interval $\lambda\in(\lambda_{\min},\lambda_{\max})$
is mapped to the lower portion of $\mathbb T,$
the interval $\lambda\in(\lambda_{\max},+\infty)$
is mapped to the real interval $z\in(0,1),$
while $\lambda=\lambda_{\min}$ is mapped to $z=-1$ and
$\lambda=\lambda_{\max}$ is mapped to $z=1.$ Thus, the upper-half complex $\lambda$-plane is mapped to the
lower portion of $\mathbb D$ 
by preserving the orientation.

Since $z$ and $z^{-1}$ appear together as the sum $(z+z^{-1})$ in \eqref{2.17},
we see that if $\psi(z,n)$ is a solution to \eqref{2.17} then $\psi(z^{-1},n)$ is also a solution to
\eqref{2.17}. 
 Thus, any solution to \eqref{2.17}
defined on either the upper or lower portion of the unit circle $\mathbb T$ is extended
to the $z$-values on the whole unit circle.
Note that we have
\begin{equation}\label{2.19}
z^{-1}=z^\ast,\qquad z\in\mathbb T,
\end{equation}
where the asterisk denotes complex conjugation. 
Thus, any solution $\psi(z,n)$ to \eqref{2.17} satisfies
\begin{equation*}
\psi(z^{-1},n)=\psi(z^\ast,n),\qquad n\in\mathbb Z,\quad z\in\mathbb T.
\end{equation*}
When the coefficient matrices in \eqref{2.15} are replaced by their limiting values specified in \eqref{2.8}, we obtain the
unperturbed $q\times q$ system
\begin{equation}\label{2.21}
\accentset\circ\psi(z,n+1)+\accentset\circ\psi(z,n-1)=
\left(z+z^{-1}\right)\accentset\circ\psi(z,n),\qquad n\in\mathbb Z.
\end{equation}
We observe that \eqref{2.21} has the two $q\times q$ matrix-valued solutions given by
\begin{equation*}
\accentset\circ\psi(z,n)=z^n\,I,\quad
\accentset\circ\psi(z,n)=z^{-n}\,I,
\end{equation*}
and those two solutions are the left and right Jost solutions, respectively, to the unperturbed system \eqref{2.21}.
The general  $q\times q$ matrix-valued solution to \eqref{2.21} has the form $z^n\,c_3+z^{-n}\,c_4,$ where
$c_3$ and $c_4$ are two arbitrary complex-valued $q\times q$ matrices
that may contain $z$ but not $n.$

\section{The direct scattering problem}
\label{section3}

We know from Section~\ref{section1} that \eqref{1.1} containing the spectral parameter $\lambda$ and \eqref{2.17} containing the auxiliary spectral parameter $z$ are equivalent. In this section we present some basic results related to
the direct scattering problem for \eqref{2.17} when the coefficients there belong to the class $\mathcal A$ specified in Definition~\ref{definition1.1}. 
In particular, 
we introduce the pair of $q\times q$ matrix-valued Jost solutions 
$f_{\rm{l}}(z,n)$ and $f_{\rm{r}}(z,n),$ the pair of $q\times q$ matrix-valued Jost solutions 
$g_{\rm{l}}(z,n)$ and $g_{\rm{r}}(z,n),$ the pair of $q\times q$ matrix-valued physical solutions
$\Psi_{\rm{l}}(z,n)$ and $\Psi_{\rm{r}}(z,n),$ the four $q\times q$ matrix-valued
scattering coefficients
$T_{\rm{l}}(z),$ $T_{\rm{r}}(z),$ $L(z),$ $R(z),$ and
the $2q\times 2q$ scattering matrix $S(z).$
After introducing those relevant quantities, we establish their basic properties that are needed to obtain the factorization results presented
in Section~\ref{section5}. While introducing the aforementioned quantities and establishing their properties, we exploit the 
connection between the solutions to \eqref{2.15} and \eqref{2.17}. This is done by using the transformation given in \eqref{2.6}.
With the help of the third asymptotics in \eqref{1.4}, from \eqref{2.6} we observe that the leading spacial asymptotics of the solution to
\eqref{2.15} and \eqref{2.17} coincide.

In the following theorem, we establish the existence of
the Jost solutions
$\tilde f_{\rm{l}}(z,n)$ and $\tilde f_{\rm{r}}(z,n)$ to \eqref{2.15} and present their
basic properties.
We recall that the sets $\mathbb T,$ $\mathbb D,$ and $\overline{\mathbb D}$
are defined in Section~\ref{section1} and that
they denote the unit circle, the open unit disk, and the closed unit disk,
respectively, in the complex plane.
The Jost solutions play a significant role in the
direct scattering theory because a certain combination of them forms
a fundamental set, which allows any other solution to be expressed as a 
linear combination of the fundamental set of solutions.
Furthermore, the Jost solutions are defined by using rather simple
spacial asymptotics, and hence their properties are not extremely difficult to determine.

\begin{theorem} \label{theorem3.1}
When the $q\times q$ matrix-valued coefficients $\tilde a(n)$ and $\tilde b(n)$ in \eqref{2.15} belong to
the class $\tilde{\mathcal A}$ specified in Definition~\ref{definition2.1}, we have the following:

\begin{enumerate}

\item[\text{\rm(a)}] For any $z\in \overline{\mathbb D},$ 
there exists a unique $q\times q$ matrix-valued solution to \eqref{2.15}, known as the
left Jost solution and denoted by $\tilde f_{\rm{ l}}(z,n),$ 
satisfying the spacial asymptotics
\begin{equation}
\label{3.1} 
\tilde f_{\rm{l}}(z,n)=z^n\left[I+o(1)\right],\qquad n\to+\infty.
\end{equation}
Moreover, for each fixed $n\in\mathbb Z,$ the quantity
$\tilde f_{\rm{ l}}(z,n)$ 
is analytic in $z \in \mathbb D$ and continuous in $z\in\overline{\mathbb D},$ 
with the understanding that a matrix satisfies a property if and only if each entry of that matrix satisfies the
aforementioned property.

\item[\text{\rm(b)}] Similarly,
for any $z\in \overline{\mathbb D},$ 
there exists a unique $q\times q$ matrix-valued solution to \eqref{2.15}, known as the
right Jost solution and denoted by $\tilde f_{\rm{ r}}(z,n),$ 
satisfying the spacial asymptotics
\begin{equation}
\label{3.2} 
\tilde f_{\rm{r}}(z,n)=z^{-n}\left[I+o(1)\right],\qquad n\to-\infty.
\end{equation}
Furthermore, for each fixed $n\in\mathbb Z,$ the quantity
$\tilde f_{\rm{ r}}(z,n)$ 
is analytic in $z \in \mathbb D$ and continuous in $z\in\overline{\mathbb D}.$

\item[\text{\rm(c)}] For any fixed $z \in \mathbb T\setminus \{-1,1\},$ the solution set 
$\{\tilde f_{\rm{l}}(z,n),\tilde f_{\rm{l}}(z^{-1},n)\}_{n\in\mathbb Z}$ and
the solution set 
$\{\tilde f_{\rm{r}}(z,n),\tilde f_{\rm{r}}(z^{-1},n)\}_{n\in\mathbb Z}$ each form a fundamental set of $q\times q$ matrix solutions to \eqref{2.15}. 

 \end{enumerate}
\end{theorem}

\begin{proof} For the results presented in the theorem, we refer the reader to
\cite{S1985} and \cite{S1987}.
\end{proof}

The results presented in Theorem~\ref{theorem3.1} for the matrix-valued system \eqref{2.15}
have analogs for some related systems. We refer the reader to \cite{G1976a,G1976b}
for the corresponding results in the scalar case on the half-line lattice,
to \cite{G1976b,G1976c} in the scalar case on the full-line lattice,
to \cite{S1980} in the matrix case on the half-line lattice,
and to \cite{SSVW2025} on the full-line lattice when the coefficients in the Jacobi system 
are bounded operators on a Hilbert space.

For $z\in\mathbb T\setminus\{-1,1\},$ any solution to \eqref{2.15} can be expressed as a linear combination of
either of the two fundamental sets of solutions specified in Theorem~\ref{theorem3.1}(c).
This fact allows us to introduce the scattering coefficients for \eqref{2.15} as follows. From \cite{S1987} we know that
there exist $q\times q$ matrix-valued functions $\tilde A(z),$ $\tilde B(z),$ $\tilde C(z),$ $\tilde D(z)$ that are continuous in
$z\in\mathbb T\setminus\{-1,1\}$ in such a way that
\begin{equation}
\label{3.3} 
\tilde f_{\rm{l}}(z,n)= \tilde f_{\rm{r}}(z^{-1},n)\, \tilde A(z)+\tilde f_{\rm{r}}(z,n)\, \tilde B(z), \qquad z \in \mathbb T\setminus \{-1,1\},
\end{equation}
\begin{equation}
\label{3.4} 
\tilde f_{\rm{r}}(z,n)= \tilde f_{\rm{l}}(z,n)\, \tilde C(z)+\tilde f_{\rm{l}}(z^{-1},n)\, \tilde D(z), \qquad z \in \mathbb T\setminus \{-1,1\}.
\end{equation}
It is also known \cite{S1987} that the matrices $\tilde A(z)$ and
$\tilde D(z)$ are invertible
when $z\in\mathbb T\setminus\{-1,1\}.$ 
We then introduce
the left transmission coefficient
$\tilde T_{\rm{l}}(z),$
the left reflection coefficient $\tilde L(z),$
the right transmission coefficient
$\tilde T_{\rm{r}}(z),$ and the right reflection coefficient $\tilde R(z)$
as 
\begin{equation}
\label{3.5} 
\tilde T_{\rm{l}}(z):=\tilde A(z)^{-1},\quad
\tilde L(z):=\tilde B(z)\,\tilde A(z)^{-1},\quad
\tilde T_{\rm{r}}(z):=\tilde D(z)^{-1},\quad
\tilde R(z):=\tilde C(z)\,\tilde D(z)^{-1}.
\end{equation}
We remark that \eqref{3.5} holds when
$z\in\mathbb T\setminus\{-1,1\}.$ 
We use $\tilde T_{\rm{l}}(z)^{-1}$ and $\tilde T_{\rm{r}}(z)^{-1}$ to denote the
inverses of the matrices $\tilde T_{\rm{l}}(z)$ and $\tilde T_{\rm{r}}(z),$ respectively.
From \eqref{3.3}--\eqref{3.5} it follows that
\begin{equation}
\label{3.6} 
\tilde f_{\rm{l}}(z^{-1},n)= \tilde f_{\rm{r}}(z,n)\, \tilde T_{\rm{r}}(z)-\tilde f_{\rm{l}}(z,n)\, \tilde R(z), \qquad z \in \mathbb T\setminus \{-1,1\},
\end{equation}
\begin{equation}
\label{3.7} 
\tilde f_{\rm{r}}(z^{-1},n)= \tilde f_{\rm{l}}(z,n)\, \tilde T_{\rm{l}}(z)-\tilde f_{\rm{r}}(z,n)\, \tilde L(z), \qquad z \in \mathbb T\setminus \{-1,1\}.
\end{equation}

In terms of the Jost solutions $\tilde f_{\rm{l}}(z,n)$ and $\tilde f_{\rm{r}}(z,n)$ to \eqref{2.15}, 
with the help of \eqref{2.6}
we introduce the particular solutions $f_{\rm{l}}(z,n)$ and $f_{\rm{r}}(z,n)$ to \eqref{2.17} as
\begin{equation}
\label{3.8} 
f_{\rm{l}}(z,n):= w_\infty^{1/2}\, w(n)^{-1/2}\tilde f_{\rm{l}}(z,n),
\end{equation}
\begin{equation}
\label{3.9} 
f_{\rm{r}}(z,n):= w_\infty^{1/2}\, w(n)^{-1/2}\tilde f_{\rm{r}}(z,n).
\end{equation}
In the next theorem, we present the analog of Theorem~\ref{theorem3.1} by introducing
the Jost solutions to 
\eqref{2.17} and describing their basic properties. In the theorem, we also introduce the scattering coefficients for \eqref{2.17}, present their basic properties, and relate them to
the scattering coefficients for \eqref{2.15}.
The significance of the theorem comes from the fact that it shows
that the scattering coefficients for \eqref{2.15} coincide with
the corresponding scattering coefficients for \eqref{2.17}.

\begin{theorem} \label{theorem3.2}
When the $q\times q$ matrix-valued coefficients $a(n),$ $b(n),$ $w(n)$ in \eqref{2.17} belong to
the class $\mathcal A$ specified in Definition~\ref{definition1.1}, we have the following:

\begin{enumerate}

\item[\text{\rm(a)}] The $q\times q$ matrix-valued quantity $f_{\rm{l}}(z,n)$ defined in \eqref{3.8} is the unique solution to \eqref{2.17} satisfying the spacial
asymptotics 
\begin{equation}\label{3.10} 
 f_{\rm{l}}(z,n)=z^n\left[I+o(1)\right],\qquad n\to+\infty.
\end{equation}
Similarly, the $q\times q$ matrix-valued quantity $f_{\rm{r}}(z,n)$ defined in \eqref{3.9} is the unique solution to \eqref{2.17} satisfying the spacial
asymptotics 
\begin{equation}\label{3.11} 
f_{\rm{r}}(z,n)=z^{-n}\left[I+o(1)\right],\qquad n\to-\infty.
\end{equation}
Since \eqref{3.1} and \eqref{3.10} are analogous and that \eqref{3.2} and \eqref{3.11}
are analogous, 
it is appropriate to refer to $f_{\rm{l}}(z,n)$ and $f_{\rm{r}}(z,n)$ as the left and right Jost solutions, respectively, to \eqref{2.17}. 

\item[\text{\rm(b)}] For each fixed $n\in \mathbb Z,$ the Jost solutions $f_{\rm{l}}(z,n)$ and $f_{\rm{r}}(z,n)$ are analytic in $z\in\mathbb D$ and continuous in $\overline{\mathbb D}.$

\item[\text{\rm(c)}] For any $z \in \mathbb T\setminus \{-1,1\},$ the two solution sets $\{f_{\rm{l}}(z,n),f_{\rm{l}}(z^{-1},n)\}$ and
$\{f_{\rm{r}}(z,n),f_{\rm{r}}(z^{-1},n)\}$ each form a fundamental set of $q\times q$ matrix-valued solutions to \eqref{2.17}.

\item[\text{\rm(d)}] The solutions $f_{\rm{l}}(z^{-1},n)$ and $f_{\rm{r}}(z^{-1},n)$ are expressed as linear combination of $f_{\rm{l}}(z,n)$ and
$f_{\rm{r}}(z,n)$ as
\begin{equation}
\label{3.12} 
f_{\rm{l}}(z^{-1},n)= f_{\rm{r}}(z,n)\, T_{\rm{r}}(z)-f_{\rm{l}}(z,n)\, R(z), \qquad z \in \mathbb T\setminus \{-1,1\},
\end{equation}
\begin{equation}
\label{3.13} 
f_{\rm{r}}(z^{-1},n)= f_{\rm{l}}(z,n)\, T_{\rm{l}}(z)-f_{\rm{r}}(z,n)\, L(z), \qquad z \in \mathbb T\setminus \{-1,1\},
\end{equation}
where the coefficients $T_{\rm{l}}(z),$ $L(z),$ $T_{\rm{r}}(z),$ $R(z)$ are some $q\times q$ matrix-valued functions of
$z.$ Furthermore, we have
\begin{equation}
\label{3.14} 
T_{\rm{l}}(z)=\tilde T_{\rm{l}}(z),\quad L(z)=\tilde L(z),\quad T_{\rm{r}}(z)=\tilde T_{\rm{r}}(z),\quad R(z)=\tilde R(z),
\qquad z \in \mathbb T\setminus \{-1,1\},
\end{equation}
where $\tilde T_{\rm{l}}(z),$
$\tilde L(z),$
$\tilde T_{\rm{r}}(z),$ and $\tilde R(z)$
are the scattering coefficients appearing in \eqref{3.6} and \eqref{3.7}
associated with \eqref{2.15}.
We refer to 
the coefficient $T_{\rm{l}}(z)$ as the left transmission coefficient for \eqref{2.17}, the coefficient $L(z)$ as the left reflection
coefficient, the coefficient $T_{\rm{r}}(z)$ as the right transmission coefficient, and the coefficient $R(z)$ as the right reflection coefficient.
Hence, the equalities in \eqref{3.14} show that the scattering coefficients for \eqref{2.15} and
the corresponding scattering coefficients for \eqref{2.17} coincide.

\item[\text{\rm(e)}] The matrices $T_{\rm{l}}(z)$ and $T_{\rm{r}}(z)$ are invertible, i.e. the $q\times q$ matrices $T_{\rm{l}}(z)^{-1}$ and
$T_{\rm{r}}(z)^{-1}$ exist for $z\in\mathbb T\setminus\{-1,1\}$.

\item[\text{\rm(f)}] For each fixed $z \in \mathbb T\setminus \{-1,1\},$ the Jost solutions $f_{\rm{l}}(z,n)$ and $f_{\rm{r}}(z,n)$ have their respective spacial asymptotics given by 
\begin{equation}\label{3.15} 
f_{\rm{l}}(z,n)=z^{n}\left[T_{\rm{l}}(z)^{-1}+o(1)\right]+z^{-n}\left[L(z)\, T_{\rm{l}}(z)^{-1}+o(1)\right],\qquad n\to-\infty,
\end{equation}
\begin{equation}\label{3.16} 
f_{\rm{r}}(z,n)=z^{-n}\left[T_{\rm{r}}(z)^{-1}+o(1)\right]+z^{n}\left[R(z)\, T_{\rm{r}}(z)^{-1}+o(1)\right],\qquad n\to+\infty.
\end{equation}

\end{enumerate}
\end{theorem}

\begin{proof}
The proofs for (a)--(c) follow from Theorem~\ref{theorem3.1} by using \eqref{3.8} and \eqref{3.9} and the third spacial asymptotics in \eqref{1.4}.
We obtain the proof of (d) by using \eqref{3.6}--\eqref{3.9}.
The result in (e) follows from \eqref{3.14} and  
the fact that $\tilde T_{\rm{l}}(z)$ and
$\tilde T_{\rm{r}}(z)$ are themselves invertible matrices for $z \in \mathbb T\setminus \{-1,1\}.$ 
In order to prove (f), we proceed as follows.
Since $T_{\rm{l}}(z)$ is invertible for $z\in\mathbb T\setminus\{-1,1\},$
we can postmultiply \eqref{3.13} by
$T_{\rm{l}}(z)^{-1}$ and write the resulting matrix equality as
\begin{equation}\label{3.17} 
  f_{\rm{l}}(z,n)=  f_{\rm{r}}(z^{-1},n)\, T_{\rm{l}}(z)^{-1}+ f_{\rm{r}}(z,n)\, L(z)\, T_{\rm{l}}(z)^{-1}, \qquad z \in \mathbb T\setminus \{-1,1\}.
\end{equation}
Similarly, since $T_{\rm{r}}(z)$ is invertible for
$z\in\mathbb T\setminus\{-1,1\},$ we can postmultiply \eqref{3.12} by
$T_{\rm{r}}(z)^{-1}$ and write the resulting matrix equality as
\begin{equation}\label{3.18} 
  f_{\rm{r}}(z,n)=  f_{\rm{l}}(z^{-1},n)\, T_{\rm{r}}(z)^{-1}+ f_{\rm{l}}(z,n)\, R(z)\, T_{\rm{r}}(z)^{-1}, \qquad z \in \mathbb T\setminus \{-1,1\}.
\end{equation}
Using the spacial asymptotics \eqref{3.11} in \eqref{3.17}, we obtain \eqref{3.15}.
Similarly, using the spacial asymptotics in \eqref{3.10} and \eqref{3.18}, we obtain \eqref{3.16}.
Hence, the proof of (f) is complete.
\end{proof}

The scattering coefficients $\tilde T_{\rm{l}}(z),$ $\tilde L(z),$
$\tilde T_{\rm{r}}(z),$ $\tilde R(z)$ defined in \eqref{3.5}
for the Jacobi system \eqref{2.15} appear in the coefficients in \eqref{3.3} and \eqref{3.4}
as well as \eqref{3.6} and \eqref{3.7}.
Similarly, the scattering coefficients $T_{\rm{l}}(z),$ $L(z),$
$T_{\rm{r}}(z),$ $R(z)$
for the Jacobi system \eqref{2.17} appear in the
coefficients in \eqref{3.12} and \eqref{3.13}.
In both cases, in order to introduce the scattering coefficients,
we have used a particular fundamental set of matrix-valued solutions
to the corresponding Jacobi system
described in Theorem~\ref{theorem3.1}(c) and Theorem~\ref{theorem3.2}(c), respectively.
 Next, we have expressed another particular matrix-valued solution
as a linear combination of the two solutions in the fundamental set.
Then, we have used the coefficients in the corresponding linear combinations
to introduce the scattering coefficients.
The scattering coefficients can alternatively be introduced with the help of the Wronskians
of some particular solutions to each corresponding Jacobi system.
For example, in Theorem~\ref{theorem3.4} in this section we express the 
scattering coefficients for \eqref{2.17}
in terms of the Wronskians of  certain particular solutions to \eqref{2.17}.
Another way to introduce the scattering coefficients is through the use of the spacial asymptotics of
the left and right Jost solutions. For example, using \eqref{3.15} and \eqref{3.16}, we can first
obtain
$T_{\rm{l}}(z)^{-1},$ $L(z)\,T_{\rm{l}}(z)^{-1},$
$T_{\rm{r}}(z)^{-1},$ $R(z)\,T_{\rm{r}}(z)^{-1},$ and then we can form
the scattering coefficients $T_{\rm{l}}(z),$ $L(z),$
$T_{\rm{r}}(z),$ $R(z).$

Letting
\begin{equation}\label{3.19}
g_{\rm{l}}(z,n):=f_{\rm{l}}(z^{-1},n),\quad g_r(z,n):=f_{\rm{r}}(z^{-1},n),
\end{equation}
with the help of \eqref{3.10}, \eqref{3.11}, \eqref{3.15}, and \eqref{3.16}, we observe that 
$g_{\rm{l}}(z,n)$ and $g_{\rm{r}}(z,n)$ are the $q\times q$ matrix-valued solutions 
to \eqref{2.17} satisfying the respective asymptotics
\begin{equation}\label{3.20}
g_{\rm{l}}(z,n)=\begin{cases}z^{-n}\left[I+o(1)\right],\qquad n\to+\infty,\\
\noalign{\medskip}
z^{-n}\left[T_{\rm{l}}(z^{-1})^{-1}+o(1)\right]+ z^n\left[L(z^{-1})\,T_{\rm{l}}(z^{-1})^{-1}+o(1)\right],\qquad n\to-\infty,
\end{cases}
\end{equation}
\begin{equation}\label{3.21}
g_{\rm{r}}(z,n)=\begin{cases} z^n\left[T_{\rm{r}}(z^{-1})^{-1}+o(1)\right]+ z^{-n}\left[R(z^{-1})\,T_{\rm{r}}(z^{-1})^{-1}+o(1)\right],\qquad n\to+\infty,\\
\noalign{\medskip}
z^n\left[I+o(1)\right],\qquad n\to-\infty.
\end{cases}
\end{equation}
Since $f_{\rm{l}}(z,n)$ and $f_{\rm{r}}(z,n)$ are analytic in $z\in\mathbb D$ and continuous in $z\in\overline{\mathbb D},$ the
quantities $g_{\rm{l}}(z,n)$ and $g_{\rm{r}}(z,n)$ are analytic when we have $|z| >1$ and continuous when we have $|z|\ge 1.$

The $2q\times 2q$ scattering matrix $S(z)$ associated with \eqref{2.17} is defined as
\begin{equation}
\label{3.22}
S(z):=\begin{bmatrix}T_{\rm{l}}(z)&R(z) \\ 
\noalign{\medskip}
 L(z)& T_{\rm{r}}(z)\end{bmatrix},\qquad z\in\mathbb T\setminus\{-1,1\}.
\end{equation}
The left and right physical solutions to \eqref{2.17}, denoted by $\Psi_{\rm{l}}(z,n)$ and $\Psi_{\rm{r}}(z,n),$ are the
two particular
$q\times q$ matrix-valued solutions that are related to the Jost solutions
$f_{\rm{l}}(z,n)$ and $f_{\rm{r}}(z,n),$ respectively, as
\begin{equation*}
\Psi_{\rm{l}}(z,n):=f_{\rm{l}}(z,n)\,T_{\rm{l}}(z),\quad \Psi_{\rm{r}}(z,n):=f_{\rm{r}}(z,n)\,T_{\rm{r}}(z),
\end{equation*}
and, as seen from \eqref{3.10}, \eqref{3.11}, \eqref{3.15}, and  \eqref{3.16}, they satisfy the respective spacial asymptotics
\begin{equation}
\label{3.24}
\Psi_{\rm{l}}(z,n)=z^n\left[T_{\rm{l}}(z)+o(1)\right],\quad \Psi_{\rm{r}}(z,n)=z^{-n}\left[I+o(1)\right] +z^n\left[ R(z)+o(1)\right],\qquad n\to+\infty,
\end{equation}
\begin{equation}
\label{3.25}
\Psi_{\rm{l}}(z,n)=z^n\left[ I+o(1)\right]+z^{-n}\left[ L(z)+o(1)\right],\quad \Psi_{\rm{r}}(z,n)=z^{-n}\left[ T_{\rm{r}}(z)+o(1)\right],\qquad n\to-\infty.
\end{equation}
With the help of \eqref{3.24} and \eqref{3.25}, we can interpret $\Psi_{\rm{l}}(z,n)$ in terms of the matrix-valued plane wave $z^n I$ of unit amplitude
sent from $n=-\infty,$ the matrix-valued reflected plane wave $z^{-n} L(z)$ of amplitude
$L(z)$ at $n=-\infty,$ and the matrix-valued transmitted plane wave $z^n\,T_{\rm{l}}(z)$ of amplitude $T_{\rm{l}}(z)$ at $n=+\infty.$
Similarly, the physical solution $\Psi_{\rm{r}}(z,n)$ can be interpreted 
 in terms of the matrix-valued plane wave $z^{-n} I$ of unit amplitude
sent from $n=+\infty,$ the matrix-valued reflected plane wave $z^n R(z)$ of amplitude
$R(z)$ at $n=+\infty,$ and the matrix-valued transmitted plane wave $z^{-n}\,T_{\rm{r}}(z)$ of amplitude $T_{\rm{r}}(z)$ at $n=-\infty.$
Thus, we have a physical interpretation of the direct scattering problem for \eqref{2.17} by using the left and right scattering coefficients.

Let us use $[\alpha(n);\beta(n)]$ to denote the Wronskian of two $q\times q$ matrix-valued functions of $n,$ where we have defined
\begin{equation}
\label{3.26}
[\alpha(n);\beta(n)]:=\alpha(n)\,a(n+1)\,\beta(n+1)-\alpha(n+1)\,a(n+1)^\dagger\,\beta(n),
\end{equation}
with $a(n+1)$ being the $q\times q$ matrix-valued coefficient appearing in \eqref{2.17}.
We can write the right-hand side of \eqref{3.26} as a matrix product, and hence \eqref{3.26} is equivalently expressed as
\begin{equation*}
[\alpha(n);\beta(n)]=\begin{bmatrix}
\alpha(n)&\alpha(n+1)\end{bmatrix}
\begin{bmatrix}a(n+1)&0\\
0&-a(n+1)^\dagger\end{bmatrix}
\begin{bmatrix}\beta(n+1)\\
\beta(n)\end{bmatrix}.
\end{equation*}
We have introduced the scattering coefficients for \eqref{2.17}
by using the spacial asymptotics of the Jost solutions. We can alternatively express those scattering coefficients
with the help of the Wronskian defined in \eqref{3.26}. For this, we need the result stated in the following proposition.
This result will be used later in the proof of Theorem~\ref{theorem3.4}.

\begin{proposition}\label{proposition3.3} Assume that the $q\times q$ matrix-valued coefficients in \eqref{1.1} belong to the class
$\mathcal A$ described in Definition~\ref{definition1.1}. Let $\phi(z,n)$ and $\varphi(z,n)$ be any two $q\times q$ matrix-valued solutions
to \eqref{2.17} for $z\in\mathbb T$ and $n\in\mathbb Z.$ Then, the Wronskian
 $[\phi(z^\ast,n)^\dagger;\varphi(z,n)]$ is independent of $n.$
\end{proposition}

\begin{proof}
From \eqref{1.5} we know that $b(n)$ and $w(n)$ are each selfadjoint for $n\in\mathbb Z,$ and from
Definition~\ref{definition1.1}(c) we know that the scalars $a_\infty,$ $b_\infty,$ and $w_\infty$ are each real.
Thus, replacing $z$ with $z^\ast$ in \eqref{2.17}, taking the adjoint of both sides of the resulting equality,
and using \eqref{2.19}, we obtain
\begin{equation}\label{3.28}
\phi(z^\ast,n+1)^\dagger\,a(n+1)^\dagger +\phi(z^\ast,n)^\dagger\,b(n)+
\phi(z^\ast,n-1)^\dagger\, a(n)=
\displaystyle
\frac{a_\infty(z+z^{-1})+b_\infty}{w_\infty}
\,
\phi(z^\ast,n)^\dagger\,w(n).
\end{equation}
On the other hand, since $\varphi(z,n)$ satisfies \eqref{2.17} we have
\begin{equation}\label{3.29}
a(n+1)\,\varphi(z,n+1)+b(n)\,\varphi(z,n)+
a(n)^\dagger\,\varphi(z,n-1)=
\displaystyle
\frac{a_\infty(z+z^{-1})+b_\infty}{w_\infty}
\,
w(n)\,\varphi(z,n).
\end{equation}
We multiply both sides of \eqref{3.28} on the right with $\varphi(z,n)$ and multiply both sides of \eqref{3.29} on the left with
$\phi(z^\ast,n)^\dagger.$ We then subtract the resulting equations side by side. This yields the equality
\begin{equation}\label{3.30}
\begin{split}
\phi(z^\ast,n+1)^\dagger\,a(n+1)^\dagger& \,\varphi(z,n)-\phi(z^\ast,n)^\dagger\,a(n+1)\,\varphi(z,n+1)\\
&+
\phi(z^\ast,n-1)^\dagger\, a(n)\,\varphi(z,n)-
\phi(z^\ast,n)^\dagger\, a(n)^\dagger \varphi(z,n-1)=
0.
\end{split}
\end{equation}
With the help of \eqref{3.26}, we can write \eqref{3.30} as
\begin{equation*}
-[\phi(z^\ast,n)^\dagger;\varphi(z,n)]+[\phi(z^\ast,n-1);\varphi(z,n-1)]=0,\qquad n\in\mathbb Z,
\end{equation*}
from which we see that the Wronskian evaluated at $n$ and the Wronskian 
evaluated at $n-1$ are equal to each other.
Hence, the proof is complete.
\end{proof}

In the next theorem, 
we present the Wronskians
of various pairs of solutions to the Jacobi system \eqref{2.17} as $n\to\pm\infty,$ and 
we express those spacial asymptotics in terms of the corresponding scattering coefficients.
The $n$-independence of those Wronskians yield various useful properties of the 
scattering coefficients.
Those properties are later used to determine various properties of the 
corresponding scattering matrix such as its unitarity.

\begin{theorem}\label{theorem3.4} Assume that the $q\times q$ matrix-valued coefficients in \eqref{1.1} belong to the class
$\mathcal A$ described in Definition~\ref{definition1.1}. Let $f_{\rm{l}}(z,n),$ $f_{\rm{r}}(z,n),$ $g_{\rm{l}}(z,n),$ and $g_{\rm{r}}(z,n)$
be the solutions to
\eqref{2.17} with the respective spacial asymptotics 
\eqref{3.10},  \eqref{3.11}, \eqref{3.15}, \eqref{3.16}, \eqref{3.20}, and \eqref{3.21}, and let
$T_{\rm{l}}(z),$ $L(z),$ $T_{\rm{r}}(z),$ $R(z)$ be the scattering coefficients
appearing in \eqref{3.22}.
The aforementioned solutions satisfy the Wronskian relations for $z\in\mathbb T \setminus\{-1,1\} ,$ which are independent of $n,$ that are given by 
\begin{equation}
\label{3.32}
 [f_{\rm{l}}(z,n)^\dagger;f_{\rm{l}}(z,n)]=\left(z-z^{-1}\right)a_\infty\,I=\left(z-z^{-1}\right)a_\infty\,[T_{\rm{l}}(z)^\dagger]^{-1}\left[I-L(z)^\dagger L(z)\right]T_{\rm{l}}(z)^{-1},
\end{equation}
\begin{equation}
\label{3.33}
 [f_{\rm{l}}(z,n)^\dagger;g_{\rm{l}}(z,n)]=0=
\left(z-z^{-1}\right)a_\infty\,T_{\rm{l}}(z^{-1})^\dagger\left[L(z^{-1})-L(z)^\dagger\right]T_{\rm{l}}(z^{-1})^{-1},
\end{equation}
\begin{equation}
\label{3.34}
 [g_{\rm{l}}(z,n)^\dagger;f_{\rm{l}}(z,n)]=0=\left(z-z^{-1}\right)a_\infty\, [T_{\rm{l}}(z^\ast)^\dagger]^{-1}\left[L(z^\ast)^\dagger-L(z)\right] T_{\rm{l}}(k)^{-1},
\end{equation}
\begin{equation}
\label{3.35}
 [g_{\rm{l}}(z,n)^\dagger;g_{\rm{l}}(z,n)]=\left(z-z^{-1}\right)a_\infty\,I=
\left(z^{-1}-z\right)a_\infty\, [T_{\rm{l}}(z^{-1})^\dagger]^{-1}\left[I-L(z^{-1})^\dagger\,L(z^{-1})\right] T_{\rm{l}}(z^{-1})^{-1},
\end{equation}
\begin{equation}
\label{3.36}
 [f_{\rm{r}}(z,n)^\dagger;f_{\rm{r}}(z,n)]=
 \left(z^{-1}-z\right)a_\infty\,[T_{\rm{r}}(z)^\dagger]^{-1}\left[I-R(z)^\dagger R(z)\right]T_{\rm{r}}(z)^{-1}=\left(z^{-1}-z\right)a_\infty\,I,
\end{equation}
\begin{equation}
\label{3.37}
 [g_{\rm{r}}(z,n)^\dagger;f_{\rm{r}}(z,n)]=
\left(z^{-1}-z\right)a_\infty\, [T_{\rm{r}}(z^\ast)^\dagger]^{-1}\left[R(z^\ast)^\dagger-R(z)\right] T_{\rm{r}}(z)^{-1}=0,
\end{equation}
\begin{equation}
\label{3.38}
 [f_{\rm{l}}(z,n)^\dagger;f_{\rm{r}}(z,n)]=\left(z-z^{-1}\right)a_\infty\,R(z)\,T_{\rm{r}}(z)^{-1}=\left(z^{-1}-z\right)a_\infty\,[T_{\rm{l}}(z)^\dagger]^{-1}
 L(z)^\dagger,
\end{equation}
 \begin{equation}
\label{3.39}
 [f_{\rm{l}}(z^\ast,n)^\dagger;f_{\rm{r}}(z,n)]=-\left(z-z^{-1}\right)a_\infty\,T_{\rm{r}}(z)^{-1}
=-\left(z-z^{-1}\right)a_\infty\,[T_{\rm{l}}(z^\ast)^\dagger]^{-1},
\end{equation}
 \begin{equation}
\label{3.40}
 [f_{\rm{r}}(z^\ast,n)^\dagger;f_{\rm{l}}(z,n)]=\left(z-z^{-1}\right)a_\infty\,[T_{\rm{r}}(z^\ast)^\dagger]^{-1}
=\left(z-z^{-1}\right)a_\infty\,T_{\rm{l}}(z)^{-1},
\end{equation}
where the middle value in each set equalities represents the asymptotic value of the Wronskian
as $n\to+\infty$ and the value on the right
represents the asymptotic value as $n\to-\infty.$

\end{theorem}

\begin{proof}
With the help of \eqref{2.19} and  \eqref{3.19}, we obtain the Wronskian values as $n\to\pm\infty$ by using
the spacial asymptotics given in \eqref{3.10}, \eqref{3.11}, \eqref{3.15}, \eqref{3.16}, \eqref{3.20}, and \eqref{3.21}.
For example, we establish \eqref{3.32} as follows. Using \eqref{2.19} and the first equality of  \eqref{3.19}, we obtain
\begin{equation}
\label{3.41}
 [f_{\rm{l}}(z,n)^\dagger;f_{\rm{l}}(z,n)]= [g_{\rm{l}}(z^\ast,n)^\dagger;f_{\rm{l}}(z,n)],\qquad z\in\mathbb T \setminus\{-1,1\}.
\end{equation}
By Proposition~\ref{proposition3.3} we know that the value of the Wronskian on the right-hand side of \eqref{3.41} is
equal to its asymptotic value as $n\to+\infty.$
From \eqref{2.19} and  the first line of \eqref{3.20}, we have the spacial asymptotics
\begin{equation}
\label{3.42}
g_{\rm{l}}(z^\ast,n)^\dagger=z^{-n}\left[I+o(1)\right].
\qquad n\to+\infty.
\end{equation}
Using \eqref{3.10} and \eqref{3.42} in \eqref{3.26}, as $n\to+\infty$ we get
\begin{equation}
\label{3.43}
 [g_{\rm{l}}(z^\ast,n)^\dagger;f_{\rm{l}}(z,n)]=a_\infty \,z^{-n}\,z^{n+1}\,I-a_\infty\,z^{-n-1}\,z^{n}\,I,
\end{equation}
which yields the first equality in \eqref{3.32} as a consequence of \eqref{3.41}.
To establish the second equality in \eqref{3.32}, we evaluate the asymptotics of the right-hand side of \eqref{3.41} as 
$n\to-\infty.$ From the second line of \eqref{3.20}, with the help of \eqref{2.19}, we have
\begin{equation}
\label{3.44}
g_{\rm{l}}(z^\ast,n)^\dagger=z^{-n}\left[[T_{\rm{l}}(z)^{-1}]^\dagger+o(1)\right]+z^n\left[
[T_{\rm{l}}(z)^{-1}]^\dagger\,L(z)^\dagger+o(1)\right],
\qquad n\to-\infty.
\end{equation}
Using \eqref{3.15} and  \eqref{3.44} on the left-hand side of \eqref{3.43}, since that left-hand side is independent of $n,$
we obtain 
\begin{equation*}
\begin{split}
 [g_{\rm{l}}(z^\ast,n)^\dagger;f_{\rm{l}}(z,n)]=&a_\infty \,z\,[T_{\rm{l}}(z)^{-1}]^\dagger\,T_{\rm{l}}(z)^{-1}+
a_\infty
\,z^{-1}\,[T_{\rm{l}}(z)^{-1}]^\dagger \,L(z)^\dagger\,L(z)\,T_{\rm{l}}(z)^{-1}\\
&
-a_\infty \,z^{-1}\,[T_{\rm{l}}(z)^{-1}]^\dagger\,T_{\rm{l}}(z)^{-1}-
a_\infty
\,z\,[T_{\rm{l}}(z)^{-1}]^\dagger \,L(z)^\dagger\,L(z)\,T_{\rm{l}}(z)^{-1},
\end{split}
\end{equation*}
where the right-hand side, after some simplification, yields the right-hand side of \eqref{3.32}.
The remaining set of equalities \eqref{3.33}--\eqref{3.40} are established similarly.
\end{proof}

Let us use $\mathbf I$ to denote the $2q\times 2q$ identity matrix, i.e. we define
\begin{equation*}
\mathbf I:=
\begin{bmatrix}I&0\\
0&I\end{bmatrix},
\end{equation*}
where we recall that we use $I$ to denote the $q\times q$ identity matrix.
With the help of \eqref{3.32}, \eqref{3.36}, and \eqref{3.38}, we obtain
\begin{equation}
\label{3.47}
\begin{bmatrix}T_{\rm{l}}(z)^\dagger \,T_{\rm{l}}(z)+L(z)^\dagger\, L(z) &T_{\rm{l}}(z)^\dagger \,R(z)+L(z)^\dagger\, T_{\rm{r}}(z)\\
\noalign{\medskip}
R(z)^\dagger \,T_{\rm{l}}(z)+T_{\rm{r}}(z)^\dagger\, L(z)&T_{\rm{r}}(z)^\dagger \,T_{\rm{r}}(z)+R(z)^\dagger\, R(z)\end{bmatrix}=\mathbf I,\qquad z\in\mathbb T,
\end{equation}
which is equivalent to 
\begin{equation}
\label{3.48}
S(z)^\dagger\, S(z)=\mathbf I,\qquad z\in\mathbb T,
\end{equation}
where $S(z)$ is the scattering matrix defined in \eqref{3.22}.
From \eqref{3.48} we conclude that  $S(z)$ is unitary, and hence we also have
\begin{equation}
\label{3.49}
S(z)\,S(z)^\dagger=\mathbf I,\qquad z\in\mathbb T.
\end{equation}
Since $S(k)$ is a unitary matrix, the absolute value of its determinant is equal to $1.$
We note that \eqref{3.49} is equivalent to
\begin{equation}
\label{3.50}
\begin{bmatrix}T_{\rm{l}}(z)\,T_{\rm{l}}(z)^\dagger+R(z) \,R(z)^\dagger&T_{\rm{l}}(z)\, L(z)^\dagger+R(z) \,T_{\rm{r}}(z)^\dagger\\
\noalign{\medskip}
L(z)\, T_{\rm{l}}(z)^\dagger +T_{\rm{r}} (z)\,R(z)^\dagger&T_{\rm{r}} (z)\,T_{\rm{r}}(z)^\dagger+L(z) \,L(z)^\dagger \end{bmatrix}=\mathbf I,\qquad z\in\mathbb T.
\end{equation}
From \eqref{3.34}, \eqref{3.37}, \eqref{3.39}, and \eqref{3.40}, respectively, we obtain
\begin{equation}
\label{3.51}
L(z^\ast)=L(z)^\dagger,\quad R(z^\ast)=R(z)^\dagger, \quad T_{\rm{l}}(z^\ast)=T_{\rm{r}}(z)^\dagger,
\quad T_{\rm{r}}(z^\ast)=T_{\rm{l}}(z)^\dagger,
\qquad z\in\mathbb T,
\end{equation}
which can equivalently be expressed as
\begin{equation*}
S(z)^\dagger= Q\, S(z^\ast)\,Q,  \qquad z \in \mathbb T,
\end{equation*}
where $Q$ is the constant $2q\times 2q$ matrix defined as
\begin{equation}
\label{3.53}
Q:=\begin{bmatrix}0&I\\
I&0\end{bmatrix}.
\end{equation}
We note that the matrix $Q$ is equal to its own inverse.

\section{The transition matrices}
\label{section4}
In preparation for the factorization results in the next section, in this section we introduce
the three relevant $2q\times 2q$ matrices
$F_{\rm{l}}(z,n),$ $F_{\rm{r}}(z,n),$ $G(z,n),$ and
the pair of $2q\times 2q$ transition matrices $\Lambda(z)$
and $\Sigma(z)$ for \eqref{2.17}. We also analyze the basic properties of those five
quantities relevant to the factorization formulas established later in
Section~\ref{section5}.

In terms of the four solutions $f_{\rm{l}}(z,n),$ $f_{\rm{r}}(z,n),$
$g_{\rm{l}}(z,n),$ $g_{\rm{r}}(z,n)$
to \eqref{2.17}, we 
introduce the three $2q\times 2q$ matrices defined as
\begin{equation}
\label{4.1}
F_{\rm{l}}(z,n):=\begin{bmatrix}f_{\rm{l}}(z,n)&g_{\rm{l}}(z,n) \\
\noalign{\medskip}
a(n+1)\,f_{\rm{l}}(z,n+1)&a(n+1)\,g_{\rm{l}}(z,n+1)\end{bmatrix},  \qquad n \in \mathbb Z,
\end{equation}
\begin{equation}
\label{4.2}
F_{\rm{r}}(z,n):=\begin{bmatrix}g_{\rm{r}}(z,n)&f_{\rm{r}}(z,n)\\
\noalign{\medskip}
a(n+1)\,g_{\rm{r}}(z,n+1)&a(n+1)\,f_{\rm{r}}(z,n+1)\end{bmatrix}, \qquad n \in \mathbb Z, 
\end{equation}
\begin{equation}
\label{4.3}
G(z,n):=\begin{bmatrix}f_{\rm{l}}(z,n)&f_{\rm{r}}(z,n)\\
\noalign{\medskip}
a(n+1)\,f_{\rm{l}}(z,n+1)&a(n+1)\,f_{\rm{r}}(z,n+1)\end{bmatrix},  \qquad n \in \mathbb Z.
\end{equation}
From Theorem~\ref{theorem3.2}(b) we know that 
the $z$-domains of 
$f_{\rm{l}}(z,n)$ and $f_{\rm{r}}(z,n)$ are $\overline{\mathbb D},$
where we recall that $\overline{\mathbb D}$ is the closed unit disk in the complex plane.
Consequently, with the help of \eqref{3.19}, we conclude that
the $z$-domains of 
$g_{\rm{l}}(z,n)$ and $g_{\rm{r}}(z,n)$ are $\mathbb C\setminus \mathbb D.$
Thus, from \eqref{4.1} and \eqref{4.2} we see that,
for each fixed $n\in\mathbb Z,$ the matrices
$F_{\rm{l}}(z,n)$ and $F_{\rm{r}}(z,n)$ have the common $z$-domain given by $z\in\mathbb T,$
and from \eqref{3.36} we see that, for each fixed $n\in\mathbb Z,$ the matrix $G(z,n)$ is defined when 
$z\in\overline{\mathbb D}.$

In the next theorem we present the explicit expressions for the 
matrix inverses of $F_{\rm{l}}(z,n),$ $F_{\rm{r}}(z,n),$ $G(z,n).$
The invertibility of the aforementioned matrices is needed later
in the proof of Theorem~\ref{theorem4.2}.

\begin{theorem}\label{theorem4.1}
Assume that the $q\times q$ matrix-valued coefficients in \eqref{2.17}
belong to the class $\mathcal A.$ 
We then have the following:

\begin{enumerate}
\item[\text{\rm(a)}] The $2q\times 2q$ matrix $F_{\rm{l}}(z,n)$ defined in \eqref{4.1} is invertible
when $z\in\mathbb T\setminus\{-1,1\},$ and we have
\begin{equation}
\label{4.4}
F_{\rm{l}}(z,n)^{-1}=\displaystyle\frac{1}{a_\infty\left(z-z^{-1}\right)}
\begin{bmatrix}-f_{\rm{l}}(z,n+1)^\dagger\,a(n+1)^\dagger&f_{\rm{l}}(z,n)^\dagger\\
\noalign{\medskip}
g_{\rm{l}}(z,n+1)^\dagger\,a(n+1)^\dagger&-g_{\rm{l}}(z,n)^\dagger\end{bmatrix},
\qquad n\in\mathbb Z,
\end{equation}
where we recall that $f_{\rm{l}}(z,n)$ and $g_{\rm{l}}(z,n)$
are the solution to \eqref{2.17} appearing in \eqref{3.10} and \eqref{3.20}, respectively.

\item[\text{\rm(b)}] The $2q\times 2q$ matrix $F_{\rm{r}}(z,n)$ defined in \eqref{4.2} is invertible
when $z\in\mathbb T\setminus\{-1,1\},$ and we have
\begin{equation*}
F_{\rm{r}}(z,n)^{-1}=\displaystyle\frac{1}{a_\infty\left(z-z^{-1}\right)}
\begin{bmatrix}-g_{\rm{r}}(z,n+1)^\dagger\,a(n+1)^\dagger&g_{\rm{r}}(z,n)^\dagger\\
\noalign{\medskip}
f_{\rm{r}}(z,n+1)^\dagger\,a(n+1)^\dagger&-f_{\rm{r}}(z,n)^\dagger\end{bmatrix},
\qquad n\in\mathbb Z,
\end{equation*}
where we recall that $f_{\rm{r}}(z,n)$ and $g_{\rm{r}}(z,n)$ are the solutions to \eqref{2.17} appearing in  \eqref{3.11} and \eqref{3.21}, respectively.

\item[\text{\rm(c)}] For each $n\in\mathbb Z,$ the $2q\times 2q$ matrix $G(z,n)$ defined in \eqref{4.3} is invertible
when $z\in\mathbb T\setminus\{-1,1\},$ and
we have
\begin{equation*}
G(z,n)^{-1}=\displaystyle\frac{1}{a_\infty\left(z-z^{-1}\right)}
\begin{bmatrix}T_{\rm{l}} (z)&0\\
\noalign{\medskip}
0&T_{\rm{r}} (z)\end{bmatrix}
\begin{bmatrix}-f_{\rm{r}}(z^\ast,n+1)^\dagger\,a(n+1)^\dagger&f_{\rm{r}}(z^\ast,n)^\dagger\\
\noalign{\medskip}
f_{\rm{l}}(z^\ast,n+1)^\dagger\,a(n+1)^\dagger&-f_{\rm{l}}(z^\ast,n)^\dagger\end{bmatrix}.
\end{equation*}
\end{enumerate}

\end{theorem}

\begin{proof}
We confirm \eqref{4.4} by direct verification. This is done
by postmultiplying the right-hand side of \eqref{4.4} with the matrix
$F_{\rm{l}}(z,n)$ given in \eqref{4.1}. The product yields the $2q\times 2q$ matrix given by
\begin{equation}
\label{4.7}
\displaystyle\frac{1}{a_\infty\left(z-z^{-1}\right)}
\begin{bmatrix}[f_{\rm{l}}(z,n)^\dagger;f_{\rm{l}}(z,n)]&[f_{\rm{l}}(z,n)^\dagger;g_{\rm{l}}(z,n)]\\
\noalign{\medskip}
-[g_{\rm{l}}(z,n)^\dagger;f_{\rm{l}}(z,n)]&-[g_{\rm{l}}(z,n)^\dagger;g_{\rm{l}}(z,n)]\end{bmatrix},
\end{equation}
where we have expressed the entries in terms of the relevant Wronskians with the help of
\eqref{3.26}. By comparing the block entries of \eqref{4.7} with \eqref{3.32}--\eqref{3.35}, we confirm that
the matrix in \eqref{4.7} is equal to the $2q\times 2q$ identity matrix. We exclude the points $z=1$ and $z=-1$
because of the terms $(z-z^{-1})$ appearing in the denominator on the right-hand side of \eqref{4.7}. Hence, the proof of (a) is complete.
The proofs for (b) and (c) are obtained by direct verification as in the proof of (a).
\end{proof}

We introduce the $2q\times 2q$ left transition matrix $\Lambda(z)$ for \eqref{2.17} as
\begin{equation}
\label{4.8}
\Lambda(z):=\begin{bmatrix}T_{\rm{l}}(z)^{-1}&L(z^{-1})\,T_{\rm{l}}(z^{-1})^{-1} \\
\noalign{\medskip}
L(z)\,T_{\rm{l}}(z)^{-1}&T_{\rm{l}}(z^{-1})^{-1}\end{bmatrix},\qquad z\in\mathbb T \setminus\{-1,1\},
\end{equation}
where we recall that $T_{\rm{l}}(z)$ and $L(z)$ are the left scattering coefficients appearing in \eqref{3.22}.
Similarly, 
we introduce the $2q\times 2q$ right transition matrix $\Sigma(z)$ for \eqref{2.17} as
\begin{equation}
\label{4.9}
\Sigma(z):=\begin{bmatrix}T_{\rm{r}}(z^{-1})^{-1}&R(z)\,T_{\rm{r}}(z)^{-1} \\
\noalign{\medskip}
R(z^{-1})\,T_{\rm{r}}(z^{-1})^{-1}&T_{\rm{r}}(z)^{-1}\end{bmatrix},\qquad
z\in\mathbb T \setminus\{-1,1\},
\end{equation}
where we recall that $T_{\rm{r}}(z)$ and $R(z)$ are the right scattering coefficients appearing in \eqref{3.22}.

In the next theorem we show that the transition matrices $\Lambda(z)$ and $\Sigma(z)$ relate the matrices
$F_{\rm{l}}(z,n)$ and $F_{\rm{r}}(z,n)$ to each other, and we present some properties 
of those two transition matrices.
In particular, we show that $\Lambda(z)$ and $\Sigma(z)$ are the matrix inverses of each other. We express
 the determinants of those two transition matrices in terms of the determinants of the left and right transmission 
 coefficients. The significance of our formulas for the determinants of the transition matrices is the following.
 We establish that the determinant of a transition matrix is equal to $1$ if and only if the determinants of the left and right transmission coefficients are equal to each other.

\begin{theorem}\label{theorem4.2}
Assume that the $q\times q$ matrix-valued coefficients in \eqref{2.17} belong to the class $\mathcal A.$
We then have the following:

\begin{enumerate}
\item[\text{\rm(a)}] 
The matrices $F_{\rm{l}}(z,n)$ and $F_{\rm{r}}(z,n)$ 
defined in \eqref{4.1} and \eqref{4.2}, respectively, are related to each other as
\begin{equation}
\label{4.10}
F_{\rm{l}}(z,n)=F_{\rm{r}}(z,n)\,\Lambda(z),\qquad z\in\mathbb T\setminus\{-1,1\},
\end{equation}
where $\Lambda(z)$ is the left transition matrix defined in \eqref{4.8}.

\item[\text{\rm(b)}] 
The matrices $F_{\rm{l}}(z,n)$ and $F_{\rm{r}}(z,n)$ are also related to each 
other via the right transition matrix $\Sigma(z)$ defined in \eqref{4.9}
as
\begin{equation}
\label{4.11}
F_{\rm{r}}(z,n)=F_{\rm{l}}(z,n)\,\Sigma(z),\qquad z\in\mathbb T\setminus\{-1,1\}.
\end{equation}

\item[\text{\rm(c)}] 
The matrices 
$\Lambda(z)$ and $\Sigma(z)$ are inverses of each other for each $z\in\mathbb T\setminus\{-1,1\},$ i.e. we have
\begin{equation}
\label{4.12}
 \Lambda(z)\,\Sigma(z)=\Sigma(z)\,\Lambda(z)=\mathbf I.
\end{equation}

\item[\text{\rm(d)}] The determinant of the left transition matrix $\Lambda(z)$ is related to the determinants of 
the left and right transmission coefficients as
\begin{equation}
\label{4.13}
\det[ \Lambda(z)]=\displaystyle\frac{\det[T_{\rm{r}}(z)]}{\det[T_{\rm{l}}(z)]},\qquad z\in\mathbb T\setminus\{-1,1\}.
\end{equation}

\item[\text{\rm(e)}] The determinant of the right transition matrix $\Lambda(z)$ is related to the determinants of 
the left and right transmission coefficients as
\begin{equation}
\label{4.14}
\det[ \Sigma(z)]=\displaystyle\frac{\det[T_{\rm{l}}(z)]}{\det[T_{\rm{r}}(z)]},\qquad z\in\mathbb T\setminus\{-1,1\}.
\end{equation}

\end{enumerate}

\end{theorem}

\begin{proof}
For the proof of (a), we proceed as follows. 
With the help  of
\eqref{3.17} and the second equality of \eqref{3.19}
we obtain
\begin{equation}\label{4.15}
f_{\rm{l}}(z,n)=g_{\rm{r}}(z,n)\, T_{\rm{l}}(z)^{-1}+ f_{\rm{r}}(z,n) \,L(z)\,T_{\rm{l}}(z)^{-1} , \qquad z \in \mathbb T\setminus\{-1,1\}.
\end{equation}
Replacing $z$ by $z^{-1}$ in \eqref{4.15} and using \eqref{3.19} in the
resulting matrix equality, we get
\begin{equation}\label{4.16}
g_{\rm{l}}(z,n)=g_{\rm{r}}(z,n)\,L(z^{-1})\, T_{\rm{l}}(z^{-1})^{-1}+ f_{\rm{r}}(z,n) \,T_{\rm{l}}(z^{-1})^{-1} , \qquad z \in \mathbb T\setminus\{-1,1\}.
\end{equation}
With the help of  \eqref{4.16}, we express the entries of the $2q\times 2q$ matrix appearing
on the right-hand side of \eqref{4.1}. That yields
\begin{equation}
\label{4.17}
\begin{split}
&\begin{bmatrix}f_{\rm{l}}(z,n)&g_{\rm{l}}(z,n) \\
\noalign{\medskip}
a(n+1)\,f_{\rm{l}}(z,n+1)&a(n+1)\,g_{\rm{l}}(z,n+1)\end{bmatrix}\\
&\phantom{xxxx}=\begin{bmatrix}g_{\rm{r}}(z,n)&f_{\rm{r}}(z,n)\\
\noalign{\medskip}
a(n+1)\,g_{\rm{r}}(z,n+1)&a(n+1)\,f_{\rm{r}}(z,n+1)\end{bmatrix}
\begin{bmatrix}T_{\rm{l}}(z)^{-1}&L(z^{-1})\,T_{\rm{l}}(z^{-1})^{-1} \\
\noalign{\medskip}
L(z)\,T_{\rm{l}}(z)^{-1}&T_{\rm{l}}(z^{-1})^{-1}\end{bmatrix}.
\end{split}
\end{equation}
With the help of \eqref{4.1} and \eqref{4.2},
we observe that \eqref{4.17} is the same as the equality in \eqref{4.10}. Thus, the proof of (a) is complete.
The proof of (b) is obtained in a similar manner.
Replacing $z$ by $z^{-1}$ in \eqref{3.18} and using \eqref{3.19}, we obtain
\begin{equation}\label{4.18}
g_{\rm{r}}(z,n)=f_{\rm{l}}(z,n)\, T_{\rm{r}}(z^{-1})^{-1}+ g_{\rm{l}}(z,n) \,R(z^{-1})\,T_{\rm{r}}(z^{-1})^{-1} , \qquad z \in \mathbb T\setminus\{-1,1\}.
\end{equation}
Using the first equality of \eqref{3.19} in \eqref{3.18}, we get
\begin{equation}\label{4.19}
f_{\rm{r}}(z,n)=f_{\rm{l}}(z,n)\, R(z)\,T_{\rm{r}}(z)^{-1}+ g_{\rm{l}}(z,n) \,T_{\rm{r}}(z)^{-1} , \qquad z \in \mathbb T\setminus\{-1,1\}.
\end{equation}
With the help of \eqref{4.18} and \eqref{4.19}, we express the entries of the $2q\times 2q$ matrix appearing
on the right-hand side of \eqref{4.2}. That yields
\begin{equation}
\label{4.20}
\begin{split}
&\begin{bmatrix}g_{\rm{r}}(z,n)&f_{\rm{r}}(z,n)\\
\noalign{\medskip}
a(n+1)\,g_{\rm{r}}(z,n+1)&a(n+1)\,f_{\rm{r}}(z,n+1)\end{bmatrix}
\\
&\phantom{xxxx}=\begin{bmatrix}f_{\rm{l}}(z,n)&g_{\rm{l}}(z,n) \\
\noalign{\medskip}
a(n+1)\,f_{\rm{l}}(z,n+1)&a(n+1)\,g_{\rm{l}}(z,n+1)\end{bmatrix}
\begin{bmatrix}T_{\rm{r}}(z^{-1})^{-1}&R(z)\,T_{\rm{r}}(z)^{-1} \\
\noalign{\medskip}
R(z^{-1})\,T_{\rm{r}}(z^{-1})^{-1}&T_{\rm{r}}(z)^{-1}\end{bmatrix}.
\end{split}
\end{equation}
With the help of \eqref{4.1} and \eqref{4.2}
we see that \eqref{4.20} is the same as the equality in \eqref{4.11}. Thus, the proof of (b) is also complete.
For the proof of (c) we proceed as follows. From Theorem~\ref{theorem4.1} we know that
the matrices 
$F_{\rm{l}}(z,n)$ and $F_{\rm{r}}(z,n)$ are both invertible when $z\in\mathbb T\setminus\{-1,1\}.$ Thus, from \eqref{4.10} and \eqref{4.11} we obtain
\begin{equation}\label{4.21}
\Lambda(z)=F_{\rm{r}}(z,n)^{-1}\,F_{\rm{l}}(z,n),\quad
\Sigma(z)=F_{\rm{l}}(z,n)^{-1}\,F_{\rm{r}}(z,n),\qquad 
z\in\mathbb T \setminus\{-1,1\}.
\end{equation}
From the two equalities in \eqref{4.21} we obtain \eqref{4.12}, and hence the proof of (c) is complete.
For the proof of (d) we can use the 
block
matrix factorization formula \cite{D2006} involving a Schur complement, i.e.
\begin{equation}
\label{4.22}
\begin{bmatrix}M_1&M_2 \\
\noalign{\medskip}
M_3&M_4\end{bmatrix}=
\begin{bmatrix}I&0\\
\noalign{\medskip}
M_3\, M_1^{-1}&I\end{bmatrix}
\begin{bmatrix}M_1&0\\
\noalign{\medskip}
0&M_4-M_3\, M_1^{-1} \,M_2\end{bmatrix}
\begin{bmatrix}I&M_1^{-1}\,M_2\\
\noalign{\medskip}
0&I\end{bmatrix},
\end{equation}
where each of $M_1,$ $M_2,$ $M_3,$ and $M_4$ is any $q\times q$ matrix.
From \eqref{4.22} we get the determinant identity given by
\begin{equation}
\label{4.23}
\det \begin{bmatrix}M_1&M_2 \\
\noalign{\medskip}
M_3&M_4\end{bmatrix}=\det[M_1]\,\det[M_4-M_3\, M_1^{-1} \,M_2].
\end{equation}
With the help of \eqref{2.19} and \eqref{3.51}, we can write $\Lambda(z)$ given in \eqref{4.8} as
\begin{equation}\label{4.24}
\Lambda(z)=\begin{bmatrix}T_{\rm{l}}(z)^{-1}&L(z)^\dagger\,[T_{\rm{r}}(z)^\dagger]^{-1} \\
\noalign{\medskip}
L(z)\,T_{\rm{l}}(z)^{-1}&[T_{\rm{r}}(z)^\dagger]^{-1}\end{bmatrix},\qquad z\in\mathbb T \setminus\{-1,1\}.
\end{equation}
Applying \eqref{4.23} on the matrix appearing on the right-hand side of \eqref{4.24}, we get
\begin{equation}
\label{4.25}
\det[\Lambda(z)]=\det[T_{\rm{l}}(z)^{-1}]\,\det[I-L(z)\,L(z)^\dagger]\,\det[[T_{\rm{r}}(z)^\dagger]^{-1} ].
\end{equation}
From the $(2,2)$-entries of the matrix equality in \eqref{3.50}, we get
\begin{equation}
\label{4.26}
I-L(z)\,L(z)^\dagger=T_{\rm{r}}(z)\,T_{\rm{r}}(z)^\dagger.
\end{equation}
Using \eqref{4.26} on the right-hand side of \eqref{4.25}, we simplify the resulting equation and get
\begin{equation*}
\det[\Lambda(z)]=\det[T_{\rm{l}}(z)^{-1}]\,\det[T_{\rm{r}}(z)],
\end{equation*}
which yields \eqref{4.13}. Thus, the proof of (d) is complete.
From (c) we know that $\Sigma(z)$ is equal to $\Lambda(z)^{-1}.$ Hence, \eqref{4.14} directly follows from \eqref{4.13}.
Thus, the proof of (e) is also complete.
\end{proof}

In the next theorem we explore further properties of the matrices $F_{\rm{l}}(z,n),$
$F_{\rm{r}}(z,n),$ and 
$G(z,n)$ defined in \eqref{4.1}, \eqref{4.2}, and \eqref{4.3}, respectively.
We relate each of the matrices $F_{\rm{l}}(z,n)$ and
$F_{\rm{r}}(z,n)$ to the matrix $G(z,n)$ and the transmission coefficients. We also relate each of the
 determinants of $F_{\rm{l}}(z,n)$ and
$F_{\rm{r}}(z,n)$ to the determinant of $G(z,n)$ and the determinants of the transmission coefficients.
Furthermore, we express the spacial asymptotics of the determinants of $F_{\rm{l}}(z,n),$
$F_{\rm{r}}(z,n),$ and $G(z,n)$ in terms of the determinants of the transmission coefficients.
The results presented are later used in establishing Theorem~\ref{theorem4.5}.

\begin{theorem}\label{theorem4.3}
Assume that the $q\times q$ matrix-valued coefficients in \eqref{2.17} belong to the class $\mathcal A.$
We then have the following:

\begin{enumerate}

\item[\text{\rm(a)}] The $2q\times 2q$ matrices
$F_{\rm{l}}(z,n)$
and 
$G(z,n)$ are related to each other via the right scattering coefficients as
\begin{equation}
\label{4.28}
G(z,n)=F_{\rm{l}}(z,n)
 \begin{bmatrix}I& R(z)\,T_{\rm{r}}(z)^{-1}\\
\noalign{\medskip}
0&T_{\rm{r}}(z)^{-1}\end{bmatrix},  \qquad z \in \mathbb T\setminus\{-1,1\},
\quad n\in\mathbb Z.
\end{equation}

\item[\text{\rm(b)}] The $2q \times 2q$ matrices
$F_{\rm{r}}(z,n)$
and 
$G(z,n)$ are related to each other via the left scattering coefficients as
\begin{equation}
\label{4.29}
G(z,n)=F_{\rm{r}}(z,n)
 \begin{bmatrix}T_{\rm{l}}(z)^{-1}&0\\
\noalign{\medskip}
L(z)\,T_{\rm{l}}(z)^{-1}&I\end{bmatrix},  \qquad z \in \mathbb T\setminus\{-1,1\},
\quad n\in\mathbb Z.
\end{equation}

\item[\text{\rm(c)}] When $z\in\mathbb T\setminus\{-1,1\}$ and $n\in\mathbb Z,$
the determinants of $F_{\rm{l}}(z,n),$ $F_{\rm{r}}(z,n),$
and 
$G(z,n)$ are related to each other via the transmission coefficients as
\begin{equation}
\label{4.30}
\det[F_{\rm{l}}(z,n)]=
\det[G(z,n)]\,\det[T_{\rm{r}}(z)],\quad
\det[F_{\rm{r}}(z,n)]=
\det[G(z,n)]\,\det[T_{\rm{l}}(z)].
\end{equation}

\item[\text{\rm(d)}] 
The determinants of $F_{\rm{l}}(z,n),$ $F_{\rm{r}}(z,n),$ $G(z,n)$ are independent of $n$ 
if and only if the determinant of
the $q\times q$ matrix $a(n)$ appearing in 
\eqref{2.17} is real for all $n\in\mathbb Z.$ This is a consequence of 
the determinant equalities given by
\begin{equation}
\label{4.31}
\det[F_{\rm{l}}(z,n)]=\displaystyle\frac{\left(\det[a(n)]\right)^\ast}
{\det[a(n)]}\,\det[F_{\rm{l}}(z,n-1)],
\qquad z \in \mathbb T\setminus\{-1,1\},\quad n\in\mathbb Z,
\end{equation}
\begin{equation}
\label{4.32}
\det[F_{\rm{r}}(z,n)]=\displaystyle\frac{\left(\det[a(n)]\right)^\ast}
{\det[a(n)]}\,\det[F_{\rm{r}}(z,n-1)],
\qquad z \in \mathbb T\setminus\{-1,1\},\quad n\in\mathbb Z,
\end{equation}
\begin{equation}
\label{4.33}
\det[G(z,n)]=\displaystyle\frac{\left(\det[a(n)]\right)^\ast}
{\det[a(n)]}\,\det[G(z,n-1)],
\qquad z \in \mathbb T\setminus\{-1,1\},\quad n\in\mathbb Z.
\end{equation}
Consequently, in the special case where the matrix $a(n)$ is selfadjoint for
all $n\in\mathbb Z,$ it follows that the determinant of
each of three matrices $F_{\rm{l}}(z,n),$ $F_{\rm{r}}(z,n),$ $G(z,n)$ is independent of $n.$

\item[\text{\rm(e)}] We have the spacial asymptotics 
\begin{equation}\label{4.34}
\lim_{n\to+\infty}\det\left[F_{\rm{l}}(z,n)\right]=\left(z^{-1}-z\right)^q\,a_\infty^q, \qquad z \in \mathbb T\setminus\{-1,1\},
\end{equation}
\begin{equation}\label{4.35}
\lim_{n\to-\infty}\det\left[F_{\rm{r}}(z,n)\right]=\left(z^{-1}-z\right)^q\,a_\infty^q, \qquad z \in \mathbb T\setminus\{-1,1\},
\end{equation}
\begin{equation}\label{4.36}
\lim_{n\to-\infty}\det\left[F_{\rm{l}}(z,n)\right]=\left(z^{-1}-z\right)^q\,a_\infty^q\,
\displaystyle\frac{\det[T_{\rm{r}}(z)]}
{\det[T_{\rm{l}}(z)]}
,\qquad z \in \mathbb T\setminus\{-1,1\},
\end{equation}
\begin{equation}\label{4.37}
\lim_{n\to+\infty}\det\left[F_{\rm{r}}(z,n)\right]=\left(z^{-1}-z\right)^q\,a_\infty^q\,
\displaystyle\frac{\det[T_{\rm{l}}(z)]}
{\det[T_{\rm{r}}(z)]}
,\qquad z \in \mathbb T\setminus\{-1,1\},
\end{equation}
\begin{equation}\label{4.38}
\lim_{n\to+\infty}\det\left[G(z,n)\right]= \frac{\left(z^{-1}-z\right)^q\,a_\infty^q}{\det\left[T_{\rm{r}}(z)\right]}, \qquad z \in\mathbb T\setminus\{-1,1\},
\end{equation}
\begin{equation}\label{4.39}
\lim_{n\to-\infty}\det\left[G(z,n)\right]= \frac{\left(z^{-1}-z\right)^q\,a_\infty^q}{\det\left[T_{\rm{l}}(z)\right]}, \qquad z \in\mathbb T\setminus\{-1,1\}.
\end{equation}

\end{enumerate}

\end{theorem}

\begin{proof}
To establish \eqref{4.28}, we proceed as follows. By using \eqref{4.19} we express $f_{\rm{r}}(z,n)$
appearing on the right-hand side of \eqref{4.3} in terms of
$f_{\rm{l}}(z,n)$ and $g_{\rm{l}}(z,n).$ For $n\in\mathbb Z$ this yields
\begin{equation}
\label{4.40}
\begin{split}
G&(z,n)\\
&=\begin{bmatrix}f_{\rm{l}}(z,n)&f_{\rm{l}}(z,n)\,R(z)\,T_{\rm{r}}(z)^{-1}+g_{\rm{l}}(z,n)\,T_{\rm{r}}(z)^{-1}
       \\
\noalign{\medskip}
a(n+1)\,f_{\rm{l}}(z,n+1)&a(n+1)\,[ f_{\rm{l}}(z,n+1)\,R(z)\,T_{\rm{r}}(z)^{-1}+g_{\rm{l}}(z,n+1)\,T_{\rm{r}}(z)^{-1}  ]\end{bmatrix}. 
\end{split}\end{equation}
The matrix on the right-hand side of \eqref{4.40} is equal to the matrix product on the right-hand side of \eqref{4.28}.
Hence, the proof of (a) is complete. To prove (b), we use \eqref{4.15} and express 
$f_{\rm{l}}(z,n)$
appearing on the right-hand side of \eqref{4.3} in terms of
$g_{\rm{r}}(z,n)$ and $f_{\rm{r}}(z,n).$ For $n\in\mathbb Z$ this yields
\begin{equation}
\label{4.41}
G(z,n)=\begin{bmatrix}g_{\rm{r}}(z,n)\,T_{\rm{l}}(z)^{-1}+f_{\rm{r}}(z,n)\,L(z)\,T_{\rm{l}}(z)^{-1}
&f_{\rm{r}}(z,n)
       \\
\noalign{\medskip}
a(n+1)[ g_{\rm{r}}(z,n+1)\,T_{\rm{l}}(z)^{-1}+f_{\rm{r}}(z,n+1)\,L(z)\,T_{\rm{l}}(z)^{-1}]&a(n+1)\,f_{\rm{r}}(z,n)   \end{bmatrix}. \end{equation}
The matrix on the right-hand side of \eqref{4.41} is equal to the matrix product on the right-hand side of \eqref{4.29}.
Hence, the proof of (b) is complete. 
By taking the determinants of both sides of \eqref{4.28}, we get
\begin{equation*}
\det[G(z,n)]=\det[F_{\rm{l}}(z,n)]\,\det[T_{\rm{r}}(z)^{-1}],
\end{equation*}
which yields the first equality in \eqref{4.30}. Similarly,
by taking the determinant of both sides of \eqref{4.29} we get
\begin{equation}
\label{4.43}
\det[G(z,n)]=\det[F_{\rm{r}}(z,n)]\,
\det[T_{\rm{l}}(z)^{-1}],
\end{equation}
which yields the second equality in \eqref{4.30}.
Hence, the proof of (c) is complete.
For the proof of \eqref{4.31} we proceed as follows.
We replace the second row block on the right-hand side of \eqref{4.1} by its equivalent obtained with the help of \eqref{1.1}
satisfied by $f_{\rm{l}}(z,n)$ and $f_{\rm{r}}(z,n),$ respectively.
For $n\in\mathbb Z$ this yields
\begin{equation}
\label{4.44}
\begin{split}
F&_{\rm{l}}(z,n)\\
&=\begin{bmatrix}f_{\rm{l}}(z,n)&g_{\rm{l}}(z,n) \\
\noalign{\medskip}
[\lambda\,w(n)-b(n)]\,f_{\rm{l}}(z,n)-a(n)^\dagger\,f_{\rm{l}}(z,n-1)
&[\lambda\,w(n)-b(n)]\,g_{\rm{l}}(z,n)-a(n)^\dagger\,g_{\rm{l}}(z,n-1)\end{bmatrix},
\end{split}\end{equation}
where we recall that $\lambda$ is related to $z$ as in \eqref{2.16}.
Using an elementary block row operation on the right-hand side of \eqref{4.44}, which does not affect the determinant, 
we get
\begin{equation}
\label{4.45}
\det[F_{\rm{l}}(z,n)]=\det\begin{bmatrix}f_{\rm{l}}(z,n)&g_{\rm{l}}(z,n) \\
\noalign{\medskip}
-a(n)^\dagger\,f_{\rm{l}}(z,n-1)
&-a(n)^\dagger\,g_{\rm{l}}(z,n-1)\end{bmatrix}.
\end{equation}
From \eqref{4.45} we have
\begin{equation}
\label{4.46}
\det[F_{\rm{l}}(z,n)]=\det[a(n)^\dagger]\,
\det\begin{bmatrix}f_{\rm{l}}(z,n)&g_{\rm{l}}(z,n) \\
\noalign{\medskip}
-f_{\rm{l}}(z,n-1)
&-g_{\rm{l}}(z,n-1)\end{bmatrix}.
\end{equation}
By interchanging the two row blocks in the second matrix on the right-hand side of \eqref{4.46},
we obtain
\begin{equation*}
\det[F_{\rm{l}}(z,n)]=\det[a(n)^\dagger]\,
\det\begin{bmatrix}f_{\rm{l}}(z,n-1)&g_{\rm{l}}(z,n-1) \\
\noalign{\medskip}
f_{\rm{l}}(z,n)
&g_{\rm{l}}(z,n)\end{bmatrix},
\end{equation*}
which is equivalent to
\begin{equation}
\label{4.48}
\det[F_{\rm{l}}(z,n)]=\det[a(n)^\dagger]\,\det[a(n)^{-1}]\,
\det\begin{bmatrix}f_{\rm{l}}(z,n-1)&g_{\rm{l}}(z,n-1) \\
\noalign{\medskip}
a(n)\,f_{\rm{l}}(z,n)
&a(n)\,g_{\rm{l}}(z,n)\end{bmatrix}.
\end{equation}
With the help of \eqref{4.1}, we recognize the third matrix on the right-hand side of \eqref{4.48} as 
$F_{\rm{l}}(z,n-1).$ Hence, we write \eqref{4.48} as
\begin{equation}
\label{4.49}
\det[F_{\rm{l}}(z,n)]=\det[a(n)^\dagger]\,\det[a(n)^{-1}]\,
\det[F_{\rm{l}}(z,n-1)].
\end{equation}
Since $\det[a(n)^\dagger]=(\det[a(n)])^\ast,$ we observe that \eqref{4.49} is equivalent to \eqref{4.31}.
Thus, the proof of \eqref{4.31} is complete.
The transition matrix $\Lambda(z)$ is independent of $n,$ and hence 
$\det[\Lambda(z)]$ is also independent of $n.$
Thus, using \eqref{4.10} in \eqref{4.31}, we establish \eqref{4.32}.
Since $\det[T_{\rm{l}}(z)]\ne 0$ for $z\in\mathbb T\setminus\{-1,1\}$ as
stated in Theorem~\ref{theorem3.2}(e),
using \eqref{4.43} in \eqref{4.32} we establish
\eqref{4.33}.
Having established \eqref{4.31}--\eqref{4.33}, we see that the proof of (d) is complete.
For the proof of (e) we proceed as follows. Since the determinant of a matrix is a continuous function
of its entries, we obtain \eqref{4.34} by using  \eqref{3.10} and  the first line of \eqref{3.20}
on the right-hand side of \eqref{4.1}.
We obtain \eqref{4.35} in a similar manner by using \eqref{3.11} and the second line of \eqref{3.21} on the right-hand side of \eqref{4.2}.
From \eqref{4.30} we have
\begin{equation}
\label{4.50}
\displaystyle\frac{\det[F_{\rm{l}}(z,n)]}
{\det[F_{\rm{r}}(z,n)]}=\displaystyle\frac{\det[T_{\rm{r}}(z)]}
{\det[T_{\rm{l}}(z)]},\qquad z \in \mathbb T\setminus\{-1,1\}, \quad n\in\mathbb Z.
\end{equation}
Thus, using \eqref{4.35} and \eqref{4.50} we obtain \eqref{4.36}. Similarly, using
\eqref{4.34} and \eqref{4.50} we get \eqref{4.37}.
We obtain \eqref{4.38} by using \eqref{4.34} and the first equality in \eqref{4.30}. Finally, we obtain
\eqref{4.39} by using \eqref{4.35} and the second equality of \eqref{4.30}.
This completes the proof of (e).
\end{proof}

In the next theorem, we relate the matrix $G(z,n)$ defined in \eqref{4.3} to
the scattering matrix $S(z)$ defined in \eqref{3.22}. In the previous theorem, we have expressed 
the spacial asymptotics of
$\det[G(z,n)]$ in terms of the determinant of a transmission coefficient. Thus, the next theorem
relates the determinant of the scattering matrix to the determinants of
the transmission coefficients. Since we already know that the scattering matrix is unitary,
the expression in the theorem for the determinant of $S(z)$ indicates that 
the determinants of the left and right scattering coefficients have the same absolute value.

\begin{theorem}\label{theorem4.4}
Assume that the $q\times q$ matrix-valued coefficients in \eqref{2.17} belong to the class $\mathcal A.$
We then have the following:

\begin{enumerate} 
\item[\text{\rm(a)}] 
The matrix $G(z,n)$ defined in \eqref{4.3} satisfies
\begin{equation}\label{4.51}
G(z^\ast,n)= G(z,n) \begin{bmatrix}- R(z)& T_{\rm{l}}(z)\\
\noalign{\medskip}
T_{\rm{r}}(z)& -L(z) \end{bmatrix},  \qquad z \in \mathbb T\setminus\{-1,1\}.
\end{equation}
where we recall that an asterisk is used to denote complex conjugation.

\item[\text{\rm(b)}] The equality in \eqref{4.51} is expressed equivalently with the help of the scattering matrix
$S(z)$ defined in \eqref{3.22} as
\begin{equation}\label{4.52}
G(z^\ast,n) = G(z,n) \,J \,S(z)\, J \,Q, \qquad z \in \mathbb T\setminus\{-1,1\},
\end{equation}
where $Q$ is the $2q\times 2q$ constant matrix defined in \eqref{3.53} and $J$ is the $2q\times 2q$ constant matrix defined as
\begin{equation}
\label{4.53}
J:=\begin{bmatrix}I&0\\
0&-I\end{bmatrix}.
\end{equation}

\item[\text{\rm(c)}] The determinant of the scattering matrix $S(z)$ is given by
\begin{equation}
\label{4.54}
\det\left[S(z)\right]= \frac{\det\left[T_{\rm{r}}(z)\right]}{\left(\det\left[T_{\rm{l}}(z)\right]\right)^\ast}, \qquad z \in \mathbb T\setminus\{-1,1\}.
\end{equation}
As a consequence of the unitarity of $S(z),$ it follows that $\det\left[T_{\rm{l}}(z)\right]$
and $\det\left[T_{\rm{r}}(z)\right]$ have the same absolute value for each $z\in \mathbb T\setminus\{-1,1\}.$
\end{enumerate}

\end{theorem}

\begin{proof}
Using \eqref{2.19} and  \eqref{3.19} in \eqref{4.3} we get
\begin{equation}
\label{4.55}
G(z^\ast,n)=\begin{bmatrix}g_{\rm{l}}(z,n)&g_{\rm{r}}(z,n)\\
\noalign{\medskip}
a(n+1)\,g_{\rm{l}}(z,n+1)&a(n+1)\,g_{\rm{r}}(z,n+1)\end{bmatrix},  \qquad n \in \mathbb Z.
\end{equation}
From \eqref{3.13} and the second equality of \eqref{3.19}, we obtain
\begin{equation}\label{4.56}
g_{\rm{r}}(z,n)=f_{\rm{l}}(z,n)\, T_{\rm{l}}(z)- f_{\rm{r}}(z,n) \,L(z), \qquad z \in \mathbb T\setminus\{-1,1\},
\end{equation}
and from \eqref{3.12} and the first equality of \eqref{3.19} we have
\begin{equation}\label{4.57}
g_{\rm{l}}(z,n)=f_{\rm{r}}(z,n)\, T_{\rm{r}}(z)- f_{\rm{l}}(z,n) \,R(z), \qquad z \in \mathbb T\setminus\{-1,1\}.
\end{equation}
Using \eqref{4.56} and \eqref{4.57} on the right-hand side of \eqref{4.55}, for $n\in\mathbb Z$ we obtain
\begin{equation}
\label{4.58}
\begin{split}
G&(z^\ast,n)\\
&=\begin{bmatrix}f_{\rm{r}}(z,n)\, T_{\rm{r}}(z)- f_{\rm{l}}(z,n) \,R(z)&f_{\rm{l}}(z,n)\, T_{\rm{l}}(z)- f_{\rm{r}}(z,n) \,L(z)\\
\noalign{\medskip}
a(n+1)\,[f_{\rm{r}}(z,n+1)\, T_{\rm{r}}(z)- f_{\rm{l}}(z,n+1) \,R(z)]&a(n+1)\,[f_{\rm{l}}(z,n)\, T_{\rm{l}}(z)- f_{\rm{r}}(z,n) \,L(z)]\end{bmatrix}.
\end{split}
\end{equation}
The matrix on the right-hand side of \eqref{4.58} is equal to the matrix product on the right-hand side of 
\eqref{4.51}. Hence, the proof of (a) is complete.
Using \eqref{3.22}, \eqref{3.53}, and \eqref{4.53} we have
\begin{equation}
\label{4.59}
J \,S(z)\, J \,Q=\begin{bmatrix}- R(z)& T_{\rm{l}}(z)\\
\noalign{\medskip}
T_{\rm{r}}(z)& -L(z) \end{bmatrix},  \qquad z \in \mathbb T\setminus\{-1,1\}.
\end{equation}
From \eqref{4.51} and \eqref{4.59} we see that \eqref{4.52} holds, and the proof of (b) is complete. For the proof of (c) we proceed as follows.
With the help of \eqref{2.19}, from \eqref{4.38} we get
\begin{equation}\label{4.60}
\lim_{n\to+\infty}\det\left[G(z^\ast,n)\right]=(-1)^q \displaystyle\frac{\left(z^{-1}-z\right)^q\,a_\infty^q}{\det\left[T_{\rm{r}}(z^\ast)\right]}, \qquad z \in\mathbb T\setminus\{-1,1\}.
\end{equation}
From \eqref{3.53} and \eqref{4.53} we have 
\begin{equation}\label{4.61}
\det[Q]=(-1)^q,\quad \det[J]=(-1)^q. 
\end{equation}
Taking the determinant of both sides of \eqref{4.52}, as $n\to+\infty$ we obtain
\begin{equation}\label{4.62}
\lim_{n\to+\infty}\det\left[G(z^\ast,n)\right]=
(\det[J])^2\,\det[Q]\, \det[S(z)]\, \lim_{n\to+\infty}\det\left[G(z,n)\right].
\end{equation}
Using \eqref{4.38}, \eqref{4.60}, and \eqref{4.61} in \eqref{4.62}, after some simplification we get
\begin{equation*}
\displaystyle\frac{1}
{\det[T_{\rm{r}}(z^\ast)]}=
\displaystyle\frac{\det[S(z)]}
{\det[T_{\rm{r}}(z)]},
\end{equation*}
which yields
\begin{equation}
\label{4.64}
\det\left[S(z)\right]= \frac{\det\left[T_{\rm{r}}(z)\right]}{\det\left[T_{\rm{r}}(z^\ast)\right]}, \qquad z \in \mathbb T\setminus\{-1,1\}.
\end{equation}
From the fourth equality of \eqref{3.51} we know that $\det[T_{\rm{r}}(z^\ast)]=\left(\det[T_{\rm{l}}(z)]\right)^\ast,$
and hence \eqref{4.64} yields \eqref{4.54}.
The unitarity of $S(z)$ is seen from \eqref{3.48} or \eqref{3.49}. Thus, the absolute value of the
left-hand side of \eqref{4.64} is equal to $1,$ which yields $|\det[T_{\rm{r}}(z)]|=|\left(\det[T_{\rm{l}}(z)]\right)^\ast|.$
This last equality implies $|\det[T_{\rm{r}}(z)]|=|\det[T_{\rm{l}}(z)]|,$ and hence the proof is complete.
\end{proof}

In the next theorem we show how the determinants of the $2q\times 2q$ matrices
$F_{\rm{l}}(z,n),$ $F_{\rm{r}}(z,n),$ $G(z,n)$
defined in \eqref{4.1}--\eqref{4.3}, respectively, are affected by the presence of the term $a(n)^\dagger$ appearing
in \eqref{2.17} when $a(n)$ is not selfadjoint.
The explicit expressions we present in that case show that those determinants are
no longer independent of $n$ when $n\in\mathbb Z.$ From the previous theorem, as a consequence of the unitarity of the scattering matrix, we already know
that the determinants of the left and right transmission coefficients
have the same absolute value.
Our theorem below confirms this result directly by relating those two determinants
to the determinants of the coefficient matrices $a(n)$ for $n\in\mathbb Z.$ The resulting relationship
also provides an explicit expression revealing the necessary and sufficient
condition for the determinants of
the left and right transmission coefficients to be equal to each other.
This latter result is later used in the proof of Theorem~\ref{theorem4.6}.

\begin{theorem}\label{theorem4.5}
Assume that the $q\times q$ matrix-valued coefficients in \eqref{2.17} belong to the class $\mathcal A.$
We then have the following:

\begin{enumerate} 

\item[\text{\rm(a)}] 
The determinant of the $2q\times 2q$ matrix
$F_{\rm{l}}(z,n)$ defined in \eqref{4.1} is given by
\begin{equation}
\label{4.65}
\det\left[F_{\rm{l}}(z,n)\right]=\left(z^{-1}-z\right)^q\,a_\infty^q\,\prod_{j=n+1}^\infty
\displaystyle\frac{\det[a(j)]}{\left(\det[a(j)]\right)^\ast},\qquad n\in\mathbb Z, \quad z\in\mathbb T\setminus\{-1,1\} .
\end{equation}

\item[\text{\rm(b)}] 
The determinant of the $2q\times 2q$ matrix
$F_{\rm{r}}(z,n)$ defined in \eqref{4.2} is given by
\begin{equation}
\label{4.66}
\det\left[F_{\rm{r}}(z,n)\right]=\left(z^{-1}-z\right)^q\,a_\infty^q\,\prod_{j=-\infty}^n
\displaystyle\frac{\left(\det[a(j)]\right)^\ast}{\det[a(j)]},\qquad n\in\mathbb Z,  \quad z\in\mathbb T\setminus\{-1,1\}.
\end{equation}

\item[\text{\rm(c)}] 
The determinant of the $2q\times 2q$ matrix
$G(z,n)$ defined in \eqref{4.3} is given by
\begin{equation}
\label{4.67}
\det\left[G(z,n)\right]=\displaystyle\frac{\left(z^{-1}-z\right)^q\,a_\infty^q}{
\det[T_{\rm{l}}(z)] }\,\prod_{j=-\infty}^n
\displaystyle\frac{\left(\det[a(j)]\right)^\ast}{\det[a(j)]},\qquad n\in\mathbb Z, \quad z\in\mathbb T\setminus\{-1,1\}.
\end{equation}
where $T_{\rm{l}}(z)$ is the left transmission coefficient appearing in \eqref{3.13}.

\item[\text{\rm(d)}] 
We have
\begin{equation}
\label{4.68}
\displaystyle\frac{\det[T_{\rm{l}}(z)]}{
\det[T_{\rm{r}}(z)] }=\prod_{j=-\infty}^\infty
\displaystyle\frac{\left(\det[a(j)]\right)^\ast}{\det[a(j)]},\qquad z\in\mathbb T\setminus\{-1,1\},
\end{equation}
where $T_{\rm{r}}(z)$ is the right transmission coefficient appearing in \eqref{3.12}.
Hence, the determinants of
$T_{\rm{l}}(z)$ and $T_{\rm{r}}(z)$ have the same absolute value and
differ from each other by a phase factor. 

\end{enumerate}

\end{theorem}

\begin{proof}
We obtain \eqref{4.65} by using iteration on the right-hand side of \eqref{4.31} and by using \eqref{4.34}.
Similarly, we get \eqref{4.66} by using iteration on the right-hand side of \eqref{4.32} and by using
\eqref{4.35}. We obtain \eqref{4.67} by using \eqref{4.66} in the second equality of \eqref{4.30}.
Finally, we get \eqref{4.68} by letting $n\to-\infty$ in \eqref{4.65} and then using \eqref{4.36}. The convergence of
the infinite products in \eqref{4.65}--\eqref{4.68} are confirmed
by using \eqref{1.6}--\eqref{1.8} and the results given in Section~10 of Chapter~8 in \cite{W1934}.  
\end{proof}

If the determinant of the coefficient matrix $a(n)$ appearing in \eqref{2.17} is
real for $n\in\mathbb Z,$ then
there are simplifications in many of the results presented in this section. 
Such a simplification is already indicated in Theorem~\ref{theorem4.3}(d).
In particular, if $a(n)$ is selfadjoint for $n\in\mathbb Z,$ then
$\det[a(n)]$ is real and those simplifications apply.
In the next theorem we indicate the resulting simplification
about the equality of the determinants of the left and right transmission coefficients and the
simplification that the determinants of the left and right transition matrices each become equal to $1.$

\begin{theorem}\label{theorem4.6}
Assume that the $q\times q$ matrix-valued coefficients in \eqref{2.17} belong to the class $\mathcal A.$
Furthermore, assume that $\det[a(n)]$ is real valued for $n\in\mathbb Z.$
We then have the following:

\begin{enumerate} 

\item[\text{\rm(a)}] The determinants of the transmission coefficients
$T_{\rm{l}}(z)$ and $T_{\rm{r}}(z)$ appearing in \eqref{3.13} and  \eqref{3.12},
respectively, are equal to each other, i.e. we have
\begin{equation*}
\det[T_{\rm{l}}(z)]=
\det[T_{\rm{r}}(z)],  \qquad z \in \mathbb T\setminus\{-1,1\}.
\end{equation*}

\item[\text{\rm(b)}] The determinants of the transition matrices
$\Lambda(z)$ and $\Sigma(z)$ appearing in \eqref{4.8} and \eqref{4.9},
respectively, are both equal to $1,$ i.e. we have
\begin{equation*}
\det[\Lambda(z)]=1,\quad
\det[\Sigma(z)]=1,\qquad  z\in\mathbb T\setminus\{1,-1\}.
\end{equation*}

\end{enumerate}

\end{theorem}

\begin{proof}
We note that (a) follows directly from \eqref{4.68}. Then, using (a) in \eqref{4.13} and \eqref{4.14}
we complete the proof of (b).
\end{proof}

\section{The factorization formulas}
\label{section5}

In this section we partition the full-line lattice $\mathbb Z$ into two fragments, and we develop
a factorization formula expressing the transition matrix for the full-line lattice as an ordered matrix product of the transition matrices corresponding to the
left and right fragments.
We then extend the result to the case where the number of fragments is
arbitrary, and we express the transition
matrix for full-line lattice as an ordered matrix product of the transition matrices for the 
fragments. 

The key element in the proof of our factorization formula is to consider the matrix
$G(z,n)$ defined in \eqref{4.3} at the partitioning point $n=m$ and to express
$G(z,m)$ in two equivalent forms, the first of which is related to
the transition matrix for the left fragment and the other is for the right fragment.
We also show how the scattering coefficients
for the full-line lattice are related to the scattering coefficients for the left and right fragments.
Our factorization formula expressed in terms of the left transition matrices 
indicates how the scattering from the full-line lattice takes place as the wave moves from the left to the right
on the full-line lattice and how the scattering from each fragment contributes to the whole scattering.
By taking the matrix inverses in our factorization formula, we obtain a factorization formula
showing how the scattering from the full-line lattice takes place
as the wave moves from the right to the left by expressing the right transition 
matrix for the full-line lattice as an ordered matrix product of the right transition matrices for the fragments.

When the matrix-valued coefficients in \eqref{2.17} belong to the class $\mathcal A,$
let us partition the full-line lattice $\mathbb Z$ into two fragments as
\begin{equation}
\label{5.1}
\mathbb Z=\mathbb Z_1\cup\mathbb Z_2,
\end{equation}
where we have defined
\begin{equation}
\label{5.2}
\mathbb Z_1:=\{\cdots,m-1,m\},\quad \mathbb Z_2:=\{m+1,m+2,\cdots\},
\end{equation}
with $m$ being an arbitrary but fixed integer. Let us define the matrix-valued coefficients
$a_1(n),$ $b_1(n),$ $w_1(n),$ $a_2(n),$ $ b_2(n),$ $w_2(n)$ for $n\in\mathbb Z$ in terms of the three coefficients
$a(n),$ $b(n),$ $w(n)$ and their limiting values appearing in \eqref{1.4} as
\begin{equation}
\label{5.3}
\left(a_1(n),b_1(n),w_1(n)\right):=\begin{cases}
\left(a(n),b(n),w(n)\right),\qquad n\in\mathbb Z_1,\\
\noalign{\medskip}
\left(a_\infty\,I,b_\infty\,I,w_\infty\,I\right),\qquad n\in\mathbb Z_2,
\end{cases}
\end{equation}
\begin{equation}
\label{5.4}
\left(a_2(n),b_2(n),w_2(n)\right):=\begin{cases}
\left(a_\infty\,I,b_\infty\,I,w_\infty\,I\right),\qquad n\in\mathbb Z_1,\\
\noalign{\medskip}
\left(a(n),b(n),w(n)\right),\qquad n\in\mathbb Z_2.
\end{cases}
\end{equation}

Let us consider the analogs of \eqref{1.1} and \eqref{2.17} when the matrix-valued coefficients 
$a(n),$ $b(n),$ $w(n)$ are replaced with
$a_1(n),$ $b_1(n),$ $w_1(n),$ respectively.
Thus, the analog of \eqref{2.17}, for $n\in\mathbb Z,$ is given by
\begin{equation}\label{5.5}
a_1(n+1)\,\phi_1(z,n+1)+b_1(n)\,\phi_1(z,n)+
a_1(n)^\dagger\,\phi_1(z,n-1)=
\displaystyle
\frac{a_\infty(z+z^{-1})+b_\infty}{w_\infty}
\,
w_1(n)\,\phi_1(z,n).
\end{equation}
The analog of \eqref{2.17}, when
the coefficients 
$a(n),$ $b(n),$ $w(n)$ are replaced with
$a_2(n),$ $b_2(n),$ $w_2(n),$ respectively, is, for $n\in\mathbb Z,$ given by
\begin{equation}\label{5.6}
a_2(n+1)\,\phi_2(z,n+1)+b_2(n)\,\phi_2(z,n)+
a_2(n)^\dagger\,\phi_2(z,n-1)=
\displaystyle
\frac{a_\infty(z+z^{-1})+b_\infty}{w_\infty}
\,
w_2(n)\,\phi_2(z,n).
\end{equation}
In analogy with \eqref{3.10}, \eqref{3.11}, \eqref{3.15}, and  \eqref{3.16}, the left and right Jost solutions $f_{\rm{l}1}(z,n)$ and
$f_{\rm{r}1}(z,n),$ respectively, to \eqref{5.5} satisfy
\begin{equation*}
f_{\rm{l}1}(z,n)=\begin{cases}z^n [I+o(1)],\qquad n\to+\infty,\\
 z^n[T_{\rm{l}1}(z)^{-1}+o(1)]+ z^{-n}[L_1(z)\,T_{\rm{l}1}(z)^{-1}+o(1)],\qquad n\to-\infty,
\end{cases}
\end{equation*}
\begin{equation}\label{5.8}
f_{\rm{r}1}(z,n)=\begin{cases} z^{-n}[T_{\rm{r}1}(z)^{-1}+o(1)]+ z^n[R_1(z)\,T_{\rm{r}1}(z)^{-1}+o(1)],\qquad n\to+\infty,\\
z^{-n}[I+o(1)],\qquad n\to-\infty,
\end{cases}
\end{equation}
where $T_{\rm{l}1}(z)$ and $L_1(z)$ are the respective left transmission and left reflection
coefficients  and $T_{\rm{r}1}(z)$ and $R_1(z)$ are the respective right transmission and right reflection
coefficients associated with \eqref{5.5}.
Similarly, in analogy with \eqref{3.10}, \eqref{3.11}, \eqref{3.15}, and  \eqref{3.16}, the left and right Jost solutions $f_{\rm{l}2}(z,n)$ and
$f_{\rm{r}2}(z,n),$ respectively, to \eqref{5.6} satisfy
\begin{equation}\label{5.9}
f_{\rm{l}2}(z,n)=\begin{cases}z^n[I+o(1)],\qquad n\to+\infty,\\
 z^n[T_{\rm{l}2}(z)^{-1}+o(1)] +z^{-n}[L_2(z)\,T_{\rm{l}2}(z)^{-1}+o(1)],\qquad n\to-\infty,
\end{cases}
\end{equation}
\begin{equation*}
f_{\rm{r}2}(z,n)=\begin{cases} z^{-n}[T_{\rm{r}2}(z)^{-1}+o(1)]+ z^n[R_2(z)\,T_{\rm{r}2}(z)^{-1}+o(1)],\qquad n\to+\infty,\\
z^{-n}[I+o(1)],\qquad n\to-\infty,
\end{cases}
\end{equation*}
where $T_{\rm{l}2}(z)$ and $L_2(z)$ are the respective left transmission and left reflection
coefficients  and $T_{\rm{r}2}(z)$ and $R_2(z)$ are the respective right transmission and right reflection
coefficients associated with \eqref{5.6}.
Analogous to \eqref{3.22} we define the $2q\times 2q$ scattering matrices
$S_1(z)$ and $S_2(z)$ corresponding to \eqref{5.5} and \eqref{5.6}, respectively, as
\begin{equation}
\label{5.11}
S_1(z):=\begin{bmatrix}T_{\rm{l}1}(z)&R_1(z) \\ 
\noalign{\medskip}
 L_1(z)& T_{\rm{r}1}(z)\end{bmatrix},\quad S_2(z):=\begin{bmatrix}T_{\rm{l}2}(z)&R_2(z) \\ 
\noalign{\medskip}
 L_2(z)& T_{\rm{r}2}(z)\end{bmatrix},\qquad z\in\mathbb T\setminus \{-1,1\}.
\end{equation}

In terms of the scattering coefficients for \eqref{5.5} and \eqref{5.6}, respectively, we use
$\Lambda_1(z)$ and $\Lambda_2(z)$ as in \eqref{4.8} and
use
$\Sigma_1(z)$ and $\Sigma_2(z)$
as in \eqref{4.9} to denote
the transition matrices corresponding to \eqref{5.5} and \eqref{5.6}, respectively. Thus, we have
\begin{equation}
\label{5.12}
\Lambda_1(z):=\begin{bmatrix}T_{\rm{l}1}(z)^{-1}&L_1(z^{-1})\,T_{\rm{l}1}(z^{-1})^{-1} \\
\noalign{\medskip}
L_1(z)\,T_{\rm{l}1}(z)^{-1}&T_{\rm{l}1}(z^{-1})^{-1}\end{bmatrix},\qquad z\in\mathbb T \setminus\{-1,1\},\end{equation}
\begin{equation}
\label{5.13}
 \Lambda_2(z):=\begin{bmatrix}T_{\rm{l}2}(z)^{-1}&L_2(z^{-1})\,T_{\rm{l}2}(z^{-1})^{-1} \\
\noalign{\medskip}
L_2(z)\,T_{\rm{l}2}(z)^{-1}&T_{\rm{l}2}(z^{-1})^{-1}\end{bmatrix},\qquad z\in\mathbb T\setminus\{-1,1\},
\end{equation}
\begin{equation}
\label{5.14}
\Sigma_1(z):=\begin{bmatrix}T_{\rm{r}1}(z^{-1})^{-1}&R_1(z)\,T_{\rm{r}1}(z)^{-1} \\
\noalign{\medskip}
R_1(z^{-1})\,T_{\rm{r}1}(z^{-1})^{-1}&T_{\rm{r}1}(z)^{-1}\end{bmatrix},\qquad z\in\mathbb T\setminus\{-1,1\},\end{equation}
\begin{equation}
\label{5.15}
 \Sigma_2(z):=\begin{bmatrix}T_{\rm{r}2}(z^{-1})^{-1}&R_2(z)\,T_{\rm{r}2}(z)^{-1} \\
\noalign{\medskip}
R_2(z^{-1})\,T_{\rm{r}2}(z^{-1})^{-1}&T_{\rm{r}2}(z)^{-1}\end{bmatrix},\qquad z\in\mathbb T\setminus\{-1,1\}.
\end{equation}

In the next proposition we present some relevant properties of the Jost solutions to
\eqref{5.5} and \eqref{5.6}, respectively.
Those properties are needed later on in the proof of Proposition~\ref{proposition5.2}.

\begin{proposition}
\label{proposition5.1}
Assume that the $q\times q$ matrix-valued coefficients in \eqref{2.17} belong to the class
$\mathcal A$ described in Definition~\ref{definition1.1}, and let $m$ be the integer
appearing in the partitioning specified in \eqref{5.1}--\eqref{5.4}.
We then have the following:

\begin{enumerate}

\item[\text{\rm(a)}]  For $z\in\mathbb T,$ the right Jost solution $f_{\rm{r}1}(z,n)$ to \eqref{5.5} is related to the right Jost solution $f_{\rm{r}}(z,n)$ to \eqref{2.17} as
\begin{equation}
\label{5.16}
f_{\rm{r}1}(z,n)=f_{\rm{r}}(z,n),\qquad n\le m,
\end{equation} 
\begin{equation}
\label{5.17}
f_{\rm{r}1}(z,m+1)=\displaystyle\frac{a(m+1)}{a_\infty}\,f_{\rm{r}}(z,m+1).
\end{equation}

\item[\text{\rm(b)}]  The right Jost solution $f_{\rm{r}1}(z,n)$ to \eqref{5.5} satisfies
\begin{equation}
\label{5.18}
f_{\rm{r}1}(z,n)=z^{-n}\,T_{\rm{r}1}(z)^{-1}+z^n\,R_1(z)\,T_{\rm{r}1}(z)^{-1},\qquad n\ge m,\quad z\in\mathbb T\setminus\{-1,1\},
\end{equation}
where $T_{\rm{r}1}(z)$ and $R_1(z)$ are the right transmission and reflection coefficients, respectively, for \eqref{5.5}.

\item[\text{\rm(c)}] For $z\in\mathbb T,$ the left Jost solution $f_{\rm{l}2}(z,n)$ to \eqref{5.6} is related to the left Jost solution $f_{\rm{l}}(z,n)$ to \eqref{2.17} as
\begin{equation}
\label{5.19}
f_{\rm{l}2}(z,n)=f_{\rm{l}}(z,n),\qquad n\ge m.
\end{equation}

\item[\text{\rm(d)}]  The left Jost solution $f_{\rm{l}2}(z,n)$ to \eqref{5.6} satisfies
\begin{equation}
\label{5.20}
f_{\rm{l}2}(z,n)=z^n\,T_{\rm{l}2}(z)^{-1}+z^{-n}\,L_2(z)\,T_{\rm{l}2}(z)^{-1},\qquad n\le m,\quad z\in\mathbb T\setminus\{-1,1\},
\end{equation}
\begin{equation}
\label{5.21}
\displaystyle\frac{a(m+1)}{a_\infty}\,f_{\rm{l}2}(z,m+1)=z^{m+1}\,T_{\rm{l}2}(z)^{-1}+z^{-m-1}\,L_2(z)\,T_{\rm{l}2}(z)^{-1},\qquad z\in\mathbb T\setminus\{-1,1\},
\end{equation}
where $T_{\rm{l}2}(z)$ and $L_2(z)$ are the left transmission and reflection coefficients, respectively, for \eqref{5.6}.

\end{enumerate}

\end{proposition}

\begin{proof}
We establish \eqref{5.16} with the help of the uniqueness of
the right Jost solution to \eqref{2.17}.
From \eqref{5.1}--\eqref{5.3} we know that each side of \eqref{5.16}
satisfies the same discrete system \eqref{2.17} when $n\le m-1$ and
that each side has the same spacial asymptotics as $n\to-\infty$ given in \eqref{3.11}.
We extend $f_{\rm{r}1}(z,n)$ specified for $n\le m$ to a solution of \eqref{2.17} for $n\in\mathbb Z,$
and we use 
$\hat f_{\rm{r}1}(z,n)$
 to denote that extension.
We first determine
$\hat f_{\rm{r}1}(z,m+1)$ 
in terms of 
$f_{\rm{r}1}(z,m)$ 
and $f_{\rm{r}1}(z,m-1),$ 
and we do this by evaluating \eqref{2.17} when $n=m.$
Then, we iterate this process by evaluating 
\eqref{2.17} first when $n=m+1$ and each time by increasing the value of $n$ by $1.$
Hence, we establish that each of 
$\hat f_{\rm{r}1}(z,n)$ and
$f_{\rm{r}}(z,n)$ satisfies \eqref{2.17} for
$n\in\mathbb Z$
and has the spacial asymptotics as $n\to-\infty$ given
in \eqref{3.11}. From Theorem~\ref{theorem3.2}(a) we know that the right Jost solution to
\eqref{2.17} is unique.
Thus, we have
$\hat f_{\rm{r}1}(z,n)=f_{\rm{r}}(z,n)$ for $n\in\mathbb Z.$
Since we already have
$\hat f_{\rm{r}1}(z,n)=f_{\rm{r}1}(z,n)$ for $n\le m,$
we conclude that \eqref{5.16} holds for $n\le m.$
Next, we establish \eqref{5.17} by proceeding as follows.
With the help of \eqref{5.3}, we evaluate each of \eqref{2.17} and \eqref{5.5} at $n=m$ satisfied by their
 respective right Jost solutions 
 $f_{\rm{r}}(z,n)$ 
and 
$f_{\rm{r}1}(z,n),$ and we get
\begin{equation}\label{5.22}
\begin{split}
a(m+1)\,f_{\rm{r}}(z,m+1)+&b_1(m)\,f_{\rm{r}}(z,m)+
a_1(m)^\dagger\,f_{\rm{r}}(z,m-1)\\
&=
\displaystyle
\frac{a_\infty(z+z^{-1})+b_\infty}{w_\infty}
\,
w_1(m)\,f_{\rm{r}}(z,m),
\end{split}
\end{equation}
\begin{equation}\label{5.23}
\begin{split}
a_\infty\,f_{\rm{r}1}(z,m+1)+&b_1(m)\,f_{\rm{r}1}(z,m)+
a_1(m)^\dagger\,f_{\rm{r}1}(z,m-1)\\
&=
\displaystyle
\frac{a_\infty(z+z^{-1})+b_\infty}{w_\infty}
\,
w_1(m)\,f_{\rm{r}1}(z,m).
\end{split}
\end{equation}
We remark that we have used $a(m)=a_1(m),$ $b(m)=b_1(m),$ and $w(m)=w_1(m)$ in \eqref{5.22}.
Using \eqref{5.16} in \eqref{5.22} and comparing the resulting equation with
\eqref{5.23}, we observe that \eqref{5.17} holds. Hence, the proof of (a) is complete.
Let us now prove (b). 
We introduce the quantity $\hat h_{\rm{r}1}(z,n)$ for $n\ge m$ by letting
\begin{equation}\label{5.24}
\hat h_{\rm{r}1}(z,n)
:=z^{-n}\,T_{\rm{r}1}(z)^{-1}+z^n\,R_1(z)\,T_{\rm{r}1}(z)^{-1},\qquad n\ge m,\quad z\in\mathbb T\setminus\{-1,1\}.
\end{equation}
Using \eqref{5.24} in \eqref{5.5} we confirm that 
$\hat h_{\rm{r}1}(z,n)$ satisfies \eqref{5.5} when $n\ge m+1.$
Next, we extend $\hat h_{\rm{r}1}(z,n)$ to a solution of
\eqref{5.5} for $n\in\mathbb Z.$ For this, we proceed as follows.
In order for $\hat h_{\rm{r}1}(z,n)$ to be a solution to \eqref{5.5}, it must satisfy \eqref{5.5} when $n=m.$ In other words, 
we must have 
\begin{equation}\label{5.25}
\begin{split}
a_1(m+1)\,\hat h_{\rm{r}1}(z,m+1)+&b_1(m)\,\hat h_{\rm{r}1}(z,m)+
a_1(m)^\dagger\,\hat h_{\rm{r}1}(z,m-1)\\
&=
\displaystyle
\frac{a_\infty(z+z^{-1})+b_\infty}{w_\infty}
\,
w_1(m)\,\hat h_{\rm{r}1}(z,m).
\end{split}
\end{equation}
The quantities $\hat h_{\rm{r}1}(z,m+1)$ and $\hat h_{\rm{r}1}(z,m)$ are evaluated by using \eqref{5.24},
and those values can be inserted in \eqref{5.25}. We solve the resulting
equality for $\hat h_{\rm{r}1}(z,m-1),$ and hence the value of
the solution $\hat h_{\rm{r}1}(z,n)$ to \eqref{5.5} is determined also when $n=m-1.$ Having
the values of $\hat h_{\rm{r}1}(z,m)$ and $\hat h_{\rm{r}1}(z,m-1),$ this procedure
can be repeated to determine the value of the solution $\hat h_{\rm{r}1}(z,n)$ to \eqref{5.5} at $n=m-2.$
We then iterate this process by evaluating $\hat h_{\rm{r}1}(z,n)$ to \eqref{5.5} for all values
of $n$ when $n<m-2$ by reducing the value of $n$ by $1$ each time.
Hence, we establish that each of 
$\hat h_{\rm{r}1}(z,n)$ and $f_{\rm{r}1}(z,n)$ 
satisfies \eqref{5.5} for $n\in\mathbb Z.$
Furthermore, from 
\eqref{5.24} and the first line of \eqref{5.8} 
it follows that $\hat h_{\rm{r}1}(z,n)$ and $f_{\rm{r}1}(z,n)$ each
have the same spacial asymptotics as $n\to+\infty.$
 From Theorem~\ref{theorem3.2}(c) we know that the pair 
 $f_{\rm{l}1}(z,n)$ and $f_{\rm{l}1}(z^{-1},n)$
  for $n\in\mathbb Z$
 forms a fundamental set of $q\times q$ matrix-valued solutions to \eqref{5.5}.
 Furthermore, \eqref{5.5} is a linear homogeneous equation, and hence
$f_{\rm{r}1}(z,n)-\hat h_{\rm{r}1}(z,n)$ is a solution to \eqref{5.5} for $n\in\mathbb Z.$
Consequently, we have
\begin{equation}\label{5.26}
f_{\rm{r}1}(z,n)-\hat h_{\rm{r}1}(z,n)
= f_{\rm{l}1}(z,n)\,c_5(z)+ f_{\rm{l}1}(z^{-1},n)\,c_6(z),  
\qquad n\in \mathbb Z,\quad z\in\mathbb T\setminus\{-1,1\},
\end{equation}
where $c_5(z)$ and $c_6(z)$ are some $q\times q$ matrix-valued coefficients.
Since $c_5(z)$ and $c_6(z)$ are independent of $n,$ we can evaluate them
by letting $n\to+\infty$ in \eqref{5.26}. Using \eqref{5.24} and the first line of \eqref{5.8} on the left-hand side of 
\eqref{5.26}, we see that that left-hand side vanishes as $n\to+\infty.$ Hence, we
obtain $c_5(z)=0$ and $c_6(z)=0,$ which implies that the left-hand side of \eqref{5.26}
vanishes for all $n\in\mathbb Z.$ Thus, we get 
$f_{\rm{r}1}(z,n)=\hat h_{\rm{r}1}(z,n)$ for $n\in\mathbb Z.$
Consequently, \eqref{5.24} and \eqref{5.26} imply \eqref{5.18},
which completes the proof of (b).
The proof of (c) is similar to the proof of (a), and for this we proceed as follows.
 We establish \eqref{5.19} with the help
of the uniqueness result for the left Jost solution stated in Theorem~\ref{theorem3.2}(a).
We note that
each side of \eqref{5.19} satisfies \eqref{2.17} when $n\geq m+1$ and has 
the same spacial asymptotics as $n\to+\infty$ given in  \eqref{3.10}.
We first extend $f_{\rm{l}2}(z,n)$
given on the left-hand side of
\eqref{5.19} to a solution of \eqref{2.17} from $n\ge m$ 
to $n\in\mathbb Z.$ We use 
$\hat f_{\rm{l}2}(z,n)$ 
to denote that solution.
Since we already have $\hat f_{\rm{l}2}(z,n)=f_{\rm{l}2}(z,n)$
for $n\ge m,$ we need to determine  $\hat f_{\rm{l}2}(z,n)$ for 
$n\le m-1.$
We  first evaluate $\hat f_{\rm{l}2}(z,m-1)$ in terms of $f_{\rm{l}2}(z,m)$ and $f_{\rm{l}2}(z,m+1)$ by using \eqref{2.17} at
$n=m$ and then we iterate this process using \eqref{2.17} at $n=m-1$
and then keep decreasing the value of $n$ by $1.$
Thus, we establish that each of $\hat f_{\rm{l}2}(z,n)$ and
$f_{\rm{l}}(z,n)$ satisfies \eqref{2.17} for
$n\in\mathbb Z$ and has the spacial asymptotics as $n\to+\infty$ given in \eqref{3.10}.
 From Theorem~\ref{theorem3.2}(a) we know that the left Jost solution to \eqref{2.17} is unique,
 and hence we conclude that $\hat f_{\rm{l}2}(z,n)= f_{\rm{l}}(z,n)$ for $n\in\mathbb Z.$ Since we already
 have $\hat f_{\rm{l}2}(z,n)=f_{\rm{l}2}(z,n)$
 for $n\ge m,$ we confirm \eqref{5.19},
which completes the proof of (c).
Finally, we prove (d) by proceeding as in the proof of (b). For this, we introduce the quantity
$\hat h_{\rm{l}2}(z,n)$ by letting
\begin{equation}\label{5.27}
\hat h_{\rm{l}2}(z,n)
:= z^n\,T_{\rm{l}2}(z)^{-1}+z^{-n}\,L_2(z)\,T_{\rm{l}2}(z)^{-1},\qquad n\le m,\quad z\in\mathbb T\setminus\{-1,1\}.
\end{equation}
From \eqref{5.27} we observe that 
$\hat h_{\rm{l}2}(z,n)$
satisfies \eqref{5.6} for
$n\le m-1.$ We extend $\hat h_{\rm{l}2}(z,n)$
to a solution of \eqref{5.6} for $n\in\mathbb Z$ as follows.
In order for $\hat h_{\rm{l}2}(z,n)$ to be a solution to \eqref{5.6}, it is necessary that $\hat h_{\rm{l}2}(z,n)$ satisfies \eqref{5.6} when $n=m.$ In other words, 
we must have 
\begin{equation}\label{5.28}
\begin{split}
a_2(m+1)\,\hat h_{\rm{l}2}(z,m+1)+&b_2(m)\,\hat h_{\rm{l}2}(z,m)+
a_2(m)^\dagger\,\hat h_{\rm{l}2}(z,m-1)\\
&=
\displaystyle
\frac{a_\infty(z+z^{-1})+b_\infty}{w_\infty}
\,
w_2(m)\,\hat h_{\rm{l}2}(z,m).
\end{split}
\end{equation}
The quantities $\hat h_{\rm{l}2}(z,m-1)$ and $\hat h_{\rm{l}2}(z,m)$ are evaluated by using \eqref{5.27},
and those values can be inserted in \eqref{5.28}. We solve the resulting
equality for $\hat h_{\rm{l}2}(z,m+1),$ and hence the value of
the solution $\hat h_{\rm{l}2}(z,n)$ to \eqref{5.6} is determined also when $n=m+1.$ Having
the values of $\hat h_{\rm{l}2}(z,m)$ and $\hat h_{\rm{l}2}(z,m+1),$ this procedure
can be repeated to determine the value of the solution $\hat h_{\rm{l}2}(z,n)$ to \eqref{5.6} at $n=m+2.$
We then iterate this process by evaluating $\hat h_{\rm{l}2}(z,n)$ to \eqref{5.6} for all values
of $n$ when $n>m+2$ by increasing the value of $n$ by $1$ each time.
It then follows that each of $f_{\rm{l}2}(z,n)$ and
$\hat h_{\rm{l}2}(z,n)$ satisfies \eqref{5.6} for
$n\in\mathbb Z.$ Furthermore,
from \eqref{5.27} and the second line of \eqref{5.9}
we see that
 $\hat h_{\rm{l}2}(z,n)$ and
$f_{\rm{l}2}(z,n)$ each have
the same spacial asymptotics as $n\to-\infty.$
From Theorem~\ref{theorem3.2}(c) we know that
the pair 
$f_{\rm{r}2}(z,n)$ and $f_{\rm{r}2}(z^{-1},n)$ for
$n\in\mathbb Z$ forms a fundamental set of $q\times q$ matrix-valued solutions to \eqref{5.6}.
In analogy with \eqref{5.26}, we express the solution $f_{\rm{l}2}(z,n)-\hat h_{\rm{l}2}(z,n)$ to \eqref{5.6} as
\begin{equation}\label{5.29}
f_{\rm{l}2}(z,n)-\hat h_{\rm{l}2}(z,n)
= f_{\rm{r}2}(z,n)\,c_7(z)+ f_{\rm{r}2}(z^{-1},n)\,c_8(z),
\qquad n\in \mathbb Z,\quad z\in\mathbb T\setminus\{-1,1\},
\end{equation}
where $c_7(z)$ and $c_8(z)$ are the $q\times q$ matrices 
that can be determined by letting $n\to-\infty$ in \eqref{5.29}.
From \eqref{5.27} and the second line of \eqref{5.9} we observe that 
the left-hand side of \eqref{5.29} vanishes as $n\to-\infty.$ Hence, from \eqref{5.29} we
get $c_7(z)=0$ and $c_8(z)=0,$
which yields 
$f_{\rm{l}2}(z,n)=\hat h_{\rm{l}2}(z,n)$ for $n\in\mathbb Z.$
Consequently, \eqref{5.27} and \eqref{5.29} imply \eqref{5.20}.
Finally, by evaluating \eqref{5.6} at $n=m,$ using \eqref{5.20}, and observing that
$a_2(m+1)=a(m+1),$ we obtain \eqref{5.21}.
\end{proof}

For the proof of our factorization formula, we need to relate
the value at $n=m$ of
the matrix $G(z,n)$ defined in \eqref{4.3} to $F_{\rm{l}}(z,m)$ and $F_{\rm{r}}(z,m),$ where the latter two matrices are defined in 
\eqref{4.1} and \eqref{4.2}, respectively. This is achieved by using the results of
Theorem~\ref{theorem4.3} at $n=m.$ Hence, from \eqref{4.28} and \eqref{4.29}, we respectively obtain
\begin{equation}
\label{5.30}
G(z,m)=F_{\rm{l}}(z,m)
\begin{bmatrix} I&R(z)\,T_{\rm{r}}(z)^{-1}\\
\noalign{\medskip}
0&T_{\rm{r}}(z)^{-1}
\end{bmatrix},\qquad z\in\mathbb T\setminus\{-1,1\},
\end{equation}
\begin{equation}
\label{5.31}
G(z,m)=F_{\rm{r}}(z,m)
\begin{bmatrix} T_{\rm{l}}(z)^{-1}&0\\
\noalign{\medskip}
L(z)\,T_{\rm{l}}(z)^{-1}&I
\end{bmatrix},\qquad z\in\mathbb T\setminus\{-1,1\}.
\end{equation}

In the next proposition, we relate the matrix $G(z,m)$ to the matrices $\Lambda_2(z)$ and
$\Sigma_1(z)$ appearing in \eqref{5.13} and \eqref{5.14}, respectively.
The results presented in the proposition are needed later on to establish the factorization formulas
in Theorem~\ref{theorem5.3}.

\begin{proposition}
\label{proposition5.2}
Assume that the $q\times q$ matrix-valued coefficients in \eqref{2.17} belong to the class
$\mathcal A$ described in Definition~\ref{definition1.1}, and let $m$ be the integer
appearing in the partitioning specified in \eqref{5.1}--\eqref{5.4}.
Then,
the $2q\times 2q$ matrix $G(z,m)$ appearing in \eqref{5.30} and \eqref{5.31} is expressed in two different but equivalent forms as
\begin{equation}
\label{5.32}
G(z,m)=
\begin{bmatrix}I&0\\
0&a_\infty\, I\end{bmatrix}
\begin{bmatrix} z^m\,I&z^{-m}\,I\\
z^{m+1}\,I&z^{-m-1}\,I
\end{bmatrix}\,
\Lambda_2(z)
\begin{bmatrix} I&R(z)\,T_{\rm{r}}(z)^{-1}\\
\noalign{\medskip}
0&T_{\rm{r}}(z)^{-1}
\end{bmatrix},\qquad z\in\mathbb T\setminus\{-1,1\},
\end{equation}
\begin{equation}
\label{5.33}
G(z,m)=\begin{bmatrix}I&0\\
0&a_\infty\, I\end{bmatrix}
\begin{bmatrix} z^m\,I&z^{-m}\,I\\
z^{m+1}\,I&z^{-m-1}\,I
\end{bmatrix}\,
\Sigma_1(z)\,
\begin{bmatrix} T_{\rm{l}}(z)^{-1}&0\\
\noalign{\medskip}
L(z)\,T_{\rm{l}}(z)^{-1}&I
\end{bmatrix},\qquad z\in\mathbb T\setminus\{-1,1\},
\end{equation}
where $\Lambda_2(z)$ and $\Sigma_1(z)$ are the transition matrices appearing in \eqref{5.13} and
\eqref{5.14}, respectively.

\end{proposition}

\begin{proof}
Note that \eqref{5.19} is valid when $z\in\mathbb T,$ and hence it also holds if we replace $z$ by $z^{-1}$ there. Thus, from \eqref{5.19} we get
\begin{equation}
\label{5.34}
g_{\rm{l}2}(z,n)=g_{\rm{l}}(z,n),\qquad n\ge m.
\end{equation}
By using \eqref{5.19} and \eqref{5.34}
in \eqref{4.1}, we get
\begin{equation}
\label{5.35}
F_{\rm{l}}(z,m)=\begin{bmatrix}f_{\rm{l}2}(z,m)&g_{\rm{l}2}(z,m)\\
\noalign{\medskip}
a(m+1)\,f_{\rm{l}2}(z,m+1)&a(m+1)\,g_{\rm{l}2}(z,m+1)\end{bmatrix}.
\end{equation}
Using \eqref{5.20} in the first row block and using \eqref{5.21} in the second
row block in the matrix on the right-hand side of \eqref{5.35}, we get
\begin{equation}
\label{5.36}
F_{\rm{l}}(z,m)=\begin{bmatrix}I&0\\
0&a_\infty\, I\end{bmatrix}
\begin{bmatrix} z^m\,I&z^{-m}\,I\\
z^{m+1}\,I&z^{-m-1}\,I
\end{bmatrix}\,
\begin{bmatrix}T_{\rm{l}2}(z)^{-1}&L_2(z^{-1})\,T_{\rm{l}2}(z^{-1})^{-1} \\
\noalign{\medskip}
L_2(z)\,T_{\rm{l}2}(z)^{-1}&T_{\rm{l}2}(z^{-1})^{-1}\end{bmatrix}.
\end{equation}
Next, using \eqref{5.36} on the right-hand side of \eqref{5.30}, we obtain \eqref{5.32}.
Let us remark that \eqref{5.16} and \eqref{5.17} hold when $z\in\mathbb T,$ and hence they remain valid if
we replace $z$ by $z^{-1}$ in them. Thus, when $z\in\mathbb T$ we have
\begin{equation}
\label{5.37}
g_{\rm{r}1}(z,n)=g_{\rm{r}}(z,n),\qquad n\le m,
\end{equation} 
\begin{equation}
\label{5.38}
g_{\rm{r}1}(z,m+1)=\displaystyle\frac{a(m+1)}{a_\infty}\,g_{\rm{r}}(z,m+1).
\end{equation}
By using \eqref{5.37} and \eqref{5.38} in \eqref{4.2}, we obtain
in the resulting expression for $F_{\rm{r}}(z,m)$ as
\begin{equation}
\label{5.39}
F_{\rm{r}}(z,m)=\begin{bmatrix}g_{\rm{r}1}(z,m)&f_{\rm{r}1}(z,m)\\
\noalign{\medskip}
a_\infty\,g_{\rm{r}1}(z,m+1)&a_\infty\,f_{\rm{r}1}(z,m+1)\end{bmatrix}.
\end{equation}
Next, with the help of \eqref{5.17} and \eqref{5.18}, we express the right-hand side of \eqref{5.39} in terms of the 
right reflection coefficients $T_{\rm{r}1}(z)$ and $R_1(z),$ and we get
\begin{equation}
\label{5.40}
F_{\rm{r}}(z,m)=\begin{bmatrix}I&0\\
0&a_\infty\, I\end{bmatrix}
\begin{bmatrix} z^m\,I&z^{-m}\,I\\
z^{m+1}\,I&z^{-m-1}\,I
\end{bmatrix}\,
\begin{bmatrix}T_{\rm{r}1}(z^{-1})^{-1}&R_1(z)\,T_{\rm{r}1}(z)^{-1} \\
\noalign{\medskip}
R_1(z^{-1})\,T_{\rm{r}1}(z^{-1})^{-1}&T_{\rm{r}1}(z)^{-1}\end{bmatrix}.
\end{equation}
Finally, using \eqref{5.40} on the right-hand side of \eqref{5.31} we obtain
\eqref{5.33}.
\end{proof}

In the next theorem we present our factorization formula for \eqref{2.17} corresponding to the partitioning described in \eqref{5.1}--\eqref{5.4}.
The significance of the factorization formulas in \eqref{5.41} and \eqref{5.42} is the following.
If we know the scattering from the individual fragments of the full-line lattice then, by using these two factorization formulas, we can determine the scattering from the full-line lattice. It is easier to determine the scattering from a fragment than the scattering from the whole lattice. This is because the expressions for the scattering coefficients from a fragment are simpler and hence easier to compute. We remark that in the factorization formula \eqref{5.41} all the transition matrices are expressed in terms of the left scattering coefficients and in the factorization formula \eqref{5.42} the transition matrices are expressed in terms of the right scattering coefficients. Hence, \eqref{5.41} describes how the total scattering is composed of the scattering from the fragments as we move from the left to the right on the lattice. On the other hand, \eqref{5.42} describes how the total scattering is composed of the scattering from the fragments as we move from the right to the left on the lattice.

\begin{theorem}
\label{theorem5.3} Assume that the $q\times q$ matrix-valued coefficients in \eqref{2.17} belong to the class
$\mathcal A$ described in Definition~\ref{definition1.1}, and let $m$ be the integer in the partitioning specified in \eqref{5.1}--\eqref{5.4}.
Let $\Lambda(z),$ $\Lambda_1(z),$ and $\Lambda_2(z)$ be the $2q\times 2q$
transition matrices defined in \eqref{4.8}, \eqref{5.12}, and \eqref{5.13}, respectively.
Similarly, let $\Sigma(z),$ $\Sigma_1(z),$ and $\Sigma_2(z)$ be the $2q\times 2q$
transition matrices defined in \eqref{4.9}, \eqref{5.14}, and \eqref{5.15}, respectively.
Then, we have the following:

\begin{enumerate}

\item[\text{\rm(a)}]  The transition matrix $\Lambda(z)$ is equal to the ordered matrix product 
$\Lambda_1(z)\,\Lambda_2(z),$ i.e. we have
\begin{equation}
\label{5.41}\Lambda(z)=\Lambda_1(z)\,\Lambda_2(z),\qquad z\in\mathbb T\setminus\{-1,1\}.
\end{equation}

\item[\text{\rm(b)}] The factorization formula \eqref{5.41} can also be expressed in terms of the transition matrices 
$\Sigma(z),$ $\Sigma_1(z),$ and $\Sigma_2(z)$ as
 \begin{equation}
\label{5.42}
\Sigma(z)=\Sigma_2(z)\,\Sigma_1(z),\qquad z\in\mathbb T\setminus\{-1,1\}.
\end{equation}

\end{enumerate}

\end{theorem}

\begin{proof}
We remark that the first and second matrices appearing on the right-hand sides of 
\eqref{5.36} and \eqref{5.40} are both invertible. The invertibility of that first matrix
is a consequence of the fact that $a_\infty\ne 0,$ as indicated
in Definition~\ref{definition1.1}(c). Hence, equating the right-hand sides of \eqref{5.32} and \eqref{5.33}
to each other, we obtain
\begin{equation*}
\Lambda_2(z)\begin{bmatrix} I&R(z)\,T_{\rm{r}}(z)^{-1}\\
\noalign{\medskip}
0&T_{\rm{r}}(z)^{-1}
\end{bmatrix}=
\Sigma_1(z)
\begin{bmatrix} T_{\rm{l}}(z)^{-1}&0\\
\noalign{\medskip}
L(z)\,T_{\rm{l}}(z)^{-1}&I
\end{bmatrix},
\end{equation*}
which is equivalent to
\begin{equation}
\label{5.44}
\Sigma_1(z)^{-1}\,\Lambda_2(z)=
\begin{bmatrix} T_{\rm{l}}(z)^{-1}&0\\
\noalign{\medskip}
L(z)\,T_{\rm{l}}(z)^{-1}&I
\end{bmatrix}\begin{bmatrix} I&R(z)\,T_{\rm{r}}(z)^{-1}\\
\noalign{\medskip}
0&T_{\rm{r}}(z)^{-1}
\end{bmatrix}^{-1}.
\end{equation}
The inverse of the second matrix on the right-hand side of \eqref{5.44} can be expressed explicitly because
that second matrix is a block upper-triangular matrix. Furthermore, \eqref{4.12} indicates that the inverse of
$\Sigma_1(z)$ is equal to $\Lambda_1(z).$ Hence, we can write \eqref{5.44} as
\begin{equation*}
\Lambda_1(z)\,\Lambda_2(z)=
\begin{bmatrix} T_{\rm{l}}(z)^{-1}&0\\
\noalign{\medskip}
L(z)\,T_{\rm{l}}(z)^{-1}&I
\end{bmatrix}
\begin{bmatrix} I&-R(z)\\
\noalign{\medskip}
0&T_{\rm{r}}(z)
\end{bmatrix},
\end{equation*}
which is equivalent to
\begin{equation}
\label{5.46}
\Lambda_1(z)\,\Lambda_2(z)=
\begin{bmatrix} T_{\rm{l}}(z)^{-1}&-T_{\rm{l}}(z)^{-1}\,R(z)\\
\noalign{\medskip}
L(z)\,T_{\rm{l}}(z)^{-1}&-L(z)\,T_{\rm{l}}(z)^{-1}\,R(z)+T_{\rm{r}}(z)
\end{bmatrix}.
\end{equation}
By comparing the matrix on the right-hand side of \eqref{5.46}
with the right-hand side of \eqref{4.8}, we see that 
\eqref{5.41} holds provided that we have the two identities given by
\begin{equation}
\label{5.47}
-T_{\rm{l}}(z)^{-1}\,R(z)=L(z^{-1})\,T_{\rm{l}}(z^{-1})^{-1},\qquad z\in\mathbb T\setminus\{-1,1\},
\end{equation}
\begin{equation}
\label{5.48}
-L(z)\,T_{\rm{l}}(z)^{-1}\,R(z)+T_{\rm{r}}(z)=T_{\rm{l}}(z^{-1})^{-1},\qquad z\in\mathbb T\setminus\{-1,1\}.
\end{equation}
With the help of \eqref{2.19}, we observe that the identities in 
\eqref{5.47} and \eqref{5.48} are equivalent to the respective identities given by
\begin{equation}
\label{5.49}
-T_{\rm{l}}(z)^{-1}\,R(z)=L(z^\ast)\,T_{\rm{l}}(z^\ast)^{-1},\qquad z\in\mathbb T\setminus\{-1,1\},
\end{equation}
\begin{equation}
\label{5.50}
-L(z)\,T_{\rm{l}}(z)^{-1}\,R(z)+T_{\rm{r}}(z)=T_{\rm{l}}(z^\ast)^{-1},\qquad z\in\mathbb T\setminus\{-1,1\}.
\end{equation}
The proof of \eqref{5.49} is established as follows. 
Using the first and third equalities in \eqref{3.51} on the right-hand side of \eqref{5.49}, we see that
\eqref{5.49} is equivalent to
\begin{equation}
\label{5.51}
-T_{\rm{l}}(z)^{-1}\,R(z)=L(z)^\dagger\,[T_{\rm{r}}(z)^\dagger]^{-1},\qquad z\in\mathbb T\setminus\{-1,1\}.
\end{equation}
We remark that the $(1,2)$-entries of the equality in \eqref{3.50} yields \eqref{5.51} and hence \eqref{5.47} is valid.
In order to complete the proof of (a), we still need to establish \eqref{5.50}.
For this, we proceed as follows. Using \eqref{5.51} on the left-hand side of \eqref{5.50} we see that
\eqref{5.50} is equivalent to the identity
\begin{equation}
\label{5.52}
L(z)\,L(z)^\dagger\,[T_{\rm{r}}(z)^\dagger]^{-1}+T_{\rm{r}}(z)=T_{\rm{l}}(z^\ast)^{-1},\qquad z\in\mathbb T\setminus\{-1,1\}.
\end{equation}
From the $(2,2)$-entries of the equality in
\eqref{3.50} we obtain
\begin{equation}
\label{5.53}
L(z)\,L(z)^\dagger\,[T_{\rm{r}}(z)^\dagger]^{-1}+T_{\rm{r}}(z)=[T_{\rm{r}}(z)^\dagger]^{-1},\qquad z\in\mathbb T\setminus\{-1,1\}.
\end{equation}
The equivalence of the right-hand sides of \eqref{5.52} and \eqref{5.53}
directly follows from the third equality of \eqref{3.51}. Thus, we have proved
that \eqref{5.41} holds and hence the proof of (a) is complete.
We remark that \eqref{5.42} is obtained from \eqref{5.41} by taking the matrix inverses
of both sides of \eqref{5.41} and by using \eqref{4.12} for
each of the transition matrices
$\Lambda(z),$ $\Lambda_1(z),$ and $\Lambda_2(z).$
Thus, the proof of (b) is also complete.
\end{proof}

The result of Theorem~\ref{theorem5.3} can easily be extended from two fragments to any finite number of fragments. This is due to the
fact that any existing fragment
can be partitioned into further subfragments by applying the factorization formulas \eqref{5.41} and \eqref{5.42}
to each fragment and to its subfragments. Since a proof can be obtained by using an induction on the number of fragments,
we state the result as a corollary without a proof.

\begin{corollary}
\label{corollary5.4} 
Assume that the $q\times q$ matrix-valued coefficients in \eqref{2.17} belong to the class
$\mathcal A$ described in Definition~\ref{definition1.1}. Let $S(z),$ $\Lambda(z),$ and 
$\Sigma(z)$ defined in \eqref{3.22}, \eqref{4.8}, and \eqref{4.9}, respectively
be the corresponding scattering matrix and the transition matrices,
with $T_{\rm{l}}(z),$ $T_{\rm{r}}(z),$ $L(z),$ and $R(z)$
denoting the corresponding left and right transmission coefficients and the left and right reflection coefficients, respectively.
Assume that the full-line lattice $\mathbb Z$ is partitioned into $P+1$ fragments as
\begin{equation*}
\mathbb Z=\mathbb Z_1\cup\mathbb Z_2\cup \cdots \cup \mathbb Z_{P+1},
\end{equation*}
where $P$ is a fixed positive integer and we have defined
\begin{equation*}
\mathbb Z_1:=\{\cdots,m_1-1,m_1\},\quad \mathbb Z_{P+1}:=\{m_P+1,m_P+2,\cdots\},
\end{equation*}
\begin{equation*}
\mathbb Z_j:=\{m_{j-1}+1,m_{j-1}+2,\cdots,m_j\},\qquad 2\le j\le P,
\end{equation*}
with the integers $m_j$ satisfying $m_1<m_2<\cdots<m_P.$
Let the matrix-valued coefficients
$a_j(n),$ $b_j(n),$ $w_j(n)$ be defined in terms of the three matrix-valued coefficients
$a(n),$ $b(n),$ $w(n)$ and their limiting values appearing in \eqref{1.2} as
\begin{equation}
\label{5.57}
\left(a_j(n),b_j(n),w_j(n)\right):=\begin{cases}
\left(a(n),b(n),w(n)\right),\qquad n\in\mathbb Z_j,\\
\noalign{\medskip}
\left(a_\infty\,I,b_\infty\,I,w_\infty\,I\right),\qquad n\not\in\mathbb Z_j.
\end{cases}
\end{equation}
Let
$T_{\rm{l}j}(z),$
$T_{\rm{r}j}(z),$
$L_j(z),$
$R_j(z)$ be the corresponding 
left and right transmission coefficients and the left and right reflection coefficients, respectively,
for the fragment described in \eqref{5.57}. Let 
$S_j(z),$
$\Lambda_j(z),$ and
$\Sigma_j(z)$ denote the  
scattering matrix, the left transition matrix, and the right transition matrix,
respectively, for the fragment described in \eqref{5.57}, where we have let
\begin{equation*}
S_{j}(z):=\begin{bmatrix}T_{\rm{l}j}(z)&R_j(z) \\ 
\noalign{\medskip}
 L_{ j }(z)& T_{\rm{r} j}(z)\end{bmatrix}, \qquad z\in\mathbb T\setminus\{-1,1\},
\end{equation*}
\begin{equation*}
\Lambda_j(z):=\begin{bmatrix}T_{\rm{l}j}(z)^{-1}&L_j(z^{-1})\,T_{\rm{l}j}(z^{-1})^{-1} \\
\noalign{\medskip}
L_j(z)\,T_{\rm{l}j}(z)^{-1}&T_{\rm{l}j}(z^{-1})^{-1}\end{bmatrix},\qquad z\in\mathbb T \setminus\{-1,1\},
\end{equation*}
\begin{equation}
\label{5.60}
\Sigma_j(z):=\begin{bmatrix}T_{\rm{r}j}(z^{-1})^{-1}&R_j(z)\,T_{\rm{r}j}(z)^{-1} \\
\noalign{\medskip}
R_j(z^{-1})\,T_{\rm{r}j}(z^{-1})^{-1}&T_{\rm{r}j}(z)^{-1}\end{bmatrix},\qquad z\in\mathbb T\setminus\{-1,1\}.
\end{equation}
Then, the transition matrices $\Lambda(z)$ and $\Sigma(z)$ for the full-line lattice $\mathbb Z$
are expressed as the ordered matrix products of the corresponding
transition matrices for the fragments as
\begin{equation*}
\Lambda(z)=\Lambda_1(z)\, \Lambda_2(z)\cdots\Lambda_P(z)\,\Lambda_{P+1}(z),\qquad z\in\mathbb T\setminus\{-1,1\},
\end{equation*}
\begin{equation*}
\Sigma(z)=\Sigma_{P+1}(z)\, \Sigma_P(z)\cdots\Sigma_2(z)\,\Sigma_1(z),\qquad z\in\mathbb T\setminus\{-1,1\}.
\end{equation*}
\end{corollary}

In Theorem~\ref{theorem5.3} the transition matrix for the full-line lattice is expressed as an ordered matrix product of
the transition matrices for the two fragments described in \eqref{5.1}--\eqref{5.4}.
In the next theorem, we express the scattering coefficients 
for the full-line lattice in terms of the scattering coefficients for the two fragments.
The results presented in the theorem are significant because it is easier to determine the scattering coefficients from a fragment than the scattering coefficients from the full-line lattice. The formulas in the theorem show how to combine the scattering coefficients for the fragments so that we can determine the scattering coefficients for the full-line lattice.

\begin{theorem}
\label{theorem5.5} Assume that the $q\times q$ matrix-valued coefficients in \eqref{2.17} belong to the class
$\mathcal A$ described in Definition~\ref{definition1.1}, and let $m$ be the integer
in the partitioning specified in \eqref{5.1}--\eqref{5.4}.
Let $S(z)$ given in \eqref{3.22} be the scattering matrix for the full-line lattice $\mathbb Z$ and let $S_1(z)$ and $S_2(z)$
given in \eqref{5.11} be the scattering matrices corresponding to the fragments described in
\eqref{5.1} and \eqref{5.2},
respectively. Then, the scattering coefficients in $S(z)$ are related to the right scattering coefficients
in $S_1(z)$ and the left scattering coefficients in $S_2(z)$ as
\begin{equation}
\label{5.63}
T_{\rm{l}}(z)=T_{\rm{l}2}(z)\left[I-R_1(z)\,L_2(z)\right]^{-1} T_{\rm{r}1}(z^{-1})^\dagger,\qquad z\in\mathbb T\setminus\{-1,1\},
\end{equation}
\begin{equation}
\label{5.64}
L(z)=[T_{\rm{r}1}(z)^\dagger]^{-1}\left[L_2(z)-R_1(z^{-1})\right]\left[I-R_1(z)\,L_2(z)\right]^{-1} T_{\rm{r}1}(z^{-1})^\dagger,\qquad  z\in\mathbb T\setminus\{-1,1\},
\end{equation}
\begin{equation}
\label{5.65}
T_{\rm{r}}(z)=T_{\rm{r}1}(z)\left[I-L_2(z)\,R_1(z)\right]^{-1} T_{\rm{l}2}(z^{-1})^\dagger,\qquad z\in\mathbb T\setminus\{-1,1\},
\end{equation}
\begin{equation}
\label{5.66}
R(z)=T_{\rm{l}2}(z)\left[I-R_1(z)\,L_2(z)\right]^{-1}  \left[R_1(z)-L_2(z^{-1})\right] T_{\rm{l}2}(z^{-1})^{-1},\qquad z\in\mathbb T\setminus\{-1,1\},
\end{equation}
where we recall that $I$ is the $q\times q$ identity matrix and the dagger denotes the matrix adjoint.
\end{theorem}

\begin{proof}
From the $(1,1)$-entries of the equality in \eqref{5.41}, with the help of \eqref{5.12} and \eqref{5.13}, we have
 \begin{equation}
\label{5.67}T_{\rm{l}}(z)^{-1}=T_{\rm{l}1}(z)^{-1} \,T_{\rm{l}2}(z)^{-1} +L_1(z^{-1})\,T_{\rm{l}1}(z^{-1})^{-1}\,L_2(z)\,T_{\rm{l}2}(z)^{-1},
\qquad z\in\mathbb T\setminus\{-1,1\}.
\end{equation}
 From the $(2,1)$-entries of the equality in \eqref{3.47} we know that 
\begin{equation}
\label{5.68}
R_1(z)^\dagger\,T_{\rm{l}1}(z)+ T_{\rm{r}1}(z)^\dagger\,L_1(z)=0,\qquad z\in\mathbb T.
\end{equation}
Multiplying both sides of \eqref{5.68} by $[T_{\rm{r}1}(z)^\dagger]^{-1}$ from the left and 
by $T_{\rm{l}1}(z)^{-1}$ from the right, we get
\begin{equation}\label{5.69}
[T_{\rm{r}1}(z)^\dagger]^{-1} R_1(z)^\dagger + L_1(z) \,T_{\rm{l}1}(z)^{-1}=0, \qquad z \in \mathbb T\setminus\{-1,1\}.
\end{equation}
In \eqref{5.69}, after replacing $z$ by $z^{-1}$ we obtain
\begin{equation}\label{5.70}
L_1(z^{-1}) \, T_{\rm{l}1}(z^{-1})^{-1}=-[T_{\rm{r}1}(z^{-1})^\dagger]^{-1} R_1(z^{-1})^\dagger, \qquad z \in \mathbb T\setminus\{-1,1\}.
\end{equation}
Next, using \eqref{2.19}, \eqref{5.70}, and the second equality of \eqref{3.51} in \eqref{5.67}, we get
\begin{equation}
\label{5.71}T_{\rm{l}}(z)^{-1}=T_{\rm{l}1}(z)^{-1} \,T_{\rm{l}2}(z)^{-1}-[T_{\rm{r}1}(z^{-1})^\dagger]^{-1}  R_1(z)\,L_2(z)\,T_{\rm{l}2}(z)^{-1},\qquad z\in\mathbb T\setminus\{-1,1\}.
\end{equation}
Using \eqref{2.19} and the third equality of \eqref{3.51} in the first term
on the right-hand side of \eqref{5.71}, we have
\begin{equation}
\label{5.72}T_{\rm{l}}(z)^{-1}=[T_{\rm{r}1}(z^{-1})^\dagger]^{-1} \left(
I  -  R_1(z)\,L_2(z)\right) T_{\rm{l}2}(z)^{-1},\qquad z\in\mathbb T\setminus\{-1,1\}.
\end{equation}
We remark that \eqref{5.72} implies that  $I  -  R_1(z)\,L_2(z)$ is invertible
because the transmission coefficients are invertible as indicated in Theorem~\ref{theorem3.2}(e). Then,  taking the inverses of both sides
of \eqref{5.72},
we obtain \eqref{5.63}.
Let us next prove \eqref{5.64}. From the $(2,1)$-entries of the equality in \eqref{5.41}, with the help of
\eqref{4.8}, \eqref{5.12}, and \eqref{5.13}, we have
\begin{equation}\label{5.73}
L(z) \,T_{\rm{l}}(z)^{-1}= L_1(z) \,T_{\rm{l}1}(z)^{-1}\, T_{\rm{l}2}(z) ^{-1} +T_{\rm{l}1}(z^{-1})^{-1}\,L_2(z)\,
T_{\rm{l}2}(z)^{-1},  \qquad z\in\mathbb T\setminus\{-1,1\}.
\end{equation}
In \eqref{5.70} by replacing $z^{-1}$ by $z$ and using the resulting equality in \eqref{5.73}, we get
\begin{equation}\label{5.74}
L(z) \,T_{\rm{l}}(z)^{-1}= -[T_{\rm{r}1}(z)^\dagger]^{-1} R_1(z)^\dagger\, T_{\rm{l}2}(z) ^{-1} +T_{\rm{l}1}(z^{-1})^{-1}\,L_2(z)\,
T_{\rm{l}2}(z)^{-1},  \qquad z\in\mathbb T\setminus\{-1,1\}.
\end{equation}
Next, using \eqref{2.19} and the third equality of \eqref{3.51} in the second term on the right-hand side of \eqref{5.74}, we obtain
\begin{equation*}
L(z) \,T_{\rm{l}}(z)^{-1}= -[T_{\rm{r}1}(z)^\dagger]^{-1} R_1(z)^\dagger\, T_{\rm{l}2}(z) ^{-1} +[T_{\rm{r}1}(z)^\dagger]^{-1}L_2(z)\,
T_{\rm{l}2}(z)^{-1},  \qquad z\in\mathbb T\setminus\{-1,1\},
\end{equation*}
which is equivalent to
\begin{equation}\label{5.76}
L(z) \,T_{\rm{l}}(z)^{-1}=[T_{\rm{r}1}(z)^\dagger]^{-1} \left[L_2(z)-R_1(z)^\dagger\right]
T_{\rm{l}2}(z)^{-1},  \qquad z\in\mathbb T\setminus\{-1,1\}.
\end{equation}
Using \eqref{2.19} and the second equality of \eqref{3.51} in \eqref{5.76}, we have
\begin{equation}\label{5.77}
L(z) \,T_{\rm{l}}(z)^{-1}=[T_{\rm{r}1}(z)^\dagger]^{-1} \left[L_2(z)-R_1(z^{-1})\right]
T_{\rm{l}2}(z)^{-1},  \qquad z\in\mathbb T\setminus\{-1,1\}.
\end{equation}
Finally, postmultiplying the equality in \eqref{5.77} by the respective sides of \eqref{5.63}, we obtain \eqref{5.64}.
Let us now prove \eqref{5.65}. From \eqref{2.19} and the fourth equality of \eqref{3.51} we have
$T_{\rm{r}}(z)=T_{\rm{l}}(z^{-1})^\dagger.$ Thus, by taking the matrix adjoint of both sides of \eqref{5.63}, 
then replacing $z$ by $z^{-1}$ in the resulting equality, and then using \eqref{2.19} and the first two
equalities of \eqref{3.51}, we get \eqref{5.65}.
Let us finally prove \eqref{5.66}. From the $(1,2)$-entries of the equality in \eqref{3.50}, we have
\begin{equation}
\label{5.78}
R(z)=-T_{\rm{l}}(z)\,L(z)^\dagger\,[T_{\rm{r}}(z)^\dagger]^{-1},\qquad z\in\mathbb T\setminus\{-1,1\}.
\end{equation}
Using \eqref{2.19} and the first and third equalities of \eqref{3.51} in \eqref{5.78}, we get
\begin{equation}
\label{5.79}R(z)=-T_{\rm{l}}(z)\,L(z^{-1})\,T_{\rm{l}}(z^{-1})^{-1},\qquad z\in\mathbb T\setminus\{-1,1\}.
\end{equation}
Then, on the right-hand side of \eqref{5.79}, we replace $T_{\rm{l}}(z)$ by the right-hand side of \eqref{5.63} and 
we also replace $L(z^{-1})\,[T_{\rm{l}}(z^{-1})]^{-1}$ by the right-hand side of \eqref{5.77} after the substitution $z\mapsto z^{-1}.$
After simplifying the resulting modified version of \eqref{5.79}, we 
obtain \eqref{5.66}.
\end{proof}

\section{Examples}
\label{section6}

In this section we illustrate the theory developed in Section~\ref{section5} with some explicit examples. The goal is to demonstrate the usefulness of the factorization formula and help for a better understanding of the theoretical 
results with some explicit examples.

In the first example, we determine the explicit expressions for scattering coefficients for \eqref{2.17} when the nonhomogeneity is concentrated at a single point, namely when $n=m.$
We recall that we use the term an elementary fragment to refer to a fragment when the nonhomogeneity is concentrated at a single point.
Having those scattering coefficients for an elementary fragment in hand, we can use \eqref{4.8} to obtain the 
explicit expression for the left transition matrix $\Lambda(z)$ and use
\eqref{4.9} to obtain the 
explicit expression for the right transition matrix $\Sigma(z)$ corresponding to an elementary fragment. 
Once we have the transition matrices $\Lambda(z)$ and $\Sigma(z)$ for an elementary fragment,
we can evaluate the left and right transition matrices corresponding to an arbitrary nonelementary fragment, i.e. a fragment where the nonhomogeneity is
concentrated at any finite number of points, and this can be achieved by
using the factorization formulas \eqref{5.41} and \eqref{5.42}.
Having the transition matrices for an arbitrary nonelementary fragment, the corresponding scattering coefficients
can be determined by using \eqref{4.8} and \eqref{4.9} with the help of \eqref{6.27} given in our first example.
In fact, using a symbolic mathematics software system such as Mathematica, the process
can be automated to evaluate the corresponding scattering coefficients explicitly, even though
those expressions become lengthy as the number of nonhomogeneity points increases.

\begin{example}
\label{example6.1}
\normalfont
We consider the special case where the nonhomogeneity is concentrated at the point $n=m,$ i.e.
the case
\begin{equation}
\label{6.1}
\left(a(n),b(n),w(n)\right)=
\left(a_\infty\, I,b_\infty\, I,w_\infty\, I\right),\qquad n\in\mathbb Z\setminus\{m\},
\end{equation}
and we assume that
\begin{equation}
\label{6.2}
\left(a(m),b(m),w(m)\right)\ne 
\left(a_\infty I,b_\infty\, I,w_\infty\, I\right).
\end{equation}
We evaluate
the corresponding scattering coefficients and other relevant quantities.
We use the partitioning described in \eqref{5.1} and \eqref{5.2}. From \eqref{5.2} and
\eqref{5.4} we see that
\begin{equation*}
\left(a_2(n),b_2(n),w_2(n)\right)=
\left(a_\infty I,b_\infty\, I,w_\infty\, I\right),\qquad n\in\mathbb Z,
\end{equation*}
and hence we conclude that the corresponding Jost solutions are given by
\begin{equation}\label{6.4}
f_{\rm{l}2}(z,n)=z^n\,I,
\quad
f_{\rm{r}2}(z,n)=z^{-n}\,I,\qquad n\in\mathbb Z.
\end{equation}
Associated with the Jost solutions in \eqref{6.4}, we have the left transition 
matrix $\Lambda_2(z),$ the right transition matrix
$\Sigma_2(z),$ and the scattering matrix $S_2(z)$ given by
\begin{equation*}
 \Lambda_2=\mathbf I,
\quad \Sigma_2=\mathbf I,\quad S_2(z)=\mathbf I,\qquad z\in\mathbb T\setminus\{-1,1\},
\end{equation*}
where we recall that $\mathbf I$ is the $2q\times 2q$ identity matrix.
Next, we evaluate the scattering data corresponding to \eqref{6.1} and \eqref{6.2}
by writing
\begin{equation}
\label{6.6}
\left(a_1(n),b_1(n),w_1(n)\right)=
\left(a_\infty\, I,b_\infty\, I,w_\infty\, I\right),\qquad n\in\mathbb Z\setminus\{m\},
\end{equation}
\begin{equation}
\label{6.7}
\left(a_1(m),b_1(m),w_1(m)\right)=
\left(a(m) ,b(m) ,w(m)\right).
\end{equation}
We would like to determine the right
Jost solution
$f_{\rm{r}1}(z,n)$ corresponding to the partitioning in \eqref{6.6} and \eqref{6.7}
in terms of the $q\times q$ matrix-valued coefficients
$a(m),$ $b(m),$ $w(m),$ and the scalars
$a_\infty,$ $b_\infty,$ and $w_\infty.$
For this, we proceed as follows.
From \eqref{5.5} and the asymptotics given in 
 \eqref{3.11}, we have
\begin{equation}\label{6.8}
f_{\rm{r}1}(z,n)=z^{-1}\,I,\qquad n\le m-2.
\end{equation}
Evaluating \eqref{5.5} at $n=m-1$ we get
\begin{equation}\label{6.9}
a(m)\,f_{\rm{r}1}(z,m)+b_\infty\,f_{\rm{r}1}(z,m-1)+
a_\infty\,f_{\rm{r}1}(z,m-2)=
\displaystyle
\frac{a_\infty(z+z^{-1})+b_\infty}{w_\infty}
\,
w_\infty\,f_{\rm{r}1}(z,m-1).
\end{equation}
Using \eqref{6.8} in \eqref{6.9}, after some minor simplification, we determine
$f_{\rm{r}1}(z,m)$ as
\begin{equation}\label{6.10}
f_{\rm{r}1}(z,m)=
a_\infty\,a(m)^{-1}\,z^{-m}.
\end{equation}
Next, we evaluate \eqref{5.5} at $n=m$ and we obtain
\begin{equation}\label{6.11}
a_\infty\,f_{\rm{r}1}(z,m+1)+b(m)\,f_{\rm{r}1}(z,m)+
a(m)^\dagger\,f_{\rm{r}1}(z,m-1)=
\displaystyle
\frac{a_\infty(z+z^{-1})+b_\infty}{w_\infty}
\,
w(m)\,f_{\rm{r}1}(z,m).
\end{equation}
Using \eqref{6.8} and \eqref{6.10} in \eqref{6.11}, after some simplification, we obtain
$f_{\rm{r}1}(z,m+1)$ as
\begin{equation}\label{6.12}
f_{\rm{r}1}(z,m+1)=q_{1}\,z^{-m-1}
+q_{2}\, z^{-m}+
q_{3}\, z^{-m+1},
\end{equation}
where we have defined
\begin{equation}\label{6.13}
q_{1}:=
\displaystyle\frac{a_\infty}{w_\infty}\,w(m)\,a(m)^{-1},\quad
q_{2}:=
\displaystyle\frac{b_\infty}{w_\infty}\,w(m)\,a(m)^{-1}-b(m)\,a(m)^{-1},
\end{equation}
\begin{equation}\label{6.14}
q_{3}:=\displaystyle\frac{a_\infty}{w_\infty}\,w(m)\,a(m)^{-1}-\displaystyle\frac{a(m)^\dagger}{a_\infty}.
\end{equation}
From \eqref{5.16} and \eqref{5.17}, using $a(m+1)=a_\infty\,I,$ we get
\begin{equation}\label{6.15}
f_{\rm{r}}(z,m)=f_{\rm{r}1}(z,m),\quad
f_{\rm{r}}(z,m+1)=f_{\rm{r}1}(z,m+1).
\end{equation}
Similarly, from \eqref{5.19} we get
\begin{equation}\label{6.16}
f_{\rm{l}}(z,m)=f_{\rm{l}2}(z,m),\quad
f_{\rm{l}}(z,m+1)=f_{\rm{l}2}(z,m+1).
\end{equation}
Using \eqref{6.4}, \eqref{6.10}, and \eqref{6.12} in
\eqref{6.15} and \eqref{6.16}, we get
\begin{equation}\label{6.17}
f_{\rm{l}}(z,m)=z^m\,I,\quad
f_{\rm{l}}(z,m+1)=z^{m+1}\,I,\quad
f_{\rm{r}}(z,m)=
a_\infty\,a(m)^{-1}\,z^{-m},
\end{equation}
\begin{equation}\label{6.18}
f_{\rm{r}}(z,m+1)=q_{1}\,z^{-m-1}
+q_{2}\, z^{-m}+
q_{3}\, z^{-m+1}.
\end{equation}
Having the values $f_{\rm{l}}(z,m),$ $f_{\rm{l}}(z,m+1),$ $f_{\rm{r}}(z,m),$ $f_{\rm{r}}(z,m+1),$ we can evaluate
the scattering coefficients corresponding to \eqref{6.1} and \eqref{6.2}. This can be accomplished by 
using the fact that the scattering coefficients can be obtained explicitly with the help of the Wronskians appearing 
in \eqref{3.38}--\eqref{3.40} evaluated
at $n=m.$ For example, we can obtain $T_{\rm{r}}(z)$ with the help of 
\begin{equation}
\label{6.19}
a_\infty\,\left(z^{-1}-z\right)T_{\rm{r}}(z)^{-1}=
 f_{\rm{l}}(z^\ast,m)^\dagger\,a_\infty\,f_{\rm{r}}(z,m+1)-
 f_{\rm{l}}(z^\ast,m+1)^\dagger\,a_\infty\,f_{\rm{r}}(z,m),
\end{equation}
which is obtained from the first equality of \eqref{3.39}. Using \eqref{6.17} and \eqref{6.18} in \eqref{6.19}, we get
\begin{equation}
\label{6.20}
T_{\rm{r}}(z)^{-1}=\displaystyle\frac{1}{z^{-1}-z}\left[q_{1}\,z^{-1}+q_{2}+\left(q_{3}-a_\infty\,a(m)^{-1}\right)z \right].
\end{equation}
Similarly, we can obtain $R(z)\,T_{\rm{r}}(z)^{-1}$ with the help of
\begin{equation}
\label{6.21}
a_\infty\left(z-z^{-1}\right)R(z)\,T_{\rm{r}}(z)^{-1}
 =
 f_{\rm{l}}(z,m)^\dagger\,a_\infty\,f_{\rm{r}}(z,m+1)-
 f_{\rm{l}}(z,m+1)^\dagger\,a_\infty\,f_{\rm{r}}(z,m),
\end{equation}
which is obtained from the first equality of \eqref{3.38}.
Using
\eqref{6.17} and \eqref{6.18} in \eqref{6.21}, we get
\begin{equation}
\label{6.22}
R(z)\,T_{\rm{r}}(z)^{-1}=\displaystyle\frac{1}{z-z^{-1}}\left[\left(q_{1}-a_\infty\,a(m)^{-1}\right)z^{-2m-1}+
q_{2}\, z^{-2m}+q_{3}\,z^{-2m+1} \right].
\end{equation}
From the third equality of \eqref{3.51} we have
\begin{equation}
\label{6.23}
T_{\rm{l}}(z)=T_{\rm{r}}(z^\ast)^\dagger,\qquad z\in\mathbb T \setminus\{-1,1\}.
\end{equation}
Hence, using \eqref{6.23}, from \eqref{6.20} we obtain
\begin{equation}
\label{6.24}
T_{\rm{l}}(z)^{-1}=\displaystyle\frac{1}{z^{-1}-z}\left[q_{1}^\dagger\,z^{-1}+q_{2}^\dagger+\left(q_{3}^\dagger-a_\infty\,[a(m)^\dagger]^{-1}\right)z \right].
\end{equation}
From the second equality of \eqref{3.38} we know that
\begin{equation}
\label{6.25}
L(z)\,T_{\rm{l}}(z)^{-1}=-[R(z)\,T_{\rm{r}}(z)^{-1}]^\dagger,
\qquad z\in \mathbb T \setminus\{-1,1\},
\end{equation}
and hence with the help of \eqref{6.22} and \eqref{6.25}, for $z\in\mathbb T \setminus\{-1,1\}$ we obtain
\begin{equation}
\label{6.26}
L(z)\,T_{\rm{l}}(z)^{-1}
=\displaystyle\frac{1}{z^{-1}-z}\left[\left(q_{1}^\dagger-a_\infty[a(m)^\dagger]^{-1}\right)z^{2m+1}+q_{2}^\dagger\,z^{2m}+q_{3}\,z^{2m-1} \right].
\end{equation}
The expressions in \eqref{6.24} and \eqref{6.26} enable us to construct
$\Lambda(z)$ defined in \eqref{4.8}, and
the expressions in \eqref{6.20} and \eqref{6.22} enable us to construct
$\Sigma(z)$ defined in \eqref{4.9}. The corresponding left transmission coefficient $T_{\rm{l}}(z)$
can be obtained from \eqref{6.24} by using the matrix inversion, 
the corresponding transmission matrix $T_{\rm{r}}(z)$
can be obtained from \eqref{6.20} by using the matrix inversion, and the reflection coefficients
in turn can be obtained with the help of matrix products from
\begin{equation}
\label{6.27}
L(z)=
[L(z)\,T_{\rm{l}}(z)^{-1}]\,T_{\rm{l}}(z),\quad
R(z)=
[R(z)\,T_{\rm{r}}(z)^{-1}]\,T_{\rm{r}}(z).
\end{equation}
By comparing \eqref{6.20} and \eqref{6.24}, we see that
$T_{\rm{l}}(z)\not\equiv T_{\rm{r}}(z)$ in the matrix case in general, but the equality holds for $z\in\mathbb T \setminus\{-1,1\}$ if  $a(m)$ is selfadjoint and commutes both with $b(m)$ and $w(m).$

\end{example}

In the next example, we illustrate a
special case of the general Jacobi system \eqref{2.17} corresponding to the matrix Schr\"odinger
equation on the full-line lattice. We again consider the case where the nonhomogeneity is concentrated 
at the single point $n=m.$ When the matrix coefficients $a(n)$ are all equal to the asymptotic value
$a_\infty I,$ there are no nearest-neighbor interactions and the force at $n=m$ is only due to the potential $V(m).$
This simplifies the expressions for the corresponding scattering coefficients. In this case,
our example also demonstrates that
if the matrix $V(m)$ is selfadjoint then
the left and right transmission coefficients become equal to each other.

\begin{example}
\label{example6.2}
\normalfont
Using
\begin{equation}
\label{6.28}
\left(a(n),b(n),w(n)\right)\equiv
\left(-I,V(n)+2I,I\right),\quad \left(a_\infty,b_\infty,w_\infty\right)\equiv
\left(-1,2,1\right),
\end{equation}
we see that \eqref{2.17} reduces to
\begin{equation*}
\phi(z,n+1)+
\phi(z,n-1)=\left[V(n)+
z\,I+z^{-1}\,I\right]
\phi(z,n),\qquad n\in\mathbb Z,
\end{equation*}
where  the first equality of \eqref{1.5} yields
$V(m)^\dagger=V(m).$
If the nonhomogeneity is concentrated at $n=m,$ then the corresponding potential $V(n)$ is zero for
$n\ne m$ and its value is equal to $V(m)$ when $n=m.$
In this special case, using 
\eqref{6.28} in \eqref{6.13} and \eqref{6.14} we get
\begin{equation}
\label{6.30}
q_{1}=I,\quad q_{2}=0,\quad
q_{3}=V(m).
\end{equation}
Using \eqref{6.30} in 
\eqref{6.20}, \eqref{6.22}, \eqref{6.24}, and
\eqref{6.26}, we obtain
\begin{equation}
\label{6.31}
T_{\rm{l}}(z)^{-1}=I-\displaystyle\frac{V(m)}{z-z^{-1}},\quad
T_{\rm{r}}(z)^{-1}=I-\displaystyle\frac{V(m)}{z-z^{-1}},
\qquad z\in\mathbb T\setminus\{-1,1\},
\end{equation}
\begin{equation}
\label{6.32}
L(z)\,T_{\rm{l}}(z)^{-1}=\displaystyle\frac{V(m)\,z^{2m}}{z-z^{-1}},\quad
R(z)\,T_{\rm{r}}(z)^{-1}=\displaystyle\frac{V(m)\,z^{-2m}}{z-z^{-1}},
\qquad z\in\mathbb T\setminus\{-1,1\}.
\end{equation}
Using \eqref{6.31} and \eqref{6.32} in \eqref{4.8} and \eqref{4.9}, we obtain the
corresponding transition matrices as
\begin{equation}
\label{6.33}
\Lambda(z)=\begin{bmatrix}I&0\\
0&I\end{bmatrix}+\displaystyle\frac{1}{z-z^{-1}}\begin{bmatrix}-V(m)&-V(m)\,z^{-2m}\\
\noalign{\medskip}
V(m)\,z^{2m}&V(m)\end{bmatrix},
\qquad z\in\mathbb T\setminus\{-1,1\},
\end{equation}
\begin{equation*}
\Sigma(z)=\begin{bmatrix}I&0\\
0&I\end{bmatrix}+\displaystyle\frac{1}{z-z^{-1}}\begin{bmatrix}V(m)&V(m)\,z^{-2m}\\
\noalign{\medskip}
-V(m)\,z^{2m}&-V(m)\end{bmatrix},
\qquad z\in\mathbb T\setminus\{-1,1\}.
\end{equation*}

\end{example}

In the next example, we show that the matrix-valued left and right
transmission coefficients associated with \eqref{2.17} in general are not equal to each other.
On the other hand, in this example we are in the case where the coefficient matrix
$a(n)$ is selfadjoint and hence Theorem~\ref{theorem4.6} applies.
Thus, in the example, although the left and right transmission coefficients are unequal,
their determinants are equal to each other.
To illustrate our findings in a clear way, we use the special case of \eqref{2.17} corresponding to
the matrix Schr\"odinger equation on the full-line lattice with the nonhomogeneity is concentrated
at two consecutive lattice points.

\begin{example}
\label{example6.3}
\normalfont
When the coefficients in \eqref{2.17} are chosen as in \eqref{6.28}, where the $q\times q$ matrix
$V(n)$ is the zero matrix
for $n\ne m$ and is nonzero only when $n=m$ for a fixed integer $m,$ from \eqref{6.31} we
conclude that the $q\times q$ matrix-valued left and right
transmission coefficients are equal to each other. However, the case where the nonhomogeneity is concentrated
at one point is really a special case. 
In order to illustrate that in general we have $T_{\rm{l}}(z)\not\equiv T_{\rm{r}}(z),$ 
let us consider the case where the nonhomogeneity is concentrated on two consecutive
lattice points. For this,
let us use the partitioning given in \eqref{5.1} and \eqref{5.2} with $m=0,$ and let us
use \eqref{5.3} and \eqref{5.4} in the respective special cases given by
\begin{equation*}
\left(a_1(n),b_1(n),w_1(n)\right):=\begin{cases}
\left(-I,2I,I\right),\qquad n\in\mathbb Z\setminus\{0\},\\
\noalign{\medskip}
\left(-I,2I+V(0),I\right),\qquad n=0,
\end{cases}
\end{equation*}
\begin{equation*}
\left(a_2(n),b_2(n),w_2(n)\right):=\begin{cases}
\left(-I,2I,I\right),\qquad n\in\mathbb Z\setminus\{1\},\\
\noalign{\medskip}
\left(-I,2I+V(1),I\right),\qquad n=1.
\end{cases}
\end{equation*}
The transition matrices $\Lambda_1(z)$ and $\Lambda_2(z)$ appearing in \eqref{5.12}
and \eqref{5.13}, respectively, are obtained with the help of 
\eqref{6.33} as
\begin{equation}
\label{6.37}
\Lambda_1(z)=\begin{bmatrix}I&0\\
0&I\end{bmatrix}+\displaystyle\frac{1}{z-z^{-1}}\begin{bmatrix}-V(0)&-V(0)\\
\noalign{\medskip}
V(0)&V(0)\end{bmatrix},
\qquad z\in\mathbb T\setminus\{-1,1\},
\end{equation}
\begin{equation}
\label{6.38}
\Lambda_2(z)=\begin{bmatrix}I&0\\
0&I\end{bmatrix}+\displaystyle\frac{1}{z-z^{-1}}\begin{bmatrix}-V(1)&-V(1)\,z^{-2}\\
\noalign{\medskip}
V(1)\,z^{2}&V(1)\end{bmatrix},
\qquad z\in\mathbb T\setminus\{-1,1\}.
\end{equation}
In this case, Theorem~\ref{theorem4.6} applies.
By using block-row operations or block-column operations on
\eqref{6.37} and \eqref{6.38}, we can verify that
the determinants of $\Lambda_1(z)$ and $\Lambda_2(z)$ each are equal to $1$ for $z\in\mathbb T\setminus\{-1,1\}.$
Using \eqref{6.37} and \eqref{6.38} in \eqref{5.41}, for
$z\in\mathbb T\setminus\{-1,1\}$
we obtain
the transition matrix $\Lambda(z)$ as
\begin{equation}
\label{6.39}
\Lambda(z)=\begin{bmatrix}I&0\\
0&I\end{bmatrix}+\displaystyle\frac{1}{z-z^{-1}}\begin{bmatrix}-V(0)-V(1)-V(0)\,V(1)\,z&-V(0)-V(1)\,z^{-2}-V(0)\,V(1)\,z^{-1}\\
\noalign{\medskip}
V(0)+V(1)\,z^2+V(0)\,V(1)\,z&V(0)+V(1)+V(0)\,V(1)\,z^{-1}\end{bmatrix}.
\end{equation}
As a concrete case, let us use the matrix size 
 $q=2$ in \eqref{2.17} and 
\begin{equation}
\label{6.40}
V(0)=\begin{bmatrix}1&i\\
-i&2\end{bmatrix},\quad V(1)=\begin{bmatrix}3&-7i\\
7i&4\end{bmatrix}.
\end{equation}
Using \eqref{6.40} in \eqref{6.39}, we evaluate the corresponding left transition matrix $\Lambda(z)$
and determine the corresponding scattering coefficients with the help of \eqref{4.8}.
We obtain $T_{\rm{l}}(z)^{-1}$ from the
$(1,1)$-entry of the matrix $\Lambda(z).$ Then, using the $(2,1)$-entry of
$\Lambda(z)$ and the first equality of \eqref{6.27} we get $L(z).$ We then recover
$T_{\rm{r}}(z)^{-1}$ with the help of the third equality of \eqref{3.51}. Finally, we
obtain $R(z)$ with the help of \eqref{5.79}. We have
\begin{equation}
\label{6.41}
T_{\rm{l}}(z)=\displaystyle\frac{1}{\mathcal P(z)}\,\begin{bmatrix}(z-1)(z+1)(6z+1)&3i z(z-1)(z+1)(z+2)\\
\noalign{\medskip}
-iz(z-1)(z+1)(11z+6)&-(z-1)^2(z+1)(5z+1)\end{bmatrix},
\end{equation}
\begin{equation}
\label{6.42}
T_{\rm{r}}(z)=\displaystyle\frac{1}{\mathcal P(z)}\,\begin{bmatrix}(z-1)(z+1)(6z+1)&i z(z-1)(z+1)(11z+6)\\
\noalign{\medskip}
-3iz(z-1)(z+1)(z+2)&-(z-1)^2(z+1)(5z+1)\end{bmatrix},
\end{equation}
\begin{equation*}
L(z)=\displaystyle\frac{1}{\mathcal P(z)}\,\begin{bmatrix}-z(77z^4+57 z^3+28z^2-8z-1)&i z(z-1)(z+1)(4z-1)(11z+1)\\
\noalign{\medskip}
-iz(z-1)(z+1)(4z-1)(11z+1)&-z(41 z^4+64 z^3+65 z^2-15 z-2)\end{bmatrix},
\end{equation*}
\begin{equation*}
R(z)=\displaystyle\frac{1}{z\,\mathcal P(z)}\,\begin{bmatrix}3z^4-21z^3-110z^2-28z+3&i (z-1)(z+1)(6z^2+21z+7)\\
\noalign{\medskip}
-i(z-1)(z+1)(6z^2+21z+7)&z^4-24z^3-109z^2-25z+4\end{bmatrix},
\end{equation*}
where we have defined
\begin{equation*}
\mathcal P(z):=33 z^4+114 z^3+17 z^2-10 z-1.
\end{equation*}
From \eqref{6.41} and \eqref{6.42} we see that we have $T_{\rm{l}}(z)\not\equiv T_{\rm{r}}(z)$ but the third equality of \eqref{3.51} holds.
The determinants of $T_{\rm{l}}(z)$ and 
$T_{\rm{r}}(z)$ are equal to each other and we have
\begin{equation*}
\det[ T_{\rm{l}}(z)] =\det[ T_{\rm{r}}(z)] =\displaystyle\frac{-(z^2-1)^2}{\mathcal P(z)},\qquad
z\in\mathbb T\setminus\{-1,1\}.
\end{equation*}

\end{example}

In the next example, we further illustrate Example~\ref{example6.1} where the nonhomogeneity is concentrated
at a single point, namely at $n=m.$ We show that we may or may not have the equivalence of the
left and right transmission coefficients and we also demonstrate that
we may or may not have the equivalence of the determinants of those two transmission coefficients.
The result in Theorem~\ref{theorem4.5}(d)
shows that we have 
$\det[T_{\rm{l}}(z)]=\det[T_{\rm{r}}(z)]$
 if and only 
if the right-hand side of \eqref{4.68} is equal to $1,$
and hence our example illustrates various cases based on the specific
choices of the matrix $a(m).$

\begin{example}
\label{example6.4}
\normalfont
Let us consider the nonhomogeneity described in \eqref{6.1} and \eqref{6.2} further.
By comparing \eqref{6.20} and \eqref{6.24}, we see that $T_{\rm{l}}(z)=T_{\rm{r}}(z)$ for $z\in\mathbb T\setminus\{-1,1\}$ if 
we have
\begin{equation*}
q_{1}^\dagger=q_{1},\quad
q_{2}^\dagger=q_{2},\quad
q_{3}^\dagger=q_{3},
\end{equation*}
where we recall that $q_{1},$ $q_{2},$ $q_{3}$ are the quantities defined in \eqref{6.13} and \eqref{6.14}.
Furthermore, from \eqref{4.68} we observe that we have
$\det[T_{\rm{l}}(z)]=\det[T_{\rm{r}}(z)]$ for $z\in\mathbb T\setminus\{-1,1\}$  if
the determinant of $a(m)$ is real valued.
For example, if we use \eqref{6.1} and replace \eqref{6.2} with
\begin{equation}
\label{6.48}
\left(a(m),b(m),w(m)\right)=
\left(a(m)^\dagger,b_\infty\, I,w_\infty\, I\right).
\end{equation}
with $a(m)\ne a_\infty\,I,$ then from \eqref{6.20} and \eqref{6.24} we get
\begin{equation*}
T_{\rm{l}}(z)^{-1}=T_{\rm{r}}(z)^{-1}=\displaystyle\frac{1}{z^{-1}-z}\left[a(m)^{-1}\,z^{-1}-a(m)\,z \right],
\end{equation*}
which yields
\begin{equation*}
T_{\rm{l}}(z)=T_{\rm{r}}(z),\quad
\det[T_{\rm{l}}(z)]=\det[T_{\rm{r}}(z)],\qquad z\in\mathbb T\setminus\{-1,1\}.
\end{equation*}
On the other hand, if we use $a(m)\ne a(m)^\dagger$ in \eqref{6.48} then we get
\begin{equation*}
T_{\rm{l}}(z)^{-1}=\displaystyle\frac{1}{z^{-1}-z}\left[[a(m)^\dagger]^{-1}\,z^{-1}-a(m)\,z \right],\quad
T_{\rm{r}}(z)^{-1}=\displaystyle\frac{1}{z^{-1}-z}\left[a(m)^{-1}\,z^{-1}-a(m)^\dagger\,z \right],
\end{equation*}
which yields $T_{\rm{l}}(z)\ne T_{\rm{r}}(z).$ In this case
we have $\det[T_{\rm{l}}(z)]=\det[T_{\rm{r}}(z)]$ if 
$\det[a(m)]$ is a real constant. For example, if we choose
\begin{equation}
\label{6.52}
a(m)=\begin{bmatrix}1&i\\
0&1\end{bmatrix},
\end{equation}
then, for $z\in\mathbb T\setminus\{-1,1\}$ we get
\begin{equation*}
T_{\rm{l}}(z)^{-1}=\displaystyle\frac{1}{z^{-1}-z}
\begin{bmatrix}z^{-1}-z&-iz\\
\noalign{\medskip}
iz^{-1}&z^{-1}-z\end{bmatrix},
\quad
T_{\rm{r}}(z)^{-1}=\displaystyle\frac{1}{z^{-1}-z}
\begin{bmatrix}z^{-1}-z&-iz^{-1}\\
\noalign{\medskip}
iz&z^{-1}-z\end{bmatrix},
\end{equation*}
which shows that $T_{\rm{l}}(z)\ne T_{\rm{r}}(z).$ We also obtain
\begin{equation*}
\det[T_{\rm{l}}(z)]=\det[T_{\rm{r}}(z)]=\displaystyle\frac{1-2z^2+z^4}{1-3z^2+z^4},\qquad z\in\mathbb T\setminus \{-1,1\},
\end{equation*}
because $\det[a(m)]=1,$ which is real valued even though $a(m)$ is not selfadjoint.
On the other hand, instead of \eqref{6.52} if we use
\begin{equation*}
a(m)=\begin{bmatrix}i&0\\
0&1\end{bmatrix},
\end{equation*}
then for $z\in\mathbb T\setminus\{-1,1\}$ we get
\begin{equation*}
T_{\rm{l}}(z)^{-1}=
\begin{bmatrix}i&0\\
0&1\end{bmatrix},
\quad
T_{\rm{r}}(z)^{-1}=
\begin{bmatrix}-i&0\\
0&1\end{bmatrix},\quad
L(z)=\begin{bmatrix}0&0\\
0&0\end{bmatrix},\quad
R(z)=\begin{bmatrix}0&0\\
0&0\end{bmatrix},
\end{equation*}
which corresponds to the reflectionless case. In this case we have
$\det[a(m)]=i,$ which is not real valued, and hence we have
 $\det[T_{\rm{l}}(z)]\ne\det[T_{\rm{r}}(z)]$ and
 $\det[T_{\rm{l}}(z)]=-i$ and $\det[T_{\rm{r}}(z)]=i.$
Instead of \eqref{6.52}, if we use
\begin{equation*}
a(m)=\begin{bmatrix}1+i&0\\
0&1\end{bmatrix},
\end{equation*}
then for $z\in\mathbb T\setminus\{-1,1\}$ we get
\begin{equation*}
T_{\rm{l}}(z)^{-1}=\displaystyle\frac{1}{z^{-1}-z}
\begin{bmatrix}\displaystyle\frac{1+i}{2}\,z^{-1}-(1+i)\,z&0\\
\noalign{\medskip}
0&z^{-1}-z\end{bmatrix},
\end{equation*}
\begin{equation*}
T_{\rm{r}}(z)^{-1}=\displaystyle\frac{1}{z^{-1}-z}
\begin{bmatrix}\displaystyle\frac{1-i}{2}\,z^{-1}-(1-i)\,z&0\\
\noalign{\medskip}
0&z^{-1}-z\end{bmatrix},
\end{equation*}
\begin{equation*}
\det[T_{\rm{l}}(z)]=\displaystyle\frac{(1-i)(1-z^2)}{1-2z^2},\quad
\det[T_{\rm{r}}(z)]=\displaystyle\frac{(1+i)(1-z^2)}{1-2z^2},
\end{equation*}
and hence we again have
 $T_{\rm{l}}(z)\ne T_{\rm{r}}(z)$ and
 $\det[T_{\rm{l}}(z)]\ne\det[T_{\rm{r}}(z)].$
 These two inequalities are caused by the fact that we have $\det[a(m)]=1+i,$ which is not real valued.
\end{example}

\noindent {\bf Acknowledgments.}
Ricardo Weder is an Emeritus National Researcher of SNII-SECIHTI, M\'exico. He thanks SNII-SECIHTI for the support
related to this research.

\end{document}